\documentclass[11pt,a4paper]{article}
\pdfoutput=1

\usepackage[utf8]{inputenc}
\usepackage[english]{babel}
\usepackage[T1]{fontenc}
\usepackage{lmodern}
\usepackage[margin=0.78in]{geometry}
\usepackage{amsmath,amssymb,amsfonts,amsthm,mathtools,bm}
\usepackage{graphicx}
\usepackage{booktabs,array,multirow,tabularx}
\usepackage{xcolor}
\usepackage{microtype}
\usepackage[colorlinks=true,linkcolor=blue!60!black,citecolor=blue!60!black,urlcolor=blue!60!black]{hyperref}
\usepackage[numbers,sort&compress]{natbib}

\newcommand{\Tr}{\mathrm{Tr}}
\newcommand{\Var}{\mathrm{Var}}
\newcommand{\ad}{\mathrm{ad}}
\newcommand{\sinc}{\mathrm{sinc}}
\newcommand{\erf}{\mathrm{erf}}
\newcommand{\diag}{\mathrm{diag}}
\newcommand{\adj}{\mathrm{adj}}
\newcommand{\dd}{\mathrm{d}}
\newcommand{\Tone}{\mathrm{T}_1}
\newcommand{\Ttwo}{\mathrm{T}_2}
\newcommand{\ket}[1]{\lvert #1 \rangle}
\newcommand{\bra}[1]{\langle #1 \rvert}
\newcommand{\braket}[2]{\langle #1 \vert #2 \rangle}
\newcommand{\dyad}[2]{\lvert #1 \rangle\langle #2 \rvert}
\newcommand{\norm}[1]{\lVert #1 \rVert}
\newcommand{\abs}[1]{\lvert #1 \rvert}
\providecommand{\doi}[1]{doi:\href{https://doi.org/#1}{\nolinkurl{#1}}}
\providecommand{\SI}[2]{#1\,\mathrm{#2}}
\providecommand{\micro}{\mu}

\theoremstyle{plain}
\newtheorem{theorem}{Theorem}
\newtheorem{lemma}{Lemma}
\newtheorem{proposition}{Proposition}

\theoremstyle{definition}
\newtheorem{definition}{Definition}
\newtheorem{protocol}{Protocol}

\theoremstyle{remark}
\newtheorem{remark}{Remark}
\newtheorem{example}{Example}

\title{\Large\textbf{Directional gate-error asymmetry and universal selectivity bounds in non-unital open quantum systems}}

\author{Amritash Patra\\[4pt]
\normalsize\textit{Independent Researcher, Bishnupur, West Bengal 722122, India}\\
\normalsize\href{mailto:amritashpatraedu@gmail.com}{\texttt{amritashpatraedu@gmail.com}} \quad
ORCID: \href{https://orcid.org/0009-0003-5244-9387}{\texttt{0009-0003-5244-9387}}}
\date{}

\begin{document}

\maketitle
\begin{abstract}
\noindent
Clifford randomized benchmarking and average gate fidelity assign a single trajectory-independent decay rate to a noisy quantum gate at leading order in the dimensionless damping exposure $\varepsilon = \gamma T$, concealing state-dependent transition-error asymmetries in untwirled quantum circuits. While state-dependent amplitude-damping errors are well known in open quantum systems, a trajectory-level analytical framework that bounds directional transition selectivity across arbitrary pulse shapes and Markovian environments has been lacking. Using first-order toggling-frame perturbation theory for a two-level qubit coupled to a Markovian bath, we show that the antipodal survival-error slopes $\chi_+$ and $\chi_-$ along any piecewise-continuous control path depend exclusively on its time-averaged Bloch vector $\bm{m}$ and transverse residence covariance matrix $G$, obeying the complete-positivity cofactor bound $\beta(\bm{r}) \le 1$. On a standard polar $\sqrt{X}$ gate, longitudinal $\mathrm{T}_1$ relaxation induces a $12.23 : 1$ directional survival-error split ($\kappa_{\mathrm{T}_1} = (3\pi - 8)/(3\pi + 8) \approx 0.081767$), whereas time-symmetric $X$ rotations maintain an exact $1 : 1$ balance ($\kappa_{\mathrm{T}_1} = 1$). Across all completely positive Markovian generators, coupling between non-unital drift and unital diffusion enforces a selectivity bound $\kappa_\star = (\pi - 2\sqrt{2})/(\pi + 2\sqrt{2}) \approx 0.052456$ ($19.06 : 1$) for time-independent single-qubit $\pi$ swaps. We apply this framework to Gaussian and derivative-removal-by-adiabatic-gate (DRAG) waveforms on superconducting transmons and trapped-ion qubits, extend the analysis to finite damping, pulse dwell, colored thermal baths, and qutrit leakage, and construct two-branch differential lock-in protocols that isolate directional error splits immune to first-order state-preparation and measurement errors.
\end{abstract}

\section{Introduction}
\label{sec:intro}

Standard quantum gate certification protocols on superconducting and trapped-ion processors---most notably Clifford randomized benchmarking (RB)~\cite{Emerson2005,Knill2008,Magesan2011} and interleaved RB~\cite{Magesan2012}---compress a noisy quantum operation into a single scalar decay parameter: the Haar-averaged gate infidelity $r_{\mathrm{avg}}$~\cite{Horodecki1999,Nielsen2002}, or equivalently the entanglement infidelity $1 - F_e$~\cite{Schumacher1996}. By deliberately twirling gate sequences over the Clifford or Pauli group~\cite{Geller2013,Wallman2016,Katabarwa2017,Hashim2021}, these benchmarks symmetrize the physical noise channel into an effective depolarizing or Pauli channel, erasing the non-unital polarization drift vector $\bm{t}$ of the Pauli transfer matrix (PTM). At leading order in the dimensionless damping exposure $\varepsilon = \gamma T \ll 1$, both the entanglement infidelity and the average gate infidelity depend solely on the scalar trace rate $\Gamma$ of the Gorini--Kossakowski--Sudarshan--Lindblad (GKSL) dissipator~\cite{Gorini1976,Lindblad1976} and are invariant under the unitary control trajectory $U_0(s)$ implementing the gate~\cite{Pedersen2007,Abad2022}:
\begin{equation}
  1 - F_e = \frac{1}{2}\,\varepsilon\Gamma + \mathcal{O}(\varepsilon^2), \qquad
  r_{\mathrm{avg}} = \frac{1}{3}\,\varepsilon\Gamma + \mathcal{O}(\varepsilon^2).
  \label{eq:intro_scalar_invariance}
\end{equation}
In untwirled quantum algorithms, repetitive stabilizer syndrome-extraction cycles, active qubit reset sequences, and dynamical decoupling protocols, however, physical microwave and optical pulses execute without random Pauli twirling. Whenever the environment is non-unital ($\bm{t} \neq \bm{0}$), the forward survival error probability $p_+(\varepsilon)$ of $\ket{0}$ and the backward survival error probability $p_-(\varepsilon)$ of $\ket{1}$ become strongly trajectory-dependent and split at first order in $\varepsilon$.

\begin{figure*}[t]
  \centering
  \includegraphics[width=\textwidth]{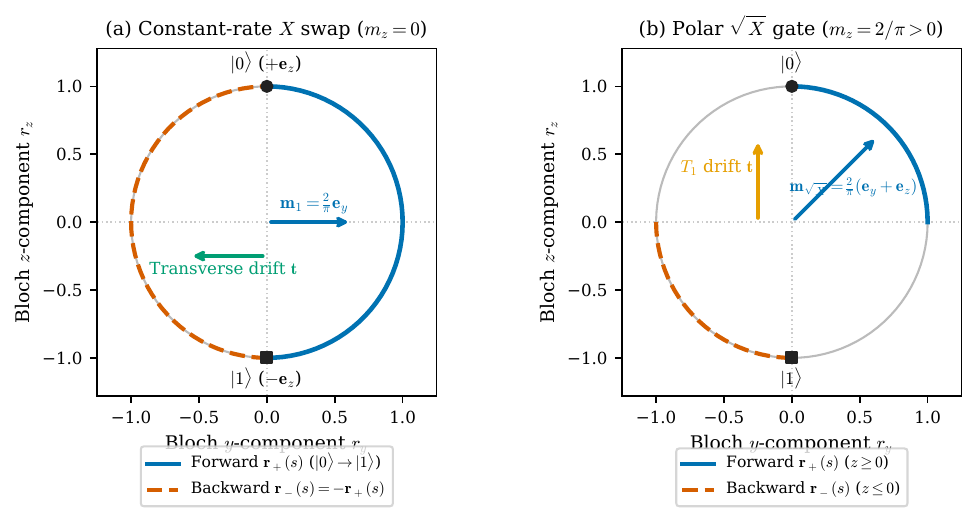}
  \caption{Trajectory mean Bloch vector $\bm{m}$ versus non-unital bath drift $\bm{t}$ in the weak-damping Markovian regime. (a) Constant-rate $X$ rotation ($H_1 = -\frac{\pi}{2}\sigma_x$, $q = +1$): the forward trajectory $\bm{r}_+(s)$ (solid blue) and backward trajectory $\bm{r}_-(s) = -\bm{r}_+(s)$ (dashed red) spend equal time in the northern and southern hemispheres ($m_z = 0$, $\bm{m}_1 = \frac{2}{\pi}\bm{e}_y$), coupling only to transverse drift $t_y \neq 0$. (b) Polar $\sqrt{X}$ gate ($H_{\sqrt{X}} = -\frac{\pi}{4}\sigma_x$, $Q = \pi/2$): the forward arc from $\ket{0}$ remains entirely in the northern hemisphere $z \ge 0$ whereas the antipodal arc from $\ket{1}$ traverses $z \le 0$ with $\bm{m}_{\sqrt{X}} = \frac{2}{\pi}(\bm{e}_y + \bm{e}_z)$, so longitudinal $T_1$ relaxation ($\bm{t} = 2\bm{e}_z$) induces a $12.23 : 1$ directional survival-error split ($\kappa_{T_1} \approx 0.081767$) while symmetric $X$ rotations remain immune ($\kappa_{T_1} = 1$).}
  \label{fig:fig1}
\end{figure*}

The physical origin of this directional gate-error asymmetry is illustrated in Figure~\ref{fig:fig1}. Consider a superconducting transmon qubit undergoing native longitudinal $T_1$ energy relaxation toward the ground state $\ket{0}$ (at the north pole $+\bm{e}_z$ of the Bloch sphere, corresponding to an upward non-unital drift vector $\bm{t} = 2\bm{e}_z$). During a symmetric constant-rate $X$ rotation ($H_1 = -\frac{\pi}{2}\sigma_x$, Figure~\ref{fig:fig1}(a)), the forward trajectory starting from $\ket{0}$ and the backward trajectory starting from $\ket{1}$ spend equal durations in the northern ($z > 0$) and southern ($z < 0$) hemispheres; the time-averaged longitudinal polarization vanishes ($m_z = 0$), and the two directional survival error slopes are identical ($\kappa_{T_1} := \chi_+/\chi_- = 1$). By contrast, during a standard $\pi/2$ rotation $\sqrt{X}$ ($H_{\sqrt{X}} = -\frac{\pi}{4}\sigma_x$, Figure~\ref{fig:fig1}(b))---the universal hardware primitive from which arbitrary single-qubit $\mathrm{SU}(2)$ operations are compiled via virtual-$Z$ phase updates~\cite{McKay2017}---the trajectory starting from $\ket{0}$ remains entirely within the northern hemisphere ($z \ge 0$, near the ground state where $T_1$ decay is suppressed), whereas the antipodal trajectory starting from $\ket{1}$ remains entirely within the southern hemisphere ($z \le 0$, near the excited state where spontaneous emission is rapid). Because the time-averaged longitudinal Bloch coordinate is $m_z = +2/\pi > 0$, ordinary $T_1$ relaxation induces a $12.23 : 1$ ratio between the backward and forward survival error rates ($\chi_\pm = \frac{3}{8} \mp \frac{1}{\pi}$, $\kappa_{T_1} = (3\pi - 8)/(3\pi + 8) \approx 0.081767$), even though scalar randomized benchmarking assigns both trajectories the exact same average gate infidelity.

\subsection{Summary of main results and physical implications}
\label{sec:intro_summary}

In this article, we establish a general analytical framework governing state-dependent directional survival errors and their universal complete-positivity bounds across open quantum systems:
\begin{enumerate}
  \item \emph{Pure-state jump-variance law and two-moment trajectory decomposition (Proposition~\ref{prop:variance_law} and Theorem~\ref{thm:two_moment}):} Using first-order toggling-frame perturbation theory for a two-level qubit ($d = 2$) coupled to a time-independent Markovian environment, we show that the instantaneous rate of pure-state survival fidelity loss equals the quantum fluctuation variance of the active traceless jump operators along the control trajectory (Proposition~\ref{prop:variance_law}). Consequently, the antipodal survival error slopes $\chi_\pm = \lim_{\varepsilon \to 0^+} p_\pm(\varepsilon)/\varepsilon$ along any piecewise-continuous Bloch trajectory $\bm{r}(s) \in S^2$ are determined exclusively by two geometric moments of the control path: its time-averaged Bloch polarization vector $\bm{m}$, which couples linearly to the non-unital bath drift $\bm{t}$, and its transverse residence covariance matrix $G$, which contracts with the unital bath damping tensor $S$ according to $\chi_\pm(\mathcal{D};\bm{r}) = \frac{1}{2}\Tr(SG) \mp \frac{1}{4}\bm{m}\cdot\bm{t}$. Maximizing the normalized directional contrast $\abs{\chi_+ - \chi_-}/(\chi_+ + \chi_-)$ over all completely positive GKSL generators yields the sharp geometric cofactor bound $\beta(\bm{r}) = \sqrt{\bm{m}^T G\bm{m}/\det G} \le 1$ and the corresponding survival-error selectivity floor $\kappa(\bm{r}) = (1 - \beta(\bm{r}))/(1 + \beta(\bm{r}))$ (Theorem~\ref{thm:two_moment}).
  \item \emph{Universal selectivity bound for autonomous swaps and constant-axis rotations (Theorem~\ref{thm:swap_floor} and Proposition~\ref{prop:midpoint_sinc}):} Specializing to time-independent single-qubit $\pi$ inversions (autonomous swaps, Theorem~\ref{thm:swap_floor}), we prove that the directional survival-error ratio $\min(\chi_+,\chi_-)/\max(\chi_+,\chi_-)$ across all Markovian environments is bounded below by the universal constant
  \begin{equation*}
    \kappa_\star = \frac{\pi - 2\sqrt{2}}{\pi + 2\sqrt{2}} \approx 0.052456 \qquad (\kappa_\star^{-1} \approx 19.0635 : 1),
  \end{equation*}
  which is uniquely saturated at odd winding $\abs{q} = 1$ by the chiral jump operators $L_\pm = (\sigma_x \pm i\sqrt{2}\sigma_z)/\sqrt{6}$. For arbitrary constant-axis rotations $H(Q,\vartheta)$, block-diagonalization in the trajectory midpoint frame yields an exact closed-form midpoint-sinc formula for $\beta(\abs{Q},\vartheta)$ (Proposition~\ref{prop:midpoint_sinc}), revealing both the $12.23 : 1$ native-$T_1$ split on polar $\sqrt{X}$ gates and a winding-independent $19 : 3 \approx 6.33 : 1$ native-$T_1$ split ($\kappa_{T_1} = 3/19 \approx 0.157895$) on single-pulse tilted Hadamard gates.
  \item \emph{Shaped AWG pulses, SPAM-invariant lock-in metrology, and analytical extensions (Sections~\ref{sec:shaped_pulses}--\ref{sec:extensions}):} We apply the framework to baseline-subtracted truncated Gaussian and DRAG waveforms~\cite{McKay2017,Motzoi2009} (Section~\ref{sec:shaped_pulses}), evaluate quantitative hardware predictions for superconducting transmons and $^{171}\mathrm{Yb}^+$ clock qubits alongside two-branch differential lock-in protocols that cancel additive state-preparation and measurement (SPAM) offsets to $\mathcal{O}(\varepsilon^2)$ for both transverse-drift swaps ($8.97\times$ plug-in variance reduction) and longitudinal-$\mathrm{T}_1$ polar $\sqrt{X}$ gates (Section~\ref{sec:experimental}), and derive analytical results for finite damping strength (where entanglement fidelity remains winding-independent through $\mathcal{O}(\eta^3)$ before splitting at $\mathcal{O}(\eta^4)$), non-autonomous pulse dwell, colored thermal baths, and multilevel qutrit cascade relaxation (Section~\ref{sec:extensions}).
\end{enumerate}

While state-dependent amplitude-damping errors ($\ket{1} \to \ket{0}$ relaxation asymmetric with respect to $\ket{0}$ versus $\ket{1}$) and the sensitivity of incoherently damped gates to pulse profiles are well known in quantum control and randomized benchmarking~\cite{Wallman2015,Sheldon2016,McKay2017,Wood2018}, our contribution lies in establishing (i) the exact geometric two-moment trajectory law $\chi_\pm = \frac{1}{2}\Tr(SG) \mp \frac{1}{4}\bm{m}\cdot\bm{t}$, (ii) the complete-positivity cofactor bound $\beta(\bm{r}) \le 1$, and (iii) the time-independent single-qubit swap selectivity bound $\kappa_\star = (\pi - 2\sqrt{2})/(\pi + 2\sqrt{2})$. These results connect three lines of research in quantum gate characterization and open-system control. First, in the theory of Lindblad fidelity relations, Pedersen \emph{et al.}~\cite{Pedersen2007} and Abad \emph{et al.}~\cite{Abad2022} showed that first-order average gate infidelity and entanglement infidelity are path-independent and proportional to the dissipator trace $\Gamma$, while subsequent studies have analyzed unitarity and non-unital noise in randomized benchmarking and leakage~\cite{Wallman2015,Sheldon2016,Wood2018}, thermodynamic dissipation bounds~\cite{VanVu2024}, higher-order nonlinearities in open qudit gates~\cite{Hartmann2025}, and Pauli-twirled Lindbladian noise symmetries~\cite{Malekakhlagh2026,Kattemolle2026}. Because scalar fidelity benchmarks and Pauli-twirled channels average over input states, the non-unital drift vector $\bm{t}$ drops out at first order; our two-moment trajectory law resolves the untwirled regime where $\bm{t}$ governs state-to-state transition asymmetry. Second, in quantum control and tomography, generalized filter-function formalisms~\cite{Green2013,PazSilva2014,Koch2016} characterize decoherence via spectral overlap integrals in the toggling frame, while Gate Set Tomography (GST)~\cite{Greenbaum2015,BlumeKohout2017,Nielsen2021,Proctor2017} reconstructs full process matrices numerically. Here we show that in the Markovian regime, the zero-frequency toggling-frame filter moments $(\bm{m}, G)$ obey the closed-form matrix cofactor inequality $\adj(G) - \bm{m}\bm{m}^T \succeq 0$, which analytically bounds directional gate selectivity for arbitrary pulse shapes. Third, in biased-noise fault-tolerant architectures such as the XZZX surface code~\cite{Aliferis2009,Tuckett2018,BonillaAtaides2021}, decoder weights are tailored to the Pauli noise bias $\eta_Z = p_Z/(p_X + p_Y)$ between dephasing and bit-flip errors. Directional transition-error asymmetry ($p_+ \neq p_-$) represents a complementary state-dependent relaxation bias within the bit-flip sector: incorporating asymmetric $\ket{0} \leftrightarrow \ket{1}$ transition weights into minimum-weight perfect-matching (MWPM) or belief-propagation decoders avoids the systematic logical-error underestimation of Pauli-twirled approximations (Appendix~\ref{app:lockin_theory}).

The remainder of this paper is organized as follows. Section~\ref{sec:master_eq} introduces the time-rescaled GKSL master equation in the qubit rotating frame, proves the pure-state jump-variance law, and illustrates the formalism with worked calculations for native $T_1$ relaxation on a $\sqrt{X}$ gate and extremal chiral dissipation on an $X$ gate. Section~\ref{sec:two_moment} establishes the general two-moment trajectory law and the geometric cofactor asymmetry bound $\beta(\bm{r}) \le 1$ (Theorem~\ref{thm:two_moment}). Section~\ref{sec:swap_bound} proves the universal selectivity bound $\kappa_\star$ for time-independent swaps (Theorem~\ref{thm:swap_floor}) and derives closed-form midpoint-sinc expressions for arbitrary constant-axis rotations, polar $\sqrt{X}$ gates, and single-pulse Hadamard gates (Proposition~\ref{prop:midpoint_sinc}). Section~\ref{sec:shaped_pulses} extends the theory to realistic microwave arbitrary-waveform-generator (AWG) pulses with truncated Gaussian and DRAG envelopes. Section~\ref{sec:experimental} evaluates quantitative hardware predictions for superconducting transmons (\texttt{ibm\_kyiv}) and $^{171}\mathrm{Yb}^+$ trapped-ion processors and presents the six-circuit signed-drive differential lock-in protocol. Section~\ref{sec:extensions} develops analytical extensions to finite damping strength, non-autonomous pulse dwell, colored thermal baths, and qutrit leakage, and Section~\ref{sec:conclusion} summarizes our conclusions. Self-contained mathematical proofs, Redfield and Slepian derivations, statistical Cram\'er--Rao analyses, and numerical verification tables across $1{,}203{,}200$ random CPTP baths are provided in Appendices~\ref{app:superoperator_trace}--\ref{app:numerical_audit}.

\section{Open-system master equation, pure-state variance law, and physical examples}
\label{sec:master_eq}

We formulate our analysis in the standard weak-damping Markovian regime relevant to high-fidelity single-qubit gates. Specifically, we consider a two-level quantum system ($\mathcal{H} \cong \mathbb{C}^2$, $d = 2$, neglecting leakage to higher excited levels such as $\ket{2}$, which is below $3.1 \times 10^{-6}$ per gate under DRAG control~\cite{McKay2017}) evolving over a gate duration $T > 0$ under a piecewise-continuous control Hamiltonian $H(s)$ and a time-independent rotating-frame Markovian Lindblad dissipator $\mathcal{D}$ with dimensionless noise exposure $\varepsilon := \gamma T \ll 1$, where $\gamma > 0$ denotes the characteristic bath relaxation rate. Throughout the analytical derivations, we set $\hbar = 1$ and parameterize evolution by the dimensionless gate time $s := t/T \in [0,1]$; when specializing to time-independent constant-amplitude drives in Section~\ref{sec:swap_bound}, the control pulse operates continuously at the quantum speed limit ($T = T_{\mathrm{QSL}}$) without idle dwell.

\subsection{Time-rescaled GKSL master equation and Dyson remainder}
\label{sec:gksl_dyson}

On an operator space over $\mathcal{H} \cong \mathbb{C}^d$ ($d \ge 2$) equipped with the Hilbert--Schmidt inner product $\langle\langle A \vert B \rangle\rangle := \Tr(A^\dagger B)$, the time-rescaled GKSL master equation~\cite{Gorini1976,Lindblad1976} for the dynamical propagator $\Phi_\varepsilon(s)$ is
\begin{equation}
  \frac{d}{ds}\Phi_\varepsilon(s) = \mathcal{L}_\varepsilon(s) \circ \Phi_\varepsilon(s), \qquad
  \mathcal{L}_\varepsilon(s) = -i\,\ad_{H(s)} + \varepsilon\,\mathcal{D}(s),
  \label{eq:gksl_rescaled}
\end{equation}
with initial condition $\Phi_\varepsilon(0) = \mathcal{I}$ (where $\mathcal{I}$ denotes the superoperator identity), commutator superoperator $\ad_{H(s)}(\rho) := [H(s),\rho]$, and Lindblad dissipator
\begin{equation}
  \begin{aligned}
    \mathcal{D}(s)(\rho) := \sum_\alpha \gamma_\alpha(s)\Biggl( &L_\alpha(s)\rho L_\alpha(s)^\dagger \\
    &- \frac{1}{2}\bigl\{L_\alpha(s)^\dagger L_\alpha(s), \rho\bigr\} \Biggr),
  \end{aligned}
  \label{eq:dissipator_def}
\end{equation}
where $\gamma_\alpha(s) \ge 0$. Decomposing each jump operator into its trace and traceless parts,
\begin{equation}
  \begin{aligned}
    L_\alpha(s) &= c_\alpha(s)\,\mathbb{I}_d + \tilde{L}_\alpha(s), \\
    c_\alpha(s) &:= \frac{1}{d}\Tr\bigl(L_\alpha(s)\bigr), \qquad \Tr\bigl(\tilde{L}_\alpha(s)\bigr) = 0,
  \end{aligned}
  \label{eq:jump_traceless_decomp}
\end{equation}
separates the generator as $\mathcal{D}(s)(\rho) = \tilde{\mathcal{D}}(s)(\rho) - i[\Delta H_{\mathrm{Lamb}}(s),\rho]$, where the traceless jump operators $\tilde{L}_\alpha(s)$ set the irreversible dissipation rate and the scalar trace parts $c_\alpha(s)\mathbb{I}_d$ generate an effective Lamb-like Hamiltonian shift
\begin{equation*}
  \Delta H_{\mathrm{Lamb}}(s) = \frac{1}{2i}\sum_\alpha \gamma_\alpha(s)\Bigl(c_\alpha^*(s)\tilde{L}_\alpha(s) - c_\alpha(s)\tilde{L}_\alpha(s)^\dagger\Bigr),
\end{equation*}
which vanishes whenever $L_\alpha$ is traceless, and is explicitly retained when analyzing non-traceless rank-$1$ nilpotent jump operators such as $L_\star := \dyad{\psi(\pi/8)}{\psi(7\pi/8)} = -\frac{\sqrt{2}}{4}\mathbb{I}_2 - \frac{i\sqrt{2}}{4}\sigma_x - \frac{1}{2}\sigma_z$ (where $\ket{\psi(\theta)} := \cos\theta\ket{0} + \sin\theta\ket{1}$, $c_\star = \frac{1}{2}\Tr(L_\star) = -\frac{\sqrt{2}}{4}$, and $\Gamma_\star = \frac{3}{4}$) in Section~\ref{sec:extensions} and Figure~\ref{fig:fig5}(a). The instantaneous traceless dissipation rate and its gate-integrated average are
\begin{equation}
  \begin{aligned}
    \Gamma(s) &:= \sum_\alpha \gamma_\alpha(s)\,\Tr\Bigl(\tilde{L}_\alpha(s)^\dagger \tilde{L}_\alpha(s)\Bigr) \ge 0, \\
    \Gamma_{\mathrm{tot}} &:= \int_0^1 \Gamma(s)\,\dd s,
  \end{aligned}
  \label{eq:gamma_trace_def}
\end{equation}
while the unshifted jump norm is $\Gamma_{\mathrm{full}}(s) := \sum_\alpha \gamma_\alpha(s)\Tr(L_\alpha(s)^\dagger L_\alpha(s)) = \Gamma(s) + d\sum_\alpha \gamma_\alpha(s)\abs{c_\alpha(s)}^2$, with $\Gamma_{\mathrm{full,tot}} := \int_0^1 \Gamma_{\mathrm{full}}(s)\,\dd s$.

Equation~\eqref{eq:gksl_rescaled} is written in the frame rotating at the qubit transition frequency $\omega_0$. When the bare transition frequency far exceeds the Rabi drive amplitude ($\omega_0 \gg \Omega_{\mathrm{Rabi}}$) and the bath correlation time satisfies $\tau_B \ll \Omega_{\mathrm{Rabi}}^{-1}$ so that the environmental noise power spectrum is flat across the Mollow triplet sidebands $\omega_0 \pm \Omega_{\mathrm{Rabi}}$, fast counter-rotating secular terms average out under the rotating-wave approximation (RWA). In this regime, spontaneous emission $L_\downarrow = \dyad{0}{1}$, thermal absorption $L_\uparrow = \dyad{1}{0}$, and pure dephasing $L_\phi = \sigma_z/\sqrt{2}$ remain time-independent in the rotating frame, while zero-duration virtual-$Z$ phase updates~\cite{McKay2017} rotate the azimuthal drive axis in software. Transforming to the unitary toggling frame $\tilde{\Phi}_\varepsilon(s) := \mathcal{U}_0(s)^\dagger \Phi_\varepsilon(s)$ with $\mathcal{U}_0(s)(\rho) := U_0(s)\rho U_0(s)^\dagger$ and $\hat{\mathcal{D}}(s) := \mathcal{U}_0(s)^\dagger \mathcal{D}(s)\mathcal{U}_0(s)$, a first-order Dyson expansion yields
\begin{equation}
  \tilde{\Phi}_\varepsilon(1) = \mathcal{I} + \varepsilon \int_0^1 \hat{\mathcal{D}}(s)\,\dd s + \mathcal{R}_2(\varepsilon).
  \label{eq:dyson_first_order}
\end{equation}
Because the induced operator norm of the dissipator obeys $\norm{\mathcal{D}(s)}_{\mathrm{op}} \le 2\Gamma_{\mathrm{full}}(s)$ (reducing to $2\Gamma(s)$ for traceless jumps $c_\alpha \equiv 0$), the second-order Dyson remainder is rigorously bounded at all finite noise exposures $\varepsilon \ge 0$ by
\begin{equation}
  \begin{aligned}
    \norm{\mathcal{R}_2(\varepsilon)}_{\mathrm{op}} &\le 2\,\varepsilon^2 \Gamma_{\mathrm{full,tot}}^2 \exp\bigl(2\varepsilon\Gamma_{\mathrm{full,tot}}\bigr) \\
    &\xrightarrow{c_\alpha \equiv 0} 2\,\varepsilon^2 \Gamma_{\mathrm{tot}}^2 \exp\bigl(2\varepsilon\Gamma_{\mathrm{tot}}\bigr).
  \end{aligned}
  \label{eq:dyson_remainder_bound}
\end{equation}

\subsection{Trajectory invariance of average gate fidelity and pure-state jump-variance law}
\label{sec:variance_law}

To contrast scalar gate benchmarks with state-resolved directional errors, recall that the entanglement fidelity of $\Phi_\varepsilon(1)$ relative to the target unitary $U_0(1)$ is $F_e(\Phi_\varepsilon, U_0(1)) := d^{-2}\Tr(\mathcal{U}_0(1)^\dagger \Phi_\varepsilon)$~\cite{Schumacher1996,Pedersen2007}, and the Haar-averaged gate infidelity is $r_{\mathrm{avg}} := \frac{d}{d+1}(1 - F_e)$~\cite{Horodecki1999,Nielsen2002}. Because the Hamiltonian superoperator $\ad_{H(s)}$ is traceless and $\Tr(\mathcal{D}(s)) = -d\,\Gamma(s)$ on the $d^2$-dimensional operator space (see Appendix~\ref{app:superoperator_trace} for the full derivation~\cite{Pedersen2007,Abad2022,Culver1966,Wolf2008}), Jacobi's determinant identity implies that the Liouville determinant of the noisy channel is locked to the integrated trace rate at all orders in $\varepsilon \ge 0$:
\begin{equation}
  \det \Phi_\varepsilon = \exp\bigl(-d\,\varepsilon\,\Gamma_{\mathrm{tot}}\bigr).
  \label{eq:liouville_determinant}
\end{equation}
Expanding Eq.~\eqref{eq:dyson_first_order} to first order in $\varepsilon$ recovers the first-order trajectory invariance of entanglement fidelity and average gate infidelity~\cite{Pedersen2007,Abad2022}:
\begin{equation}
  \begin{aligned}
    1 - F_e\bigl(\Phi_\varepsilon, U_0(1)\bigr) &= \frac{\varepsilon\,\Gamma_{\mathrm{tot}}}{d} + \mathcal{O}(\varepsilon^2), \\
    r_{\mathrm{avg}}\bigl(\Phi_\varepsilon, U_0(1)\bigr) &= \frac{\varepsilon\,\Gamma_{\mathrm{tot}}}{d + 1} + \mathcal{O}(\varepsilon^2),
  \end{aligned}
  \label{eq:first_order_scalar_invariance}
\end{equation}
both of which are completely independent of the control trajectory $U_0(s)$. Furthermore, by Schur's lemma, $\Tr(\mathcal{D}) = -d\,\Gamma$ is, up to scalar multiplication, the unique $\mathrm{SU}(d)$-conjugation-invariant linear functional on trace-preserving superoperators (Appendix~\ref{app:superoperator_trace}).

In an untwirled circuit initialized in a specific pure state $\ket{\psi_{\mathrm{in}}}$, however, the state-to-state survival error probability $p_{\mathrm{err}}(\varepsilon;\psi_{\mathrm{in}}) := 1 - \bra{\psi_{\mathrm{out}}}\Phi_\varepsilon(\dyad{\psi_{\mathrm{in}}}{\psi_{\mathrm{in}}})\ket{\psi_{\mathrm{out}}} = \varepsilon\,\chi(\psi_{\mathrm{in}}) + \mathcal{O}(\varepsilon^2)$ relative to the ideal output state $\ket{\psi_{\mathrm{out}}} = U_0(1)\ket{\psi_{\mathrm{in}}}$ is governed by a different physical law. Let $\ket{\psi(s)} := U_0(s)\ket{\psi_{\mathrm{in}}}$ and $P(s) := \dyad{\psi(s)}{\psi(s)}$ denote the evolving pure state along the control trajectory. Substituting $\dyad{\psi_{\mathrm{out}}}{\psi_{\mathrm{out}}} = \mathcal{U}_0(1)(P(0))$ into the Dyson expansion~\eqref{eq:dyson_first_order} gives
\begin{equation}
  \begin{aligned}
    \bra{\psi_{\mathrm{out}}}\Phi_\varepsilon(P(0))\ket{\psi_{\mathrm{out}}} &= \Tr\Bigl(P(0)\,\tilde{\Phi}_\varepsilon(1)(P(0))\Bigr) \\
    &\hspace{-2.2em}= 1 + \varepsilon \int_0^1 \Tr\Bigl(P(s)\,\mathcal{D}(s)(P(s))\Bigr)\,\dd s + \mathcal{O}(\varepsilon^2).
  \end{aligned}
  \label{eq:survival_dyson}
\end{equation}
Because $\Tr(P(s)[\Delta H_{\mathrm{Lamb}}(s), P(s)]) = 0$ and $P(s)^2 = P(s) = \dyad{\psi(s)}{\psi(s)}$, the recycling and anticommutator terms of each traceless jump operator $\tilde{L}_\alpha = \tilde{L}_\alpha(s)$ evaluate to
\begin{equation}
  \begin{aligned}
    \Tr\Bigl(P\tilde{L}_\alpha P\tilde{L}_\alpha^\dagger\Bigr) &= \abs{\bra{\psi(s)}\tilde{L}_\alpha\ket{\psi(s)}}^2, \\
    \frac{1}{2}\Tr\Bigl(P\{\tilde{L}_\alpha^\dagger \tilde{L}_\alpha, P\}\Bigr) &= \bra{\psi(s)}\tilde{L}_\alpha^\dagger \tilde{L}_\alpha\ket{\psi(s)}.
  \end{aligned}
  \label{eq:recycling_anticommutator}
\end{equation}
Subtracting the two terms in Eq.~\eqref{eq:recycling_anticommutator} establishes the pure-state jump-variance law.

\begin{proposition}[Pure-state jump-variance law]
\label{prop:variance_law}
For any finite-dimensional open quantum system ($\mathcal{H} \cong \mathbb{C}^d$, $d \ge 2$) evolving under the time-rescaled GKSL master equation~\eqref{eq:gksl_rescaled}, the first-order survival error slope $\chi(\psi_{\mathrm{in}}) = \lim_{\varepsilon \to 0^+} p_{\mathrm{err}}(\varepsilon;\psi_{\mathrm{in}})/\varepsilon$ of any pure input state $\ket{\psi_{\mathrm{in}}}$ along the unitary control trajectory $P(s) = U_0(s)\dyad{\psi_{\mathrm{in}}}{\psi_{\mathrm{in}}}U_0(s)^\dagger$ is given by the integrated quantum variance of the traceless jump operators:
\begin{equation}
  \begin{aligned}
    \chi(\psi_{\mathrm{in}}) &= -\int_0^1 \Tr\Bigl(P(s)\,\mathcal{D}(s)(P(s))\Bigr)\,\dd s \\
    &= \int_0^1 \sum_\alpha \gamma_\alpha(s)\,\Var_{P(s)}\bigl(\tilde{L}_\alpha(s)\bigr)\,\dd s \ge 0,
  \end{aligned}
  \label{eq:jump_variance_law}
\end{equation}
where $\Var_{P(s)}(L) := \bra{\psi(s)}L^\dagger L\ket{\psi(s)} - \abs{\bra{\psi(s)}L\ket{\psi(s)}}^2 \ge 0$ is the quantum fluctuation variance of $L$ in the instantaneous pure state $\ket{\psi(s)}$. For a two-level qubit ($d = 2$) evolving along any continuous, non-constant trajectory $s \mapsto P(s)$ on the Bloch sphere under a time-independent bath with $\Gamma > 0$, the survival error slope is strictly positive ($\chi(\psi_{\mathrm{in}}) > 0$).
\end{proposition}

\begin{proof}
Inserting Eq.~\eqref{eq:recycling_anticommutator} into Eq.~\eqref{eq:survival_dyson} immediately yields Eq.~\eqref{eq:jump_variance_law}. By the Cauchy--Schwarz inequality $\abs{\bra{\psi(s)}\tilde{L}_\alpha\ket{\psi(s)}}^2 \le \braket{\psi(s)}{\psi(s)}\bra{\psi(s)}\tilde{L}_\alpha^\dagger \tilde{L}_\alpha\ket{\psi(s)}$, the integrand $\Var_{P(s)}(\tilde{L}_\alpha(s))$ is non-negative and vanishes at time $s$ if and only if $\ket{\psi(s)}$ is a right eigenstate (instantaneous dark state) of $\tilde{L}_\alpha(s)$. For $d = 2$, any non-zero $2 \times 2$ traceless jump operator $\tilde{L}_\alpha$ has at most two projective eigenstates on the Bloch sphere $S^2$, so a continuous, non-constant trajectory $s \mapsto P(s)$ cannot remain in an eigenstate over any subinterval of positive measure, ensuring $\chi(\psi_{\mathrm{in}}) > 0$ whenever $\Gamma > 0$.
\end{proof}

Proposition~\ref{prop:variance_law} has a direct quantum-optical interpretation: the instantaneous rate at which a pure state loses survival fidelity equals the quantum fluctuation variance of the environmental jump operators along the trajectory, vanishing at time $s$ if and only if $\ket{\psi(s)}$ is an instantaneous dark state of every active jump operator $\tilde{L}_\alpha(s)$. Because a non-trivial $2 \times 2$ jump operator possesses at most two discrete dark states on the Bloch sphere, a continuous non-constant gate trajectory cannot remain trapped in a decoherence-free dark state throughout $s \in [0,1]$.

\subsection{Worked physical examples: native \texorpdfstring{$T_1$}{T1} decay on \texorpdfstring{$\sqrt{X}$}{sqrt(X)} and extremal chiral dissipation on \texorpdfstring{$X$}{X}}
\label{sec:worked_examples}

Before developing the general tensor theory, we illustrate how the jump-variance law~\eqref{eq:jump_variance_law} predicts directional error splitting in two concrete physical settings. For orthogonal computational basis states $\ket{\psi_+} = \ket{0}$ and $\ket{\psi_-} = \ket{1}$ mapped to target output states $U_0(1)\ket{0}$ and $U_0(1)\ket{1}$, the forward and backward survival error probabilities are $p_\pm(\varepsilon) := 1 - \bra{\psi_\pm}U_0(1)^\dagger \Phi_\varepsilon(\dyad{\psi_\pm}{\psi_\pm})U_0(1)\ket{\psi_\pm} = \varepsilon\,\chi_\pm + \mathcal{O}(\varepsilon^2)$.

\begin{example}[Native longitudinal $T_1$ decay on a polar $\sqrt{X}$ gate]
\label{ex:sqrt_x_t1}
Consider a standard polar $\sqrt{X}$ gate $U_{\mathrm{target}} = \exp(i\frac{\pi}{4}\sigma_x)$ driven by $H_{\sqrt{X}} = -\frac{\pi}{4}\sigma_x$ on $s \in [0,1]$ under native longitudinal $T_1$ relaxation ($L_\downarrow = \dyad{0}{1} = \frac{1}{2}(\sigma_x + i\sigma_y)$, $\Gamma = \Tr(L_\downarrow^\dagger L_\downarrow) = 1$). The evolving pure-state projectors are $P_\pm(s) = \frac{1}{2}[\mathbb{I}_2 \pm \sin(\frac{\pi s}{2})\sigma_y \pm \cos(\frac{\pi s}{2})\sigma_z]$. Since $L_\downarrow^\dagger L_\downarrow = \dyad{1}{1} = \frac{1}{2}(\mathbb{I}_2 - \sigma_z)$, we have $\Tr(P_\pm(s)L_\downarrow^\dagger L_\downarrow) = \frac{1}{2}[1 \mp \cos(\frac{\pi s}{2})]$ and $\abs{\Tr(P_\pm(s)L_\downarrow)}^2 = \frac{1}{4}\sin^2(\frac{\pi s}{2})$, giving the instantaneous jump variance
\begin{equation*}
  \Var_{P_\pm(s)}(L_\downarrow) = \frac{1}{2} \mp \frac{1}{2}\cos\left(\frac{\pi s}{2}\right) - \frac{1}{4}\sin^2\left(\frac{\pi s}{2}\right).
\end{equation*}
Integrating over $s \in [0,1]$ yields $\chi_\pm(\sqrt{X};T_1) = \frac{3}{8} \mp \frac{1}{\pi}$, corresponding to the $12.23 : 1$ native $T_1$ survival-error split ($\kappa_{T_1} = (3\pi - 8)/(3\pi + 8) \approx 0.081767$) shown in Figure~\ref{fig:fig1}(b).
\end{example}

\begin{example}[Extremal chiral dissipation on a constant-rate $X$ swap]
\label{ex:chiral_swap}
Consider a time-independent single-qubit $\pi$ inversion (swap) $U_{\mathrm{target}} = \pm i\sigma_x$ driven by $H_q = -\frac{q\pi}{2}\sigma_x$ on $s \in [0,1]$ for an odd integer winding number $q \in \{\pm 1, \pm 3, \dots\}$, where the forward and backward survival errors are
\begin{equation}
  \begin{aligned}
    p_+(\varepsilon;q) &:= 1 - \bra{1}\Phi_\varepsilon(\dyad{0}{0})\ket{1} = \Tr\Bigl(\dyad{0}{0}\,\Phi_\varepsilon(\dyad{0}{0})\Bigr), \\
    p_-(\varepsilon;q) &:= 1 - \bra{0}\Phi_\varepsilon(\dyad{1}{1})\ket{0} = \Tr\Bigl(\dyad{1}{1}\,\Phi_\varepsilon(\dyad{1}{1})\Bigr).
  \end{aligned}
  \label{eq:swap_survival_defs}
\end{equation}
Every odd $q$ has the exact same first-order entanglement infidelity $1 - F_e = \frac{1}{2}\varepsilon\Gamma + \mathcal{O}(\varepsilon^2)$. Under the engineered chiral jump operator $L_+ = (\sigma_x + i\sqrt{2}\sigma_z)/\sqrt{6}$ ($\Gamma = \Tr(L_+^\dagger L_+) = 1$), however, the survival errors split as
\begin{equation}
  p_\pm(\varepsilon;q) = \varepsilon\,\chi_\pm(q) + \mathcal{O}(\varepsilon^2), \qquad
  \chi_\pm(q) = \frac{1}{3} \pm \frac{2\sqrt{2}}{3\pi q}.
  \label{eq:chiral_swap_slopes}
\end{equation}
Indeed, for $H_q = -\frac{q\pi}{2}\sigma_x$, $U_0(s) = \cos(\frac{q\pi s}{2})\mathbb{I}_2 + i\sin(\frac{q\pi s}{2})\sigma_x$ evolves $\ket{0}, \ket{1}$ along $P_\pm(s) = \frac{1}{2}[\mathbb{I}_2 \pm \sin(q\pi s)\sigma_y \pm \cos(q\pi s)\sigma_z]$. Using $\sigma_x\sigma_z = -i\sigma_y$, we find
\begin{equation}
  L_+^\dagger L_+ = \frac{1}{6}(\sigma_x - i\sqrt{2}\sigma_z)(\sigma_x + i\sqrt{2}\sigma_z) = \frac{1}{6}\bigl(3\mathbb{I}_2 + 2\sqrt{2}\sigma_y\bigr),
  \label{eq:Lplus_product}
\end{equation}
so $\Tr(P_\pm(s)L_+^\dagger L_+) = \frac{1}{2} \pm \frac{\sqrt{2}}{3}\sin(q\pi s)$ and $\abs{\Tr(P_\pm(s)L_+)}^2 = \frac{1}{3}\cos^2(q\pi s)$. Subtracting gives the instantaneous jump variance
\begin{equation}
  \Var_{P_\pm(s)}(L_+) = \frac{1}{2} - \frac{1}{3}\cos^2(q\pi s) \pm \frac{\sqrt{2}}{3}\sin(q\pi s),
  \label{eq:Lplus_inst_variance}
\end{equation}
whose integral over $s \in [0,1]$ yields Eq.~\eqref{eq:chiral_swap_slopes}. At $q = 1$, $\chi_+(1) \approx 0.633439$, $\chi_-(1) \approx 0.033228$, and $\chi_-(1)/\chi_+(1) = (\pi - 2\sqrt{2})/(\pi + 2\sqrt{2}) = \kappa_\star \approx 0.052456$ ($19.0635 : 1$). We now prove that both examples are special cases of a universal two-moment trajectory law.
\end{example}

\section{Two-moment trajectory law and geometric cofactor bound}
\label{sec:two_moment}

In the weak-damping Markovian regime for a two-level qubit ($d = 2$) coupled to a time-independent bath, every traceless jump operator $\tilde{L}_\alpha$ expands in the Pauli operator basis $\bm{\sigma} = (\sigma_x, \sigma_y, \sigma_z)^T$ as
\begin{equation}
  \tilde{L}_\alpha = (\bm{a}_\alpha + i\,\bm{b}_\alpha)\cdot\bm{\sigma}, \qquad \bm{a}_\alpha, \bm{b}_\alpha \in \mathbb{R}^3.
  \label{eq:pauli_jump_expansion}
\end{equation}
The Pauli algebra identity $(\bm{u}\cdot\bm{\sigma})(\bm{v}\cdot\bm{\sigma}) = (\bm{u}\cdot\bm{v})\mathbb{I}_2 + i(\bm{u}\times\bm{v})\cdot\bm{\sigma}$ yields
\begin{equation}
  \begin{aligned}
    \tilde{L}_\alpha^\dagger \tilde{L}_\alpha &= \bigl(\norm{\bm{a}_\alpha}^2 + \norm{\bm{b}_\alpha}^2\bigr)\mathbb{I}_2 - 2(\bm{a}_\alpha \times \bm{b}_\alpha)\cdot\bm{\sigma}, \\
    \tilde{L}_\alpha \tilde{L}_\alpha^\dagger &= \bigl(\norm{\bm{a}_\alpha}^2 + \norm{\bm{b}_\alpha}^2\bigr)\mathbb{I}_2 + 2(\bm{a}_\alpha \times \bm{b}_\alpha)\cdot\bm{\sigma}.
  \end{aligned}
  \label{eq:pauli_jump_products}
\end{equation}

\begin{definition}[Markovian bath tensors and control trajectory moments]
\label{def:bath_and_moments}
Any time-independent Markovian qubit bath $\mathcal{D}$ with traceless jump expansion~\eqref{eq:pauli_jump_expansion} is characterized by its scalar trace rate $\Gamma \ge 0$, symmetric positive-semidefinite unital damping tensor $S \in \mathbb{R}^{3 \times 3}$, and non-unital polarization drift vector $\bm{t} \in \mathbb{R}^3$:
\begin{equation}
  \begin{aligned}
    \Gamma &= 2\sum_\alpha \gamma_\alpha\bigl(\norm{\bm{a}_\alpha}^2 + \norm{\bm{b}_\alpha}^2\bigr) = \Tr(S), \\
    S &:= 2\sum_\alpha \gamma_\alpha\bigl(\bm{a}_\alpha \bm{a}_\alpha^T + \bm{b}_\alpha \bm{b}_\alpha^T\bigr) \succeq 0, \\
    \bm{t} &:= 8\sum_\alpha \gamma_\alpha(\bm{a}_\alpha \times \bm{b}_\alpha) = \Tr\bigl(\bm{\sigma}\,\mathcal{D}(\mathbb{I}_2)\bigr).
  \end{aligned}
  \label{eq:bath_tensors_def}
\end{equation}
For orthogonal antipodal input states $\ket{\psi_+} = \ket{0}$ and $\ket{\psi_-} = \ket{1}$, the evolving pure-state projectors $P_\pm(s) := U_0(s)\dyad{\psi_\pm}{\psi_\pm}U_0(s)^\dagger$ trace out antipodal curves on the Bloch sphere:
\begin{equation}
  P_\pm(s) = \frac{1}{2}\bigl(\mathbb{I}_2 \pm \bm{r}(s)\cdot\bm{\sigma}\bigr), \qquad \norm{\bm{r}(s)} = 1.
  \label{eq:bloch_projectors}
\end{equation}
The geometry of the control trajectory $\bm{r} : [0,1] \to S^2$ is captured by two integral moments---its time-averaged Bloch polarization vector $\bm{m} \in \mathbb{R}^3$ and its transverse residence covariance matrix (covariance deficit) $G \in \mathbb{R}^{3 \times 3}$:
\begin{equation}
  \begin{aligned}
    \bm{m} &:= \int_0^1 \bm{r}(s)\,\dd s, \\
    G &:= \mathbb{I}_3 - \int_0^1 \bm{r}(s)\bm{r}(s)^T\,\dd s = \int_0^1 \bigl(\mathbb{I}_3 - \bm{r}(s)\bm{r}(s)^T\bigr)\,\dd s.
  \end{aligned}
  \label{eq:trajectory_moments_def}
\end{equation}
\end{definition}

Physically, in the affine Bloch--Redfield and Pauli transfer matrix representation (Appendix~\ref{app:ptm_proof}), the matrix $S - \Gamma\mathbb{I}_3$ is the unital Bloch relaxation damping matrix that contracts the Bloch sphere along each axis, whereas $\bm{t} = \Tr(\bm{\sigma}\,\mathcal{D}(\mathbb{I}_2))$ is the non-unital polarization drift vector that drives the maximally mixed state toward the environmental steady state. Geometrically, $\bm{m}$ represents the net vector centroid of the trajectory inside the unit Bloch ball (which couples linearly to the non-unital bath drift $\bm{t}$), while the diagonal entries $G_{ii} = \int_0^1 (1 - r_i(s)^2)\,\dd s$ measure the integrated transverse residence time spent away from each Bloch axis $\pm\bm{e}_i$.

\begin{theorem}[Two-moment trajectory law and geometric asymmetry bound]
\label{thm:two_moment}
In the weak-damping Markovian regime for a two-level qubit, let $\bm{r} : [0,1] \to S^2$ be any piecewise-continuous Bloch trajectory with moments $(\bm{m}, G)$ defined in Eq.~\eqref{eq:trajectory_moments_def}, and let $\mathcal{D}$ be any time-independent Markovian bath with tensors $(S, \bm{t})$ defined in Eq.~\eqref{eq:bath_tensors_def}. Then:
\begin{enumerate}
  \item \emph{Exact two-moment decomposition:} The first-order survival error slopes $\chi_\pm(\mathcal{D};\bm{r})$ depend on $\bm{r}(s)$ exclusively through $(\bm{m}, G)$:
  \begin{equation}
    \begin{aligned}
      \chi_\pm(\mathcal{D};\bm{r}) &= \frac{1}{2}\Tr(SG) \mp \frac{1}{4}\bm{m}\cdot\bm{t}, \\
      \chi_+(\mathcal{D};\bm{r}) + \chi_-(\mathcal{D};\bm{r}) &= \Tr(SG), \\
      \chi_+(\mathcal{D};\bm{r}) - \chi_-(\mathcal{D};\bm{r}) &= -\frac{1}{2}\bm{m}\cdot\bm{t}.
    \end{aligned}
    \label{eq:two_moment_law}
  \end{equation}
  \item \emph{Geometric asymmetry bound:} For every piecewise-continuous trajectory $\bm{r}(s)$, $\adj(G) - \bm{m}\bm{m}^T \succeq 0$. Whenever $G \succ 0$ ($\det G > 0$), the supremum of the directional contrast across all non-zero Markovian baths ($\Gamma > 0$) is
  \begin{equation}
    \begin{aligned}
      \sup_{\Gamma > 0} \frac{\abs{\chi_+(\mathcal{D};\bm{r}) - \chi_-(\mathcal{D};\bm{r})}}{\chi_+(\mathcal{D};\bm{r}) + \chi_-(\mathcal{D};\bm{r})} &= \beta(\bm{r}) \\
      &:= \sqrt{\frac{\bm{m}^T G\bm{m}}{\det G}} \in [0,1],
    \end{aligned}
    \label{eq:beta_bound}
  \end{equation}
  and the ratio of the smaller to the larger first-order survival error slope obeys the universal lower bound
  \begin{equation}
    \inf_{\Gamma > 0} \frac{\min(\chi_+,\chi_-)}{\max(\chi_+,\chi_-)} = \kappa(\bm{r}) := \frac{1 - \beta(\bm{r})}{1 + \beta(\bm{r})}.
    \label{eq:kappa_bound}
  \end{equation}
  Moreover, when $G \succ 0$ and $\bm{m} \neq \bm{0}$, the lower bound $\kappa(\bm{r})$ in Eq.~\eqref{eq:kappa_bound} is uniquely attained (up to a positive scalar rate, global phase $e^{i\phi}$, and chirality reversal $\tilde{L}_{\mathrm{ext}} \mapsto \tilde{L}_{\mathrm{ext}}^\dagger$) by a single rank-$1$ traceless jump operator $\tilde{L}_{\mathrm{ext}} = (\bm{a}_\star + i\,\bm{b}_\star)\cdot\bm{\sigma}$ satisfying $\bm{a}_\star^T G\bm{a}_\star = \bm{b}_\star^T G\bm{b}_\star$, $\bm{a}_\star^T G\bm{b}_\star = 0$, and $G^{1/2}\bm{a}_\star \times G^{1/2}\bm{b}_\star \parallel G^{1/2}\bm{m}$ (giving $(\bm{a}_\star \times \bm{b}_\star)\cdot\bm{m} = \mp\beta(\bm{r})\,\bm{a}_\star^T G\bm{a}_\star$).
\end{enumerate}
\end{theorem}

\begin{proof}[Proof sketch and physical mechanism]
Evaluating $\Tr(P_\pm(s)\tilde{L}_\alpha^\dagger \tilde{L}_\alpha) - \abs{\Tr(P_\pm(s)\tilde{L}_\alpha)}^2$ via Eq.~\eqref{eq:pauli_jump_products} and integrating over $s \in [0,1]$ yields the single-jump variance identity
\begin{equation}
  \int_0^1 \Var_{P_\pm(s)}(\tilde{L}_\alpha)\,\dd s = \bm{a}_\alpha^T G\bm{a}_\alpha + \bm{b}_\alpha^T G\bm{b}_\alpha \mp 2(\bm{a}_\alpha \times \bm{b}_\alpha)\cdot\bm{m}.
  \label{eq:single_jump_variance}
\end{equation}
Summing over $\alpha$ with weights $\gamma_\alpha$ establishes Eq.~\eqref{eq:two_moment_law} (an independent $4 \times 4$ PTM proof is given in Appendix~\ref{app:ptm_proof}). By convexity of the positive cone of Lindblad generators, the directional contrast ratio is maximized on a single jump operator:
\begin{equation}
  \frac{\abs{\chi_+ - \chi_-}}{\chi_+ + \chi_-} = \frac{2\,\abs{(\bm{a}\times\bm{b})\cdot\bm{m}}}{\bm{a}^T G\bm{a} + \bm{b}^T G\bm{b}}.
  \label{eq:single_jump_contrast}
\end{equation}
Transforming to whitened coordinates $\tilde{\bm{a}} = G^{1/2}\bm{a}$, $\tilde{\bm{b}} = G^{1/2}\bm{b}$ and applying the matrix cofactor identity $\bm{a}\times\bm{b} = (\det G)^{-1/2}G^{1/2}(\tilde{\bm{a}}\times\tilde{\bm{b}})$, the Cauchy--Schwarz inequality, and the arithmetic--geometric mean inequality gives
\begin{equation}
  \begin{aligned}
    2\,\abs{(\bm{a}\times\bm{b})\cdot\bm{m}} &= 2(\det G)^{-1/2}\abs{(\tilde{\bm{a}}\times\tilde{\bm{b}})\cdot(G^{1/2}\bm{m})} \\
    &\le \sqrt{\frac{\bm{m}^T G\bm{m}}{\det G}}\bigl(\norm{\tilde{\bm{a}}}^2 + \norm{\tilde{\bm{b}}}^2\bigr) \\
    &= \beta(\bm{r})\bigl(\bm{a}^T G\bm{a} + \bm{b}^T G\bm{b}\bigr).
  \end{aligned}
  \label{eq:whitened_cross_product_bound}
\end{equation}
The complete proof of the constrained cross-product minimum, the cofactor inequality $\adj(G) - \bm{m}\bm{m}^T \succeq 0$ implying $\beta(\bm{r}) \le 1$, and the $\mathrm{SO}(2)$ gauge uniqueness of $\tilde{L}_{\mathrm{ext}}$ is given in Appendix~\ref{app:cofactor_proof}.
\end{proof}

For an experimentalist characterizing a quantum gate, Theorem~\ref{thm:two_moment} states that regardless of how complicated the environmental jump operators are, the directional survival-error split between $\ket{0}$ and $\ket{1}$ is completely determined by the inner product $-\frac{1}{2}\bm{m}\cdot\bm{t}$ between the trajectory centroid and the bath drift vector, and cannot exceed the geometric limit $\beta(\bm{r})$ set by the pulse trajectory itself. Physically, the inequality $\beta(\bm{r}) \le 1$ is a direct manifestation of complete positivity: in a general non-CPTP affine linear map on the Bloch sphere, the polarization drift vector $\bm{t}$ could be chosen arbitrarily large relative to the damping matrix $S$, whereas GKSL complete positivity links non-unital drift $\bm{t} = 8\sum_\alpha \gamma_\alpha(\bm{a}_\alpha \times \bm{b}_\alpha)$ to unital diffusion $S = 2\sum_\alpha \gamma_\alpha(\bm{a}_\alpha \bm{a}_\alpha^T + \bm{b}_\alpha \bm{b}_\alpha^T)$ via the cross-product bound~\eqref{eq:whitened_cross_product_bound}, preventing $\min(\chi_+,\chi_-)$ from vanishing whenever $\det G > 0$ and $\beta(\bm{r}) < 1$.

\section{Universal selectivity bound for swaps and constant-axis quantum gates}
\label{sec:swap_bound}

\subsection{Time-independent single-qubit swaps and winding-number compression}
\label{sec:autonomous_swaps}

We now specialize to time-independent constant-amplitude control Hamiltonians $H \in \mathfrak{su}(2)$ that invert the computational basis states $\ket{0} \leftrightarrow \ket{1}$ up to phase. Up to an azimuthal rotation of the transverse axes around $\bm{e}_z$, any such time-independent swap Hamiltonian takes the form $H_q := -\frac{q\pi}{2}\sigma_x$ for an odd integer winding number $q \in 2\mathbb{Z} + 1$ (corresponding to total rotation angle $Q = q\pi$), which generates the meridian Bloch trajectory $\bm{r}_q(s) = (0, \sin(q\pi s), \cos(q\pi s))^T$.

\begin{theorem}[Universal selectivity bound of single-qubit swaps]
\label{thm:swap_floor}
In the weak-damping Markovian regime, let $H_q = -\frac{q\pi}{2}\sigma_x$ ($q \in 2\mathbb{Z} + 1$) be any time-independent single-qubit swap Hamiltonian. Then for each odd winding number $q$, the trajectory moments~\eqref{eq:trajectory_moments_def}, geometric asymmetry parameter $\beta_q$, and selectivity bound $\kappa_q$ are
\begin{equation}
  \begin{aligned}
    \bm{m}_q &= \left(0, \frac{2}{q\pi}, 0\right)^T, \qquad
    G_q = \diag\left(1, \frac{1}{2}, \frac{1}{2}\right), \\
    \det G_q &= \frac{1}{4}, \qquad
    \bm{m}_q^T G_q\bm{m}_q = \frac{2}{q^2\pi^2},
  \end{aligned}
  \label{eq:swap_moments_q}
\end{equation}
\begin{equation}
  \beta_q = \frac{2\sqrt{2}}{\pi\abs{q}}, \qquad
  \kappa_q := \inf_{\Gamma > 0} \frac{\min(\chi_+,\chi_-)}{\max(\chi_+,\chi_-)} = \frac{\pi\abs{q} - 2\sqrt{2}}{\pi\abs{q} + 2\sqrt{2}}.
  \label{eq:beta_kappa_q}
\end{equation}
Over all time-independent single-qubit swap Hamiltonians and all Markovian baths ($\Gamma > 0$), the survival-error ratio obeys the universal lower bound
\begin{equation}
  \begin{aligned}
    \frac{\min\bigl(\chi_+(\mathcal{D};H_q), \chi_-(\mathcal{D};H_q)\bigr)}{\max\bigl(\chi_+(\mathcal{D};H_q), \chi_-(\mathcal{D};H_q)\bigr)} &\ge \kappa_\star := \frac{\pi - 2\sqrt{2}}{\pi + 2\sqrt{2}}, \\
    \kappa_\star \approx 0.052456, \qquad \kappa_\star^{-1} &\approx 19.0635,
  \end{aligned}
  \label{eq:universal_kappa_star}
\end{equation}
with equality at $\abs{q} = 1$ attained by a single traceless jump operator if and only if, up to a positive rate and global phase $e^{i\phi}$, it equals one of the two chiral partners $L_\pm := \frac{1}{\sqrt{6}}(\sigma_x \pm i\sqrt{2}\sigma_z)$.
\end{theorem}

\begin{proof}
For $H = \frac{Q}{2}\bm{n}\cdot\bm{\sigma}$, the swap condition $\bra{0}e^{-iH}\ket{0} = \cos(Q/2) - i\sin(Q/2)(\bm{n}\cdot\bm{e}_z) = 0$ forces $Q = q\pi$ ($q \in 2\mathbb{Z} + 1$) and $\bm{n}\cdot\bm{e}_z = 0$. Choosing $\bm{n} = -\bm{e}_x$ yields $\bm{r}_q(s) = (0, \sin(q\pi s), \cos(q\pi s))^T$, whose time integrals over $s \in [0,1]$ give Eq.~\eqref{eq:swap_moments_q}. Substituting Eq.~\eqref{eq:swap_moments_q} into Eqs.~\eqref{eq:beta_bound}--\eqref{eq:kappa_bound} of Theorem~\ref{thm:two_moment} yields Eqs.~\eqref{eq:beta_kappa_q}--\eqref{eq:universal_kappa_star} since $\kappa_q$ is strictly increasing in $\abs{q} \ge 1$. Finally, the extremality conditions of Theorem~\ref{thm:two_moment} in $(G_q^{1/2}\bm{m}_q)^\perp = \mathrm{span}\{\bm{e}_x,\bm{e}_z\}$ select $\tilde{\bm{a}} = \frac{1}{\sqrt{6}}\bm{e}_x$ and $\tilde{\bm{b}} = \pm\frac{1}{\sqrt{6}}\bm{e}_z$, giving $\bm{a}_\star = \frac{1}{\sqrt{6}}\bm{e}_x$, $\bm{b}_\star = \pm\frac{1}{\sqrt{3}}\bm{e}_z$, or $L_\pm = \frac{1}{\sqrt{6}}(\sigma_x \pm i\sqrt{2}\sigma_z)$. For the tilted jump family $L_\theta := \frac{1}{\sqrt{6}}[\sigma_x + i\sqrt{2}(\cos\theta\,\sigma_z - \sin\theta\,\sigma_y)]$, the directional contrast is $\abs{\chi_+ - \chi_-}/(\chi_+ + \chi_-) = \beta_q\abs{\cos\theta}$, attaining $\beta_q$ only at $\theta \in \{0,\pi\}$.
\end{proof}

In plain physical terms, Theorem~\ref{thm:swap_floor} proves that no Markovian environment can cause one computational basis state to fail more than $19.0635$ times as often as the other during a constant-rate single-qubit $\pi$ swap; observing a larger directional ratio in an experiment immediately diagnoses departures from the autonomous two-level Markovian regime (such as non-autonomous pole idling or multilevel leakage).

\begin{figure*}[t]
  \centering
  \includegraphics[width=\textwidth]{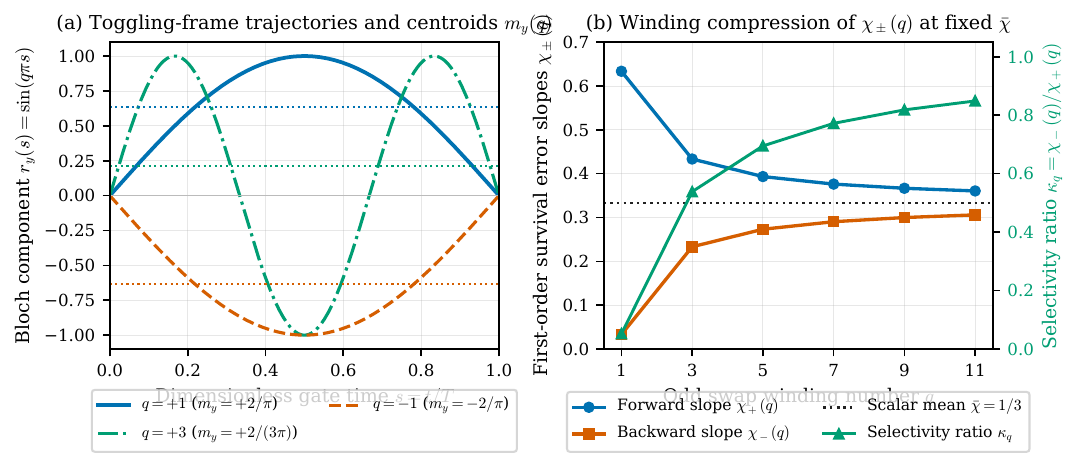}
  \caption{Winding-number compression of directional survival errors for time-independent swaps. (a) Toggling-frame Bloch component $r_y(s) = \sin(q\pi s)$ for constant-amplitude swaps $H_q = -\frac{q\pi}{2}\sigma_x$ with odd winding numbers $q = +1$ (solid blue), $q = +3$ (dash-dotted green), and $q = -1$ (dashed red); horizontal dotted lines mark the trajectory centroids $m_y(q) = 2/(q\pi)$. (b) Directional survival error slopes $\chi_+(q)$ (blue circles) and $\chi_-(q)$ (red squares) under the extremal chiral jump $L_+ = (\sigma_x + i\sqrt{2}\sigma_z)/\sqrt{6}$ ($\Gamma = 1$), their constant average $\bar{\chi} = 1/3$ (black dotted), and the selectivity ratio $\kappa_q = \chi_-(q)/\chi_+(q)$ (green triangles, right axis), demonstrating that higher odd windings cancel opposing-hemisphere drift contributions and compress directional error asymmetry as $1/\abs{q}$.}
  \label{fig:fig2}
\end{figure*}

Figure~\ref{fig:fig2} illustrates the physical mechanism behind the $1/\abs{q}$ suppression of directional asymmetry in Eq.~\eqref{eq:beta_kappa_q}. A multi-winding swap ($\abs{q} \ge 3$) executes $\abs{q} - 1$ complete $2\pi$ Rabi nutation cycles whose eastern ($y > 0$) and western ($y < 0$) hemisphere segments cancel identically in the time integral for $\bm{m}_q$, leaving only the final $\pi$ half-cycle to contribute net drift coupling $\propto 1/\abs{q}$. Meanwhile, the unital covariance deficit $G_q = \diag(1, \frac{1}{2}, \frac{1}{2})$ is completely independent of $q$, so the average survival error slope $\bar{\chi} = \frac{1}{2}(\chi_+ + \chi_-) = \frac{1}{3}$ remains fixed while $\chi_+(q)$ and $\chi_-(q)$ pinch symmetrically toward $\bar{\chi}$ as $\abs{q}$ increases (Figure~\ref{fig:fig2}(b)).

\subsection{Arbitrary constant-axis rotations, polar \texorpdfstring{$\sqrt{X}$}{sqrt(X)}, and Hadamard gates}
\label{sec:constant_axis_rotations}

Theorem~\ref{thm:two_moment} applies equally to non-swap constant-axis single-qubit gates generated by $H(Q,\vartheta) = \frac{Q}{2}(\sin\vartheta\,\sigma_x + \cos\vartheta\,\sigma_z)$ with total rotation angle $\abs{Q} > 0$ and polar axis tilt $\vartheta \in (0,\pi)$. Defining $\sinc_0 := \sinc(\abs{Q}/2)$ and $\sinc_1 := \sinc(\abs{Q})$ (where $\sinc(x) := \sin(x)/x$), block-diagonalizing $G$ in the trajectory midpoint frame (Appendix~\ref{app:midpoint_frame}) yields an exact closed-form midpoint-sinc formula.

\begin{proposition}[Midpoint-sinc formula for constant-axis rotations]
\label{prop:midpoint_sinc}
For any constant-axis single-qubit gate generated by $H(Q,\vartheta) = \frac{Q}{2}(\sin\vartheta\,\sigma_x + \cos\vartheta\,\sigma_z)$ with $\abs{Q} > 0$ and $\vartheta \in (0,\pi)$, the geometric asymmetry parameter~\eqref{eq:beta_bound} evaluates in closed form to
\begin{equation}
  \begin{aligned}
    \beta(\abs{Q},\vartheta) &= \sqrt{\frac{2\,N(\abs{Q},\vartheta)}{D(\abs{Q},\vartheta)}}, \\
    N(\abs{Q},\vartheta) &:= 2(1 - \sinc_0^2)\cos^2\vartheta + \sinc_0^2(1 - \sinc_1)\sin^2\vartheta, \\
    D(\abs{Q},\vartheta) &:= \bigl[2 - (1 - \sinc_1)\sin^2\vartheta\bigr] \\
    &\quad \times \bigl[2(1 - \sinc_0^2)\cos^2\vartheta + (1 - \sinc_1)\sin^2\vartheta\bigr].
  \end{aligned}
  \label{eq:midpoint_sinc}
\end{equation}
\end{proposition}

\begin{figure*}[t]
  \centering
  \includegraphics[width=\textwidth]{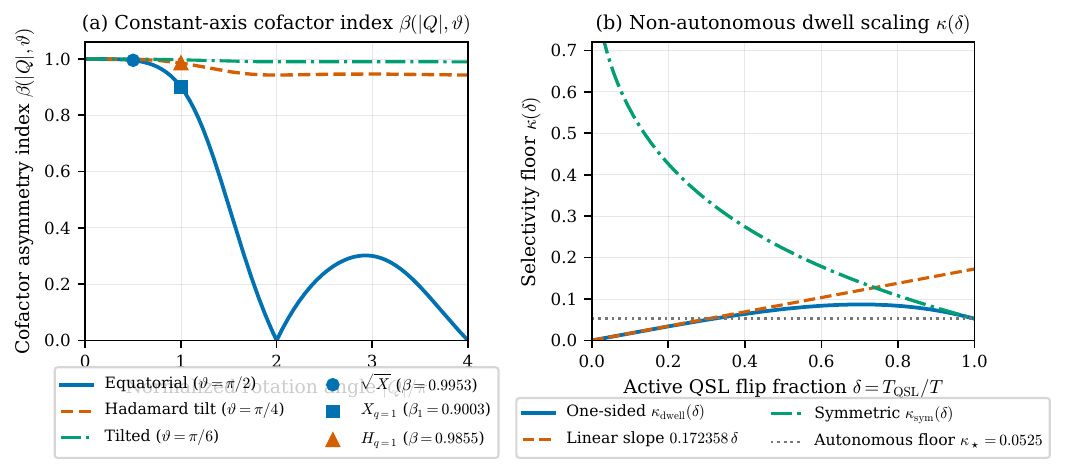}
  \caption{Constant-axis geometric asymmetry parameter $\beta(\abs{Q},\vartheta)$ and non-autonomous pulse-dwell scaling $\kappa(\delta)$. (a) Geometric asymmetry parameter $\beta(\abs{Q},\vartheta)$ from Eq.~\eqref{eq:midpoint_sinc} versus normalized rotation angle $\abs{Q}/\pi$ for equatorial drive ($\vartheta = \pi/2$, solid blue), Hadamard tilt ($\vartheta = \pi/4$, dashed red), and $\vartheta = \pi/6$ (dash-dotted green), with markers at polar $\sqrt{X}$, $X_{q=1}$, and $H_{q=1}$. (b) Selectivity bound $\kappa(\delta)$ for wait-flip-wait pulse schedules with active quantum-speed-limit flip fraction $\delta = T_{\mathrm{QSL}}/T \in (0,1]$ (Section~\ref{sec:extensions}): one-sided wait-then-flip $\kappa_{\mathrm{dwell}}(\delta)$ (solid blue) with linear slope $(3\pi^2 - 16)/(8\pi^2)\delta \approx 0.172358\,\delta$ (dashed red), contrasted against symmetric wait-flip-wait $\kappa_{\mathrm{sym}}(\delta)$ (dash-dotted green) and the autonomous bound $\kappa_\star \approx 0.052456$ (dotted gray), showing that asymmetric pole idling degrades directional error bounds linearly in $\delta$.}
  \label{fig:fig3}
\end{figure*}

Evaluating Eq.~\eqref{eq:midpoint_sinc} (plotted in Figure~\ref{fig:fig3}(a)) for specific hardware gate families reveals a sharp physical contrast between equatorial swaps and hemisphere-biased rotations:

\paragraph{Equatorial rotations ($\vartheta = \pi/2$) and polar $\sqrt{X}$ gates.}
For any rotation around an equatorial axis ($\vartheta = \pi/2$), the factor $(1 - \sinc_1)$ cancels in Eq.~\eqref{eq:midpoint_sinc}, leaving
\begin{equation}
  \beta(\abs{Q},\pi/2) = \sqrt{\frac{2\,\sinc_0^2}{1 + \sinc_1}} = \sqrt{\frac{4(1 - \cos\abs{Q})}{\abs{Q}\bigl(\abs{Q} + \sin\abs{Q}\bigr)}}.
  \label{eq:equatorial_beta}
\end{equation}
For the principal polar $\sqrt{X}$ gate ($H_{\sqrt{X}} = -\frac{\pi}{4}\sigma_x$, $\abs{Q} = \pi/2$, $k = 0$), the trajectory remains entirely in the northern hemisphere ($z \ge 0$, Figure~\ref{fig:fig1}(b)), producing a non-diagonal covariance deficit $G_{\sqrt{X}}$ with $G_{yz} = -1/\pi$:
\begin{equation}
  \begin{aligned}
    \bm{m}_{\sqrt{X}} &= \left(0, \frac{2}{\pi}, \frac{2}{\pi}\right)^T, \qquad
    G_{\sqrt{X}} = \begin{pmatrix} 1 & 0 & 0 \\ 0 & \frac{1}{2} & -\frac{1}{\pi} \\ 0 & -\frac{1}{\pi} & \frac{1}{2} \end{pmatrix}, \\
    \det G_{\sqrt{X}} &= \frac{1}{4} - \frac{1}{\pi^2}, \qquad
    \bm{m}_{\sqrt{X}}^T G_{\sqrt{X}}\bm{m}_{\sqrt{X}} = \frac{4(\pi - 2)}{\pi^3}.
  \end{aligned}
  \label{eq:sqrtX_moments}
\end{equation}
Consequently, while a full $X$ rotation ($\abs{Q} = \pi$) has $m_z = 0$ and zero directional split under native longitudinal $T_1$ decay ($\kappa_{T_1} = 1$), the polar $\sqrt{X}$ gate has $m_z = +2/\pi$ and experiences a $12.23 : 1$ native-$T_1$ directional survival ratio ($\kappa_{T_1} = (3\pi - 8)/(3\pi + 8) \approx 0.081767$), along with a universal complete-positivity bound $\beta(\pi/2,\pi/2) = 4/\sqrt{\pi(\pi + 2)} \approx 0.995259$ and $\kappa_{\sqrt{X}} = (\sqrt{\pi(\pi + 2)} - 4)/(\sqrt{\pi(\pi + 2)} + 4) \approx 0.002376$ ($420.87 : 1$). Conversely, implementing the same target unitary $\sqrt{X}$ via a reversed $-3\pi/2$ over-rotation ($\abs{Q} = 3\pi/2$, $k = 1$) drives the state through the opposite hemisphere for two-thirds of the pulse duration, reducing $m_z$ to $-2/(3\pi)$ and moderating the native $T_1$ ratio to $\kappa_{T_1} = (9\pi - 8)/(9\pi + 8) \approx 0.558917$ (with $\beta(3\pi/2,\pi/2) = 4/\sqrt{3\pi(3\pi - 2)} \approx 0.478170$ and $\kappa \approx 0.353024$), which provides a direct experimental calibration knob to modulate directional $T_1$ asymmetry.

\paragraph{Single-pulse tilted Hadamard gates ($\vartheta = \pi/4$).}
For a single-pulse Hadamard gate driven around the tilted axis $\bm{n} = (\bm{e}_x + \bm{e}_z)/\sqrt{2}$ ($\vartheta = \pi/4$, $\abs{Q} = q\pi$ for odd $q \ge 1$), the $q = 1$ moments are $\bm{m}_{\mathrm{Had}} = (\frac{1}{2}, -\frac{\sqrt{2}}{\pi}, \frac{1}{2})^T$ and
\begin{equation}
  G_{\mathrm{Had}} = \begin{pmatrix} \frac{5}{8} & \frac{\sqrt{2}}{2\pi} & -\frac{1}{8} \\ \frac{\sqrt{2}}{2\pi} & \frac{3}{4} & \frac{\sqrt{2}}{2\pi} \\ -\frac{1}{8} & \frac{\sqrt{2}}{2\pi} & \frac{5}{8} \end{pmatrix}, \qquad
  \det G_{\mathrm{Had}} = \frac{9}{32} - \frac{3}{4\pi^2},
  \label{eq:hadamard_moments}
\end{equation}
with $\bm{m}_{\mathrm{Had}}^T G_{\mathrm{Had}}\bm{m}_{\mathrm{Had}} = (\pi^2 - 2)/(4\pi^2)$. For arbitrary odd winding $q \ge 1$, Eq.~\eqref{eq:midpoint_sinc} gives
\begin{equation}
  \begin{aligned}
    \beta_{\mathrm{Had}}(q) &:= \beta(q\pi,\pi/4) = \sqrt{\frac{8(q^2\pi^2 - 2)}{3(3q^2\pi^2 - 8)}}, \\
    \beta_{\mathrm{Had}}(1) &\approx 0.985474, \qquad \kappa_{\mathrm{Had}}(1) \approx 0.007316,
  \end{aligned}
  \label{eq:hadamard_beta_q}
\end{equation}
approaching $\beta_{\mathrm{Had}}(\infty) = 2\sqrt{2}/3 \approx 0.942809$ and $\kappa_{\mathrm{Had}}(\infty) = 17 - 12\sqrt{2} \approx 0.029437$ as $q \to \infty$. For any odd winding $q \in 2\mathbb{Z} + 1$, the trajectory starting from $\ket{0}$ under $H(q\pi,\vartheta)$ is $\bm{r}_q(s) = \bigl(\sin\vartheta\cos\vartheta(1 - \cos q\pi s),\, -\sin\vartheta\sin q\pi s,\, \cos^2\vartheta + \sin^2\vartheta\cos q\pi s\bigr)^T$. Because $\int_0^1 \cos(q\pi s)\,\dd s = 0$ and $\int_0^1 \cos^2(q\pi s)\,\dd s = 1/2$ for every odd $q$, both the time-averaged longitudinal polarization $m_z = \int_0^1 r_z(s)\,\dd s = \cos^2\vartheta$ and the mean squared longitudinal projection $\langle r_z^2 \rangle := \int_0^1 r_z(s)^2\,\dd s = \cos^4\vartheta + \frac{1}{2}\sin^4\vartheta$ are independent of $q$. Using $\norm{\bm{r}_q(s)}^2 = 1$, the transverse residence trace evaluates to
\begin{equation}
  \begin{aligned}
    G_{xx} + G_{yy} &= \int_0^1 \bigl(2 - r_x(s)^2 - r_y(s)^2\bigr)\,\dd s = 1 + \langle r_z^2 \rangle \\
    &= 1 + \cos^4\vartheta + \frac{1}{2}\sin^4\vartheta = 2 - \sin^2\vartheta\left(1 - \frac{3}{8}\sin^2\vartheta\right).
  \end{aligned}
  \label{eq:hadamard_Gxx_Gyy}
\end{equation}
At the Hadamard tilt $\vartheta = \pi/4$ ($\cos^2\vartheta = \sin^2\vartheta = 1/2$), Eq.~\eqref{eq:hadamard_Gxx_Gyy} gives $m_z = 1/2$, $\langle r_z^2 \rangle = 1/4 + 1/8 = 3/8$, and $G_{xx} + G_{yy} = 5/8 + 3/4 = 11/8$ for all odd $q$. Under longitudinal $\mathrm{T}_1$ relaxation ($S_{\mathrm{T}_1} = \diag(1/2, 1/2, 0)$ and $\bm{t}_{\mathrm{T}_1} = 2\bm{e}_z$), Theorem~\ref{thm:two_moment} yields $\chi_\pm = \frac{1}{4}(G_{xx} + G_{yy}) \mp \frac{1}{2}m_z = \frac{11}{32} \mp \frac{1}{4}$ ($\chi_+ = 3/32$, $\chi_- = 19/32$), establishing a winding-independent $19 : 3 \approx 6.33 : 1$ directional error split ($\kappa_{\mathrm{T}_1} = 3/19 \approx 0.157895$).

\section{Shaped Gaussian and DRAG control pulses}
\label{sec:shaped_pulses}

In experimental superconducting and semiconductor quantum processors, microwave arbitrary-waveform generators (AWGs) never apply ideal rectangular envelopes because sharp turn-on and turn-off edges have broad Fourier spectra that excite off-resonant transitions. Instead, hardware control stacks (such as IBM Qiskit Pulse) shape single-qubit rotations using zero-pedestal truncated Gaussian and derivative-removal-by-adiabatic-gate (DRAG) waveforms~\cite{McKay2017,Motzoi2009}. Because Theorem~\ref{thm:two_moment} holds for any piecewise-continuous trajectory $\bm{r}(s)$, it applies directly to time-dependent AWG pulse envelopes (Table~\ref{tab:pulse_shapes}). In the language of generalized open-loop filter-function theory~\cite{Green2013,PazSilva2014,Koch2016}, the moments $(\bm{m}, G)$ represent the exact zero-frequency (DC) first- and second-order toggling-frame filter moments for non-unital and unital Markovian noise.

\begin{table*}[t]
  \centering
  \caption{Geometric asymmetry parameter $\beta(\bm{r})$, universal selectivity bound $\kappa(\bm{r}) = (1-\beta)/(1+\beta)$, maximum directional contrast $\kappa^{-1}$, and native longitudinal-$T_1$ survival ratio $\kappa_{T_1}$ across rectangular, baseline-subtracted truncated Gaussian ($\sigma \in \{T/6, T/4, T/3\}$), and tilted-quadrature DRAG ($\sigma = T/4$, $\eta_{\mathrm{DRAG}} = 0.02$) single-qubit pulses in the weak-damping Markovian regime ($d = 2$). Time-symmetric $X$ inversions maintain exact native-$T_1$ symmetry ($\kappa_{T_1} = 1$) across all pulse envelopes while narrowing the Gaussian width suppresses transverse chiral noise coupling; hemisphere-biased $\sqrt{X}$ and tilted Hadamard pulses exhibit pronounced native-$T_1$ directional asymmetry ($\kappa_{T_1} \in [0.081767, 0.157895]$).}
  \label{tab:pulse_shapes}
  \small
  \setlength{\tabcolsep}{3pt}
  \begin{tabular}{@{}lcccc@{}}
    \toprule
    Single-qubit gate and pulse envelope ($d = 2$) & $\beta(\bm{r})$ & $\kappa(\bm{r})$ & Max contrast $\kappa^{-1}$ & Native $\kappa_{\mathrm{T}_1}$ \\
    \midrule
    \multicolumn{5}{@{}l}{\textit{Bit-flip ($X$) swap family ($\Theta = \pi$, equatorial axis $\vartheta = \pi/2$)}} \\
    Pauli $X$ ($q = \pm 1$, rectangular) & 0.900316 & 0.052456 & 19.0635 & 1.000000 \\
    Pauli $X$ ($q = \pm 3$, rectangular) & 0.300105 & 0.538337 & 1.8576 & 1.000000 \\
    Pauli $X$ ($q = \pm 5$, rectangular) & 0.180063 & 0.694824 & 1.4392 & 1.000000 \\
    Gaussian $X$ ($\sigma = T/6$, truncated) & 0.696010 & 0.179238 & 5.5792 & 1.000000 \\
    Gaussian $X$ ($\sigma = T/4$, truncated) & 0.764029 & 0.133768 & 7.4756 & 1.000000 \\
    Gaussian $X$ ($\sigma = T/3$, truncated) & 0.785575 & 0.120088 & 8.3273 & 1.000000 \\
    DRAG $X$ ($\sigma = T/4$, $\eta_{\mathrm{DRAG}} = 0.02$) & 0.763394 & 0.134177 & 7.4529 & 1.000000 \\
    \midrule
    \multicolumn{5}{@{}l}{\textit{Polar $\sqrt{X}$ rotation family ($\Theta = \pi/2$, equatorial axis $\vartheta = \pi/2$)}} \\
    Polar $\sqrt{X}$ ($k = 0$, rectangular, $\abs{Q} = \pi/2$) & 0.995259 & 0.002376 & 420.87 & 0.081767 \\
    Polar $\sqrt{X}$ ($k = 1$, reversed drive, $\abs{Q} = 3\pi/2$) & 0.478170 & 0.353024 & 2.8327 & 0.558917 \\
    Gaussian $\sqrt{X}$ ($\sigma = T/6$, truncated) & 0.991945 & 0.004044 & 247.30 & 0.133034 \\
    Gaussian $\sqrt{X}$ ($\sigma = T/4$, truncated) & 0.992309 & 0.003861 & 259.03 & 0.118418 \\
    Gaussian $\sqrt{X}$ ($\sigma = T/3$, truncated) & 0.992517 & 0.003756 & 266.26 & 0.113134 \\
    DRAG $\sqrt{X}$ ($\sigma = T/4$, $\eta_{\mathrm{DRAG}} = 0.02$) & 0.992276 & 0.003877 & 257.93 & 0.118569 \\
    \midrule
    \multicolumn{5}{@{}l}{\textit{Tilted single-pulse Hadamard ($H$) family ($\vartheta = \pi/4$, $\abs{Q} = q\pi$)}} \\
    Hadamard $H$ ($q = 1$, single pulse) & 0.985474 & 0.007316 & 136.69 & 0.157895 \\
    Hadamard $H$ ($q = 3$, single pulse) & 0.946450 & 0.027512 & 36.3479 & 0.157895 \\
    Hadamard $H$ ($q \to \infty$ plateau) & 0.942809 & 0.029437 & 33.9706 & 0.157895 \\
    \bottomrule
  \end{tabular}
\end{table*}

Consider a baseline-subtracted truncated Gaussian envelope on $s \in [0,1]$ with root-mean-square width $\sigma \in \{T/6, T/4, T/3\}$,
\begin{equation}
  \begin{aligned}
    g_{\mathrm{env}}(s) := Z_g^{-1}\Biggl[ &\exp\left(-\frac{(s - 1/2)^2}{2(\sigma/T)^2}\right) \\
    &- \exp\left(-\frac{1}{8(\sigma/T)^2}\right) \Biggr],
  \end{aligned}
  \label{eq:gaussian_envelope}
\end{equation}
where subtracting the pedestal $\exp[-1/(8(\sigma/T)^2)]$ enforces continuous zero-amplitude boundaries $g_{\mathrm{env}}(0) = g_{\mathrm{env}}(1) = 0$ matching experimental AWG waveforms, and $Z_g$ normalizes the unit area $\int_0^1 g_{\mathrm{env}}(s)\,\dd s = 1$. The instantaneous rotation angle $\theta_G(s) = \Theta\int_0^s g_{\mathrm{env}}(u)\,\dd u$ ($\Theta = \pi$ for $X$, $\Theta = \pi/2$ for $\sqrt{X}$) evaluates in closed form via the error function:
\begin{equation}
  \begin{aligned}
    \frac{\theta_G(s)}{\Theta} &= \frac{1}{Z_g}\Biggl\{ \sqrt{\frac{\pi}{2}}\frac{\sigma}{T}\Biggl[ \erf\left(\frac{s - 1/2}{\sqrt{2}\sigma/T}\right) \\
    &\quad + \erf\left(\frac{1}{2\sqrt{2}\sigma/T}\right) \Biggr] - s\,e^{-1/[8(\sigma/T)^2]} \Biggr\}.
  \end{aligned}
  \label{eq:theta_G_erf}
\end{equation}
When DRAG quadrature driving~\cite{McKay2017,Motzoi2009} is added via $\Omega_y(t) = \beta_{\mathrm{DRAG}}\dot{\Omega}_x(t)$ to suppress $\ket{2}$ leakage, its action within the two-level dressed qubit subspace is described by the frame-tilted unitary $U_{\mathrm{dress}}(s) = \exp(-i\frac{\phi_{\mathrm{DRAG}}(s)}{2}\sigma_y)\exp(i\frac{\theta_G(s)}{2}\sigma_x)$ with instantaneous tilt angle $\phi_{\mathrm{DRAG}}(s) = \eta_{\mathrm{DRAG}}\Theta\,g_{\mathrm{env}}(s)$, which slightly mixes $G_{xx}$ and $G_{zz}$ while preserving time-reversal symmetry on an $X$ gate ($\eta_{\mathrm{DRAG}} = 0.02$ in Table~\ref{tab:pulse_shapes}).

Evaluating the trajectory moments for Eq.~\eqref{eq:theta_G_erf} establishes three practical pulse-design principles:
\begin{enumerate}
  \item \emph{Time-symmetric $X$ envelopes guarantee exact native-$T_1$ balance:} Any time-symmetric pulse envelope satisfying $g_{\mathrm{env}}(1 - s) = g_{\mathrm{env}}(s)$ for a $\pi$ inversion ($\Theta = \pi$) enforces the reflection symmetry $\theta_G(1 - s) = \pi - \theta_G(s)$, which identically zeroes out longitudinal drift coupling ($m_z = 0$, $G_{yz} = 0$) and guarantees $\kappa_{T_1} = 1$ regardless of pulse width or DRAG quadrature (Table~\ref{tab:pulse_shapes}). Conversely, intentionally skewing a pulse envelope ($g_{\mathrm{env}}(1 - s) \neq g_{\mathrm{env}}(s)$) breaks $m_z = 0$ even on a full $\pi$ swap, allowing experimentalists to continuously tune the directional $T_1$ survival ratio $\chi_-/\chi_+$ without altering the target unitary $U_{\mathrm{target}} = -i\sigma_x$ or gate duration $T$.
  \item \emph{Pole lingering passively filters worst-case transverse chiral noise on $X$ gates:} During the smooth turn-on and turn-off ramps of a Gaussian pulse near $s = 0$ and $s = 1$, the Bloch vector dwells near the poles $\ket{0}$ and $\ket{1}$ ($\pm\bm{e}_z$) where $\sin\theta_G(s) \approx 0$, and sweeps rapidly across the equator at $s = 1/2$. Narrowing the Gaussian width therefore reduces transverse drift coupling $m_y$ from $2/\pi \approx 0.636620$ (rectangular) to $0.449869$ ($\sigma = T/3$), $0.419731$ ($\sigma = T/4$), and $0.338831$ ($\sigma = T/6$), while increasing $G_{yy}$ from $0.5$ to $0.672058$, $0.698198$, and $0.763008$. Both effects reduce $\beta$ and raise the universal selectivity bound $\kappa_{\mathrm{Gauss}}(X)$ from $\kappa_\star \approx 0.052456$ ($19.06 : 1$) to $0.120088$ ($8.33 : 1$ at $\sigma = T/3$), $0.133768$ ($7.48 : 1$ at $\sigma = T/4$), and $0.179238$ ($5.58 : 1$ at $\sigma = T/6$).
  \item \emph{Pulse-width dependence of native-$T_1$ asymmetry on $\sqrt{X}$ gates:} On a polar $\sqrt{X}$ pulse ($\Theta = \pi/2$), the initial Gaussian ramp lingers near the ground state $\ket{0}$ ($+z$) while the final ramp approaches the equator ($z = 0$). This asymmetric pole dwell increases $m_z$ and moderates the native $T_1$ directional survival ratio from $12.23 : 1$ ($\kappa_{T_1} \approx 0.081767$, rectangular) to $8.44 : 1$ ($\kappa_{T_1} \approx 0.118418$ at $\sigma = T/4$) and $7.52 : 1$ ($\kappa_{T_1} \approx 0.133034$ at $\sigma = T/6$), providing a directly testable pulse-width signature.
\end{enumerate}

\section{Experimental predictions and lock-in protocol on superconducting and trapped-ion processors}
\label{sec:experimental}

\subsection{Cross-platform hardware predictions: superconducting transmons, trapped ions, and engineered reservoirs}
\label{sec:hardware_predictions}

\begin{table*}[t]
  \centering
  \caption{Predicted first-order directional survival split $\abs{\Delta p^{(1)}} = \varepsilon\abs{\chi_+ - \chi_-}$ versus exact $4 \times 4$ Lindblad master-equation exponentiation $\abs{\Delta p^{\mathrm{exact}}} = \abs{p_+(\varepsilon) - p_-(\varepsilon)}$, finite-$\varepsilon$ error ratio $\min(p_+,p_-)/\max(p_+,p_-)$, and ideal $3\sigma$ shot budget $N_{\mathrm{tot}}^{(3\sigma)}$ across representative superconducting transmon~\cite{McKay2017,Motzoi2009,Kim2023}, trapped-ion $^{171}\mathrm{Yb}^+$~\cite{Pino2021,Moses2023}, and engineered circuit-QED reservoir scenarios. Both the single-device DRAGZ benchmark of McKay \emph{et al.}~\cite{McKay2017} ($\SI{13.3}{ns}$, $\Tone = \SI{45}{\micro s}$) and representative modern transmon combinations pairing $\SI{25}{ns}$ or $\SI{13.3}{ns}$ gate durations with the median $\Tone = \SI{288}{\micro s}$ of \texttt{ibm\_kyiv}~\cite{Kim2023} are shown.}
  \label{tab:hardware_predictions}
  \footnotesize
  \setlength{\tabcolsep}{2.5pt}
  \resizebox{\textwidth}{!}{%
  \begin{tabular}{@{}llccccc@{}}
    \toprule
    Platform / Scenario & Gate \& Bath Parameters & $\varepsilon = \gamma T$ & $\abs{\Delta p^{(1)}}$ & $\abs{\Delta p^{\mathrm{exact}}}$ & $p_{\min}/p_{\max}$ & $N_{\mathrm{tot}}^{(3\sigma)}$ \\
    \midrule
    Rep.\ Transmon (\texttt{ibm\_kyiv} $\Tone$)~\cite{Kim2023} & $\sqrt{X}$ ($\SI{25}{ns}$), $\Tone = \SI{288}{\micro s}$ & $8.6806 \times 10^{-5}$ & $5.5262 \times 10^{-5}$ & $5.5260 \times 10^{-5}$ & 0.081771 & $3.84 \times 10^5$ \\
    Rep.\ Fast Transmon~\cite{McKay2017,Kim2023} & $\sqrt{X}$ ($\SI{13.3}{ns}$), $\Tone = \SI{288}{\micro s}$ & $4.6181 \times 10^{-5}$ & $2.9399 \times 10^{-5}$ & $2.9399 \times 10^{-5}$ & 0.081769 & $7.21 \times 10^5$ \\
    McKay \emph{et al.}\ (2017) Device~\cite{McKay2017} & $\sqrt{X}$ ($\SI{13.3}{ns}$), $\Tone = \SI{45}{\micro s}$ & $2.9556 \times 10^{-4}$ & $1.8816 \times 10^{-4}$ & $1.8813 \times 10^{-4}$ & 0.081782 & $1.13 \times 10^5$ \\
    Trapped Ion $^{171}\mathrm{Yb}^+$~\cite{Pino2021,Moses2023} & $\sqrt{X}$ ($\SI{5}{\micro s}$), $\Tone = \SI{300}{s}$ & $1.6667 \times 10^{-8}$ & $1.0610 \times 10^{-8}$ & $1.0610 \times 10^{-8}$ & 0.081767 & $2.00 \times 10^9$ \\
    Eng.\ Chiral Bath ($\varepsilon = 10^{-3}$) & $X_{q=1}$, chiral jump $L_+$ ($\Gamma = 1$) & $1.0000 \times 10^{-3}$ & $6.0021 \times 10^{-4}$ & $6.0001 \times 10^{-4}$ & 0.052456 & $3.33 \times 10^4$ \\
    Eng.\ Chiral Bath ($\varepsilon = 10^{-2}$) & $X_{q=1}$, chiral jump $L_+$ ($\Gamma = 1$) & $1.0000 \times 10^{-2}$ & $6.0021 \times 10^{-3}$ & $5.9821 \times 10^{-3}$ & 0.052456 & $3.32 \times 10^3$ \\
    Eng.\ Chiral Bath ($\varepsilon = 5 \times 10^{-2}$) & $X_{q=1}$, chiral jump $L_+$ ($\Gamma = 1$) & $5.0000 \times 10^{-2}$ & $3.0011 \times 10^{-2}$ & $2.9516 \times 10^{-2}$ & 0.052456 & $6.57 \times 10^2$ \\
    \bottomrule
  \end{tabular}}
\end{table*}

Table~\ref{tab:hardware_predictions} compares the analytical first-order predictions of Theorem~\ref{thm:two_moment} against exact $4 \times 4$ matrix exponentiation $\exp(\mathcal{L}_\varepsilon)$ across superconducting transmons, optical-clock trapped ions, and engineered reservoirs. To keep experimental sources distinct, we distinguish between two hardware parameter sets: (i) in McKay \emph{et al.}~\cite{McKay2017}, a $\SI{13.3}{ns}$ Gaussian DRAGZ $X_{\pi/2} = \sqrt{X}$ pulse was benchmarked on a single-qubit fixed-frequency transmon with $\Tone \approx \SI{45}{\micro s}$ and $\Ttwo \approx \SI{50}{\micro s}$, achieving an interleaved randomized benchmarking error of $1.95(3) \times 10^{-4}$ and leakage to $\ket{2}$ of $3.1(6) \times 10^{-6}$ per gate (which corresponds to $\varepsilon_{\mathrm{McKay}} = (\SI{13.3}{ns})/(\SI{45}{\micro s}) \approx 2.9556 \times 10^{-4}$ and $\Delta p \approx 1.8816 \times 10^{-4}$, row 3 of Table~\ref{tab:hardware_predictions}); and (ii) on the 127-qubit \texttt{ibm\_kyiv} Eagle processor, Kim \emph{et al.}~\cite{Kim2023} report a device-wide median longitudinal relaxation time $\Tone = \SI{288}{\micro s}$ and median $\Ttwo = \SI{127}{\micro s}$. Pairing the median \texttt{ibm\_kyiv} relaxation time $\Tone = \SI{288}{\micro s}$ with a representative $\SI{25}{ns}$ single-qubit pulse duration (or a $\SI{13.3}{ns}$ fast DRAGZ duration) yields the representative modern transmon exposures
\begin{equation}
  \varepsilon_{\mathrm{transmon}} = \frac{T_{\mathrm{gate}}}{\Tone} = \frac{\SI{25}{ns}}{\SI{288}{\micro s}} \approx 8.680556 \times 10^{-5}
  \label{eq:eps_transmon}
\end{equation}
(and $\varepsilon_{\mathrm{fast}} = (\SI{13.3}{ns})/(\SI{288}{\micro s}) \approx 4.618056 \times 10^{-5}$ at $T_{\mathrm{gate}} = \SI{13.3}{ns}$; rows 1--2 of Table~\ref{tab:hardware_predictions}). Under longitudinal $T_1$ decay ($L_\downarrow = \dyad{0}{1}$, $\Gamma = 1$), the bath tensors $S_{T_1} = \diag(\frac{1}{2}, \frac{1}{2}, 0)$ and $\bm{t}_{T_1} = (0, 0, 2)^T$ yield
\begin{equation}
  \begin{aligned}
    \chi_+\bigl(\sqrt{X};T_1\bigr) &= \frac{3}{8} - \frac{1}{\pi} \approx 0.056690, \\
    \chi_-\bigl(\sqrt{X};T_1\bigr) &= \frac{3}{8} + \frac{1}{\pi} \approx 0.693310, \\
    \kappa\bigl(\sqrt{X};T_1\bigr) &= \frac{3\pi - 8}{3\pi + 8} \approx 0.081767,
  \end{aligned}
  \label{eq:sqrtX_T1_slopes}
\end{equation}
or a directional error ratio $\chi_-/\chi_+ \approx 12.23$, predicting a single-gate survival split $\Delta p = \frac{2}{\pi}\varepsilon_{\mathrm{transmon}} \approx 5.5262 \times 10^{-5}$ for a rectangular $\sqrt{X}$ pulse and $5.1318 \times 10^{-5}$ for a Gaussian $\sqrt{X}$ pulse ($\sigma = T/4$). Note that pure dephasing ($L_\phi = \sigma_z/\sqrt{2}$, contributing when $T_2 < 2T_1$) is unital ($\bm{t}_\phi = \bm{0}$) and leaves the directional split $\Delta p = \varepsilon\abs{\chi_+ - \chi_-}$ completely unchanged at first order, while a residual thermal excited-state population $n_{\mathrm{th}} = 0.01$ modifies the $T_1$ ratio only slightly to $\kappa_{\mathrm{th}} = [3\pi(1 + 2n_{\mathrm{th}}) - 8]/[3\pi(1 + 2n_{\mathrm{th}}) + 8] \approx 0.091594. $ Resolving $\Delta p$ at $3\sigma$ statistical significance under ideal readout requires a total shot budget
\begin{equation}
  N_{\mathrm{tot}}^{(3\sigma)} = \frac{18\bigl[p_+(1 - p_+) + p_-(1 - p_-)\bigr]}{(\Delta p)^2} \simeq \frac{18(\chi_+ + \chi_-)}{(\chi_- - \chi_+)^2\,\varepsilon},
  \label{eq:ideal_shot_budget}
\end{equation}
or $N_{\mathrm{tot}}^{(3\sigma)} = 3.84 \times 10^5$ ideal shots on \texttt{ibm\_kyiv} (Table~\ref{tab:hardware_predictions}).

Comparing superconducting transmons with $^{171}\mathrm{Yb}^+$ hyperfine clock-state trapped-ion processors~\cite{Pino2021,Moses2023} highlights how directional gate-error splitting serves as a cross-platform noise diagnostic. In $^{171}\mathrm{Yb}^+$ race-track processors, stimulated-Raman single-qubit gates have duration $T_{\mathrm{gate}} \sim 5\,\mu\text{s}$, average gate infidelity $r_{\mathrm{avg}} = 2.5(3) \times 10^{-5}$~\cite{Moses2023}, and magnetic-dipole hyperfine relaxation $T_1 \gtrsim 300\text{ s}$ ($\varepsilon_{T_1} \le 1.67 \times 10^{-8}$, $\Delta p_{T_1} \le 1.06 \times 10^{-8}$; Table~\ref{tab:hardware_predictions}). Because elastic Rayleigh and symmetric Raman photon scattering during the pulse are unital ($\bm{t} = \bm{0}$), hyperfine detuning asymmetry $\omega_{\mathrm{HF}}/\Delta_{\mathrm{laser}} \sim 10^{-3}\text{--}10^{-2}$ induces a fractional non-unital optical-pumping imbalance $\abs{\eta_{\mathrm{sc}}} \lesssim 10^{-2}$ of $\varepsilon_{\mathrm{sc}} \lesssim r_{\mathrm{avg}} \approx 2.5 \times 10^{-5}$, giving a native directional split $\Delta p_{\mathrm{sc}} \lesssim 2 \times 10^{-7}$ ($> 10^{10}$ shots); any larger observed split on $^{171}\mathrm{Yb}^+$ cleanly diagnoses residual optical-pumping beam leakage. Conversely, to test the $19.0635 : 1$ universal selectivity bound $\kappa_\star$ directly on a full $X_{q=1}$ swap in circuit QED, coupling a transmon to a lossy readout resonator via bichromatic sideband drives synthesizes the engineered chiral jump operator $L_+ = (\sigma_x + i\sqrt{2}\sigma_z)/\sqrt{6}$ at controlled exposure $\varepsilon \in [10^{-3}, 10^{-2}]$, requiring only $3.33 \times 10^4$ to $3.32 \times 10^3$ shots (Table~\ref{tab:hardware_predictions}).

\subsection{Six-circuit SPAM-invariant lock-in protocols and calibration error budget}
\label{sec:lockin_protocol}

\begin{figure*}[t]
  \centering
  \includegraphics[width=\textwidth]{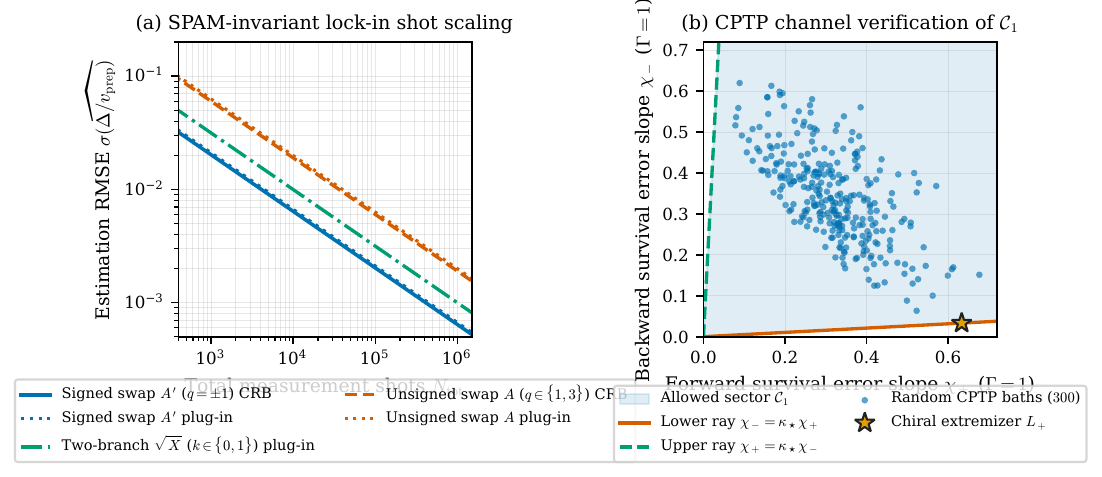}
  \caption{Two-branch differential lock-in protocols and numerical CPTP verification of the swap sector $\mathcal{C}_1$. (a) Root-mean-square estimation error $\sigma(\widehat{\Delta/v_{\mathrm{prep}}})$ versus total measurement shots $N_{\mathrm{tot}}$ under native state-preparation and measurement (SPAM) errors $\bm{\theta}_0 = (0.02, 0.06, 0.01, 0.02, 0.02, 0.018)$ for the signed-winding $X$-swap lock-in estimator $\widehat{g}_{A'}$ ($q = \pm 1$, Protocol~\ref{prot:lockin}(a), solid blue; $8.97\times$ plug-in variance reduction and $8.54\times$ Cram\'er--Rao reduction for $\Delta_1$ over unsigned $q \in \{1, 3\}$ subtraction $\widehat{g}_A$, dashed red) and the two-branch $\sqrt{X}$ longitudinal-$\mathrm{T}_1$ lock-in estimator $\widehat{g}_{\sqrt{X}}$ ($k \in \{0, 1\}$, $Q_0 = +\pi/2$ versus $Q_1 = -3\pi/2$, Protocol~\ref{prot:lockin}(b), dash-dotted green). (b) Random sample of $300$ multi-jump CPTP Markovian baths ($\Gamma = 1$, blue circles, drawn from $1{,}203{,}200$ random baths audited in Appendix~\ref{app:numerical_audit}) in the $(\chi_+, \chi_-)$ plane for an $X_{q=1}$ swap, confirming that every physical bath lies strictly inside the sector $\kappa_\star\chi_+ \le \chi_- \le \kappa_\star^{-1}\chi_+$ bounded by the chiral extremizer $L_+$ (gold star).}
  \label{fig:fig4}
\end{figure*}

On real hardware, native state-preparation errors $(e_0, e_1)$ and readout assignment errors $(f_0, f_1)$ are of order $10^{-2}$, exceeding the single-gate transmon split $\Delta p \approx 5.5 \times 10^{-5}$ by two orders of magnitude. Extracting a directional gate-error split immune to first-order SPAM asymmetry requires contrasting two control trajectories that implement the same target population transfer (up to global phase) and incur identical first-order unital damping $\Sigma = p_+ + p_- = \varepsilon\Tr(SG)$, but possess different trajectory centroids $\bm{m}$ along the active bath drift vector $\bm{t}$.

Crucially, the action of reversing the drive Hamiltonian $H \mapsto -H$ ($Q \mapsto -Q$ for $H(Q) = -\frac{Q}{2}\sigma_x$) must be distinguished from antipodal state inversion $\ket{0} \leftrightarrow \ket{1}$: whereas exchanging the initial states $\ket{0}$ and $\ket{1}$ negates the full Bloch vector $\bm{r}_-(s) = -\bm{r}_+(s)$, reversing the drive sign $H(Q) \mapsto H(-Q) = -H(Q)$ reflects the trajectory starting from $\ket{0}$ across the $x$--$z$ plane:
\begin{equation}
  \begin{aligned}
    \bm{r}_+(s; Q) &= \bigl(0,\, \sin(Qs),\, \cos(Qs)\bigr)^T \\
    \implies \bm{r}_+(s; -Q) &= \bigl(0,\, -\sin(Qs),\, +\cos(Qs)\bigr)^T.
  \end{aligned}
  \label{eq:reflection_law}
\end{equation}
Consequently, under $H \mapsto -H$, the transverse centroid component $m_y(Q) = (1 - \cos Q)/Q$ is odd in $Q$ ($m_y(-Q) = -m_y(Q)$), whereas the longitudinal centroid component $m_z(Q) = (\sin Q)/Q$ is even in $Q$ ($m_z(-Q) = +m_z(Q)$). This parity distinction dictates two distinct six-circuit SPAM-invariant lock-in protocols depending on whether the bath drift is transverse or longitudinal (Figure~\ref{fig:fig4}(a)):
\begin{enumerate}
  \item \emph{Protocol~\ref{prot:lockin}(a) (Signed-winding $X$ swaps $q = \pm 1$ for transverse bath drift $t_y \neq 0$):} For a full $\pi$ swap $H_q = -\frac{q\pi}{2}\sigma_x$ ($Q = q\pi$ with $q \in \{+1, -1\}$), both trajectories map $\ket{0} \leftrightarrow \ket{1}$ ($U(\pm\pi) = \pm i\sigma_x$) with $m_z(\pm\pi) = 0$ and opposite transverse centroids $m_y(q) = \frac{2}{q\pi}$. Reversing $q = +1 \mapsto -1$ therefore flips the transverse directional split $\Delta_q = \frac{1}{2}\varepsilon\,m_y(q)t_y = \Delta_1/q$ while preserving $\Sigma_{+1} = \Sigma_{-1}$.
  \item \emph{Protocol~\ref{prot:lockin}(b) (Two-branch lock-in $k \in \{0, 1\}$ for polar $\sqrt{X}$ under longitudinal $\mathrm{T}_1$ drift $\bm{t} = 2\bm{e}_z$):} For a polar $\sqrt{X}$ gate under longitudinal $\mathrm{T}_1$ relaxation ($\bm{t} = 2\bm{e}_z$), simple drive reversal $Q = +\pi/2 \mapsto -\pi/2$ fails for two reasons: (i) because $m_z(\pm\pi/2) = +2/\pi$ is even in $Q$, the $\mathrm{T}_1$ directional split $\frac{1}{2}\varepsilon m_z t_z$ is identical for $Q = \pm\pi/2$ and cancels in their difference, and (ii) $\exp(\pm i\frac{\pi}{4}\sigma_x)$ map $\ket{0}$ to opposite equatorial states $\ket{\mp y}$ rather than the same target state. Instead, to modulate $m_z$ while implementing the exact same target unitary up to sign, we contrast the principal branch $k = 0$ ($Q_0 = +\pi/2$, $U(Q_0) = \exp(i\frac{\pi}{4}\sigma_x)$, spending all its time in $z \ge 0$ with $m_z^{(0)} = +2/\pi$) against the reversed over-rotated branch $k = 1$ ($Q_1 = -3\pi/2$, $U(Q_1) = \exp(-i\frac{3\pi}{4}\sigma_x) = -U(Q_0)$, which traverses the southern hemisphere $z \le 0$ for two-thirds of the pulse and has $m_z^{(1)} = \frac{\sin(-3\pi/2)}{-3\pi/2} = -\frac{2}{3\pi} = -\frac{1}{3}m_z^{(0)}$) at matched pulse duration $T$. Following the test gate $U(Q_k)$ ($k \in \{0, 1\}$) with a fixed reference $\sqrt{X}$ analysis pulse $U_{\mathrm{an}} = \exp(i\frac{\pi}{4}\sigma_x)$ maps the ideal composite gate to a full population inversion $U_{\mathrm{an}}U(Q_k) = \pm i\sigma_x$ prior to $Z$-basis readout. Because both branches share the exact same diagonal covariance deficit $\diag(G^{(0)}) = \diag(G^{(1)}) = (1, 1/2, 1/2)^T$, they experience identical first-order unital damping $\Sigma^{(0)} = \Sigma^{(1)} = \frac{3}{4}\varepsilon$ under longitudinal $\mathrm{T}_1$ relaxation and pure dephasing while exhibiting opposite directional splits $\Delta_{\sqrt{X}}^{(1)} = -\frac{1}{3}\Delta_{\sqrt{X}}^{(0)}$ (giving differential lever arm $\Delta m_z = m_z^{(0)} - m_z^{(1)} = \frac{8}{3\pi}$).
\end{enumerate}

\begin{protocol}[Six-circuit two-branch differential lock-in estimators]
\label{prot:lockin}
Let $v_{\mathrm{prep}} := 1 - e_0 - e_1$ ($\eta_{\mathrm{prep}} := e_0 - e_1$) and $v_{\mathrm{ro}} := 1 - f_0 - f_1$ ($V_{\mathrm{SPAM}} := v_{\mathrm{ro}}v_{\mathrm{prep}}$) denote the preparation and readout visibilities, and let $R_0^{(\alpha)}(m)$ and $R_1^{(\alpha)}(m)$ denote the probabilities of recording readout outcome ``$1$'' when preparing $\ket{0}$ or $\ket{1}$ and executing $m = 0$ (two SPAM calibration circuits) or $m = 1$ (four gate circuits across the two trajectory branches $\alpha$). Forming $Y_m^{(\alpha)} := R_0^{(\alpha)}(m) + R_1^{(\alpha)}(m)$ and $Z_m^{(\alpha)} := R_0^{(\alpha)}(m) - R_1^{(\alpha)}(m)$ gives $Z_0 = -V_{\mathrm{SPAM}}$ and (full derivation in Appendix~\ref{app:lockin_theory}):
\begin{enumerate}
  \item \emph{Signed-winding $X$-swap lock-in ($\alpha = q \in \{+1, -1\}$, transverse bath drift $t_y \neq 0$):} With $\Sigma_q := p_+(q) + p_-(q) = \Sigma_1$ and $\Delta_q := p_-(q) - p_+(q) = \Delta_1/q + \mathcal{O}(\varepsilon^2)$,
  \begin{align}
    Y_1^{(q)} &= 2f_0 + v_{\mathrm{ro}}\bigl(1 - \eta_{\mathrm{prep}} - \Delta_q + \eta_{\mathrm{prep}}\Sigma_q\bigr), \label{eq:lockin_Y1} \\
    Z_1^{(q)} &= V_{\mathrm{SPAM}}\bigl(1 - \Sigma_q\bigr), \qquad Z_0 = -V_{\mathrm{SPAM}}, \label{eq:lockin_Z1} \\
    \widehat{g}_{A'} &:= \frac{Y_1^{(-1)} - Y_1^{(+1)}}{-2\,Z_0} = \frac{\Delta_1}{v_{\mathrm{prep}}} + \mathcal{O}(\varepsilon^2). \label{eq:lockin_estimator}
  \end{align}
  \item \emph{Two-branch $\sqrt{X}$ lock-in ($\alpha = k \in \{0, 1\}$, $Q_0 = +\pi/2$ vs.\ $Q_1 = -3\pi/2$ at matched duration $T$ followed by a fixed reference $\sqrt{X}$ analysis pulse, longitudinal $\mathrm{T}_1$ drift):} With common analysis-pulse error $(\Sigma_{\mathrm{an}}, \Delta_{\mathrm{an}})$ cancelling in the branch difference and $\Delta_{\sqrt{X}}^{(1)} = -\frac{1}{3}\Delta_{\sqrt{X}}^{(0)} + \mathcal{O}(\varepsilon^2)$,
  \begin{equation}
    \widehat{g}_{\sqrt{X}} := \frac{3\bigl(Y_1^{(k=0)} - Y_1^{(k=1)}\bigr)}{-4\,Z_0} = \frac{\Delta_{\sqrt{X}}^{(0)}}{v_{\mathrm{prep}}} + \mathcal{O}(\varepsilon^2).
    \label{eq:lockin_sqrtX_estimator}
  \end{equation}
\end{enumerate}
\end{protocol}

For transverse-drift $X$ swaps, subtracting opposite-helicity drives ($q = +1$ versus $q = -1$, lever arm $\Delta m_y = 4/\pi$) triples the directional signal compared to unsigned multi-winding subtraction ($q = 1$ versus $q = 3$, $\Delta m_y = 4/(3\pi)$), achieving an asymptotic $3^2 = 9\times$ variance reduction ($8.97\times$ plug-in variance reduction, $8.81\times$ Cram\'er--Rao reduction for $\Delta_1/v_{\mathrm{prep}}$, and $8.54\times$ Cram\'er--Rao reduction for $\Delta_1$ at the benchmark parameter vector $\bm{\theta}_0$ in Figure~\ref{fig:fig4}(a)). For polar $\sqrt{X}$ gates under longitudinal $\mathrm{T}_1$ decay, the two-branch estimator $\widehat{g}_{\sqrt{X}}$ in Eq.~\eqref{eq:lockin_sqrtX_estimator} achieves lever arm $\Delta m_z = \frac{8}{3\pi}$ (two-thirds of a full signed swap), yielding $N_{\mathrm{tot}}\Var(\widehat{g}_{\sqrt{X}}) = 0.9915$ ($4.00\times$ lower variance than unsigned $q \in \{1, 3\}$ subtraction) and agreeing with exact $4 \times 4$ PTM exponentiation on \texttt{ibm\_kyiv} parameters to $\sim 10^{-4}$ relative accuracy (Appendix~\ref{app:lockin_theory}). When the $k = 1$ branch ($Q_1 = -3\pi/2$) is implemented at fixed Rabi amplitude by tripling the pulse duration to $3T$ rather than tripling the drive amplitude at fixed duration $T$, its unital damping scales as $\Sigma^{(1)} = 3\Sigma^{(0)}$ while its directional split becomes $\Delta_{\sqrt{X}}^{(1)}(3T) = -\Delta_{\sqrt{X}}^{(0)}(T)$, and the $\eta_{\mathrm{prep}}(\Sigma^{(0)} - \Sigma^{(1)})$ cross-term is removed either by the six-circuit calibration ratio $(Z_1^{(1)} - Z_1^{(0)})/Z_0$ or by standard gate-count slope fitting (Appendix~\ref{app:lockin_theory}).

An experimental error-budget audit confirms that five potential systematic biases on a superconducting transmon remain well below $\Delta p \approx 5.5 \times 10^{-5}$:
\begin{enumerate}
  \item \emph{Coherent over-rotation ($\delta Q_\pm$):} By the skew-symmetry identity~\eqref{eq:appE_skew_symmetry} in Appendix~\ref{app:tangent_dyson}, coherent rotation-angle calibration errors $\delta Q_\pm$ vanish at first order in survival probabilities after a swap and enter at second order, $\delta p_\pm^{\mathrm{coh}} \approx \frac{1}{4}(\delta Q_\pm)^2$. Standard Rabi amplitude calibration ($\abs{\delta Q_\pm} \lesssim 10^{-3}$) limits the differential coherent bias to $\abs{\delta Q_+^2 - \delta Q_-^2}/4 \lesssim 2.5 \times 10^{-7} \ll \Delta p$.
  \item \emph{Transmon $\ket{2}$ state leakage:} Measured DRAGZ leakage per $\sqrt{X}$ gate is $3.1(6) \times 10^{-6}$~\cite{McKay2017,Wood2018}, which is $18\times$ smaller than $\Delta p \approx 5.5 \times 10^{-5}$ on \texttt{ibm\_kyiv} and $61\times$ smaller than $\Delta p \approx 1.88 \times 10^{-4}$ on the $\Tone = \SI{45}{\micro s}$ device of Ref.~\cite{McKay2017}.
  \item \emph{Spectator $ZZ$ crosstalk:} Static $ZZ$ coupling shifts the effective drive detuning and is eliminated in isolated single-qubit characterization by initializing spectator qubits in their ground state $\ket{0}$.
  \item \emph{Behavior of multi-pulse swap chains ($k = 2m + 1$):} Applying an odd train $H_q^k$ of $k = 2m + 1$ identical swaps alternates each trajectory between $\ket{0} \to \ket{1}$ on odd steps and $\ket{1} \to \ket{0}$ on even steps. Consequently, the cumulative survival error probabilities obey
  \begin{equation}
    p_\pm^{(k)} = \bar{p}^{(k)}\left(1 \pm \frac{\mu_q}{k}\right) + \mathcal{O}(k^2\varepsilon^2), \qquad
    \bar{p}^{(k)} = \frac{k}{2}(\chi_+ + \chi_-)\varepsilon,
    \label{eq:multipulse_amplification}
  \end{equation}
  where $\mu_q = (\chi_+ - \chi_-)/(\chi_+ + \chi_-)$. Because $\bar{p}^{(k)} \propto k$, the $2m$ paired forward-and-backward steps cancel in the directional difference, leaving the absolute split $p_-^{(k)} - p_+^{(k)} = \Delta_1 + \mathcal{O}(k^2\varepsilon^2)$ constant in $k$ while the relative contrast $\mu_q/k$ decreases as $1/k$. Constructive $k$-fold accumulation of the directional split ($p_-^{(k)} - p_+^{(k)} = k\Delta_1 + \mathcal{O}(k^2\varepsilon^2)$) is achieved for transverse bath drift by alternating the drive sign ($+H_1$ on odd steps, $-H_1$ on even steps so that every step traverses the same transverse hemisphere), or for longitudinal $\mathrm{T}_1$ drift by active reset between pulses.
  \item \emph{Rejection of $1/f$ $\Tone$ and readout telegraph drift:} Because transmon $\Tone$ and readout assignment fidelities fluctuate over minutes to hours, interleaving the two lock-in branch circuits shot-by-shot within each hardware repetition buffer (at $1\text{--}\SI{10}{kHz}$ repetition rates) acts as a lock-in chopper that rejects slow $1/f$ drift.
\end{enumerate}

\section{Extensions to finite damping, pulse dwell, colored baths, and multilevel systems}
\label{sec:extensions}

A natural physical question is how the leading-order autonomous qubit bounds generalize when any of the four baseline regime assumptions---weak damping ($\varepsilon \to 0^+$), continuous quantum-speed-limit driving ($T = T_{\mathrm{QSL}}$), white Markovian noise, or strictly two-level evolution ($d = 2$)---is extended to broader experimental conditions. Each extension yields an exact analytical law.

\subsection{Finite damping strength and exact all-orders damped-Bloch solution}
\label{sec:finite_damping}

When the dimensionless damping exposure $\varepsilon > 0$ is finite, let $a(\varepsilon) := \max(p_+(\varepsilon), p_-(\varepsilon))$ and $b(\varepsilon) := \min(p_+(\varepsilon), p_-(\varepsilon))$ denote the larger and smaller exact survival error probabilities of a constant-amplitude swap $H_q = -\frac{q\pi}{2}\sigma_x$. Because $\lambda_{\min}(G_q) = 1/2$ guarantees $\chi_{\max} \ge \frac{1}{4}\Gamma$, combining Theorem~\ref{thm:swap_floor} with the explicit Dyson remainder bound~\eqref{eq:dyson_remainder_bound} yields a rigorous non-perturbative lower bound valid for every Markovian bath with $\eta_{\mathrm{rem}}(\varepsilon) := 8\varepsilon\Gamma e^{2\varepsilon\Gamma} < 1$ (proof in Appendix~\ref{app:tangent_dyson}):
\begin{equation}
  \begin{aligned}
    \frac{b(\varepsilon)}{a(\varepsilon)} &\ge \frac{\kappa_q - \eta_{\mathrm{rem}}(\varepsilon)}{1 + \eta_{\mathrm{rem}}(\varepsilon)} \ge \kappa_q\Bigl(1 - C_q\,\varepsilon\Gamma\,e^{2\varepsilon\Gamma}\Bigr), \\
    C_q &:= \frac{8(1 + \kappa_q)}{\kappa_q} = \frac{16\pi\abs{q}}{\pi\abs{q} - 2\sqrt{2}},
  \end{aligned}
  \label{eq:finite_eps_dyson_bound}
\end{equation}
where $C_1 = 16\pi/(\pi - 2\sqrt{2}) \approx 160.51$ at $\abs{q} = 1$.

Moreover, for the extremal chiral jump operator $L_+ = (\sigma_x + i\sqrt{2}\sigma_z)/\sqrt{6}$ ($\Gamma = 1$) under $H_q = -\frac{q\pi}{2}\sigma_x$ ($Q = q\pi$), the full $4 \times 4$ Liouvillian $\mathcal{L}_\varepsilon = -i\,\ad_{H_q} + \varepsilon\,\mathcal{D}[L_+]$ integrates in exact closed form at arbitrary damping strength $\varepsilon \ge 0$ (Appendix~\ref{app:tangent_dyson}). With damped Rabi frequency $\Omega_{\varepsilon,q} = \sqrt{Q^2 - \varepsilon^2/9}$ and phase lag $\delta_q = Q - \Omega_{\varepsilon,q}$, the exact entanglement fidelity $F_e(\varepsilon,Q)$ and survival error probabilities $p_\pm(\varepsilon,q) = \frac{1}{2}[C(\varepsilon,Q) \pm J(\varepsilon,Q)]$ are given at all orders in $\varepsilon$ by
\begin{align}
  F_e(\varepsilon,Q) &= \frac{1}{4}\Bigl[1 + e^{-2\varepsilon/3}\bigl(1 + 2\cos\delta_q\bigr)\Bigr], \label{eq:exact_Fe} \\
  C(\varepsilon,Q) &= 1 - e^{-2\varepsilon/3}\left(\cos\delta_q - \frac{\varepsilon}{3\Omega_{\varepsilon,q}}\sin\delta_q\right), \label{eq:exact_C} \\
  J(\varepsilon,Q) &= \frac{2\sqrt{2}\varepsilon Q\Bigl[1 + e^{-\frac{2\varepsilon}{3}}\bigl(\cos\delta_q - \frac{2\varepsilon}{3\Omega_{\varepsilon,q}}\sin\delta_q\bigr)\Bigr]}{3(Q^2 + \varepsilon^2/3)}. \label{eq:exact_J}
\end{align}

\begin{figure*}[t]
  \centering
  \includegraphics[width=\textwidth]{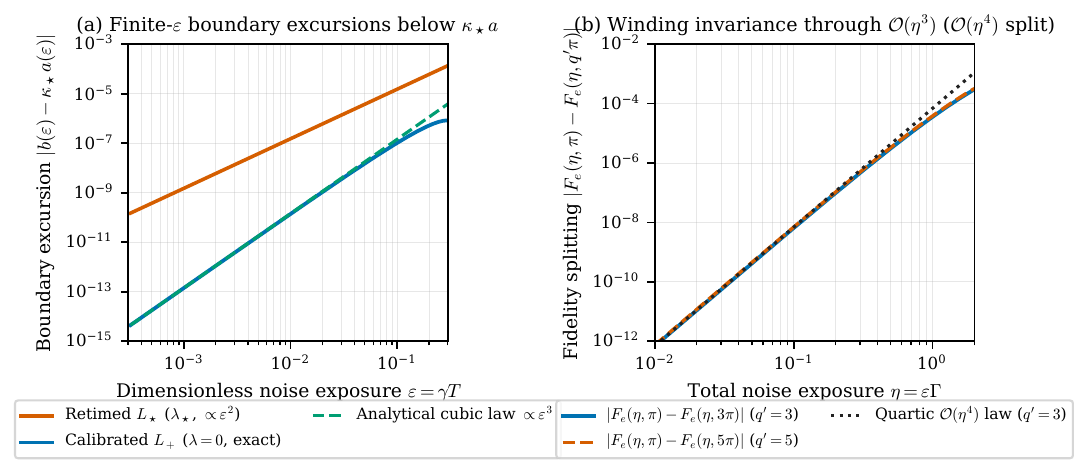}
  \caption{Exact all-orders finite-$\varepsilon$ trajectory curvature and winding-number invariance of entanglement fidelity through $\mathcal{O}(\eta^3)$. (a) Finite-damping excursion $\abs{b(\varepsilon) - \kappa_\star a(\varepsilon)}$ versus dimensionless noise exposure $\varepsilon = \gamma T$ for the retimed rank-$1$ jump $L_\star = \dyad{\psi(\pi/8)}{\psi(7\pi/8)} = -\frac{\sqrt{2}}{4}\mathbb{I}_2 - \frac{i\sqrt{2}}{4}\sigma_x - \frac{1}{2}\sigma_z$ at $\lambda_\star = (\pi - 1)/(4\pi^2)$ (solid red, $\propto \varepsilon^2$) and the calibrated fixed-drive traceless chiral jump $L_+$ at $\lambda = 0$ (solid blue, exact damped-Bloch solution matching the analytical cubic law~\eqref{eq:cubic_excursion}, dashed green), showing that coherent-dissipative frequency pulling shifts the finite-$\varepsilon$ curve by $\sim 10^{-10}$ at $\varepsilon = 10^{-2}$. (b) Exact entanglement-fidelity splitting $\abs{F_e(\eta,\pi) - F_e(\eta,q\pi)}$ from Eq.~\eqref{eq:exact_Fe} for $q = 3$ (solid blue) and $q = 5$ (dashed red) versus total noise exposure $\eta = \varepsilon\Gamma$, confirming that winding-number invariance of entanglement fidelity persists through $\mathcal{O}(\eta^3)$ before splitting at $\mathcal{O}(\eta^4)$ according to Eq.~\eqref{eq:quartic_winding_split} (dotted black).}
  \label{fig:fig5}
\end{figure*}

Expanding Eq.~\eqref{eq:exact_Fe} in powers of $\eta := \varepsilon\Gamma = \varepsilon$ shows an all-orders structural property (Figure~\ref{fig:fig5}(b)): the winding-number invariance of entanglement fidelity $F_e(\eta,q\pi)$ holds not merely at first order, but identically through third order $\mathcal{O}(\eta^3)$, splitting only at fourth order $\mathcal{O}(\eta^4)$:
\begin{equation}
  \begin{aligned}
    F_e(\eta,q\pi) &- F_e(\eta,q'\pi) \\
    &= \left(\frac{1}{q'^2} - \frac{1}{q^2}\right)\left[\frac{\eta^4}{1296\pi^2} - \frac{\eta^5}{1944\pi^2}\right] + \mathcal{O}(\eta^6).
  \end{aligned}
  \label{eq:quartic_winding_split}
\end{equation}
Simultaneously, expanding Eqs.~\eqref{eq:exact_C}--\eqref{eq:exact_J} for the calibrated $q = 1$ swap shows that non-commutativity between the Rabi drive $-i\,\ad_{H_1}$ and the dissipator $\varepsilon\mathcal{D}[L_+]$ pulls the effective Rabi frequency to $\Omega_{\varepsilon,1} = \sqrt{\pi^2 - \varepsilon^2/9}$, bending the exact survival-error curve slightly below the first-order ray $\kappa_\star a(\varepsilon)$ at third order (Figure~\ref{fig:fig5}(a)):
\begin{equation}
  \begin{aligned}
    b(\varepsilon) - \kappa_\star\,a(\varepsilon) &= -\frac{\sqrt{2}\,(4\pi^2 - 39)}{81\pi^2(\pi + 2\sqrt{2})}\,\varepsilon^3 + \mathcal{O}(\varepsilon^4) \\
    &\approx -1.417626 \times 10^{-4}\,\varepsilon^3 < 0.
  \end{aligned}
  \label{eq:cubic_excursion}
\end{equation}
Because the cubic coefficient is $-1.417626 \times 10^{-4}$, at circuit-QED exposures $\varepsilon \le 10^{-2}$ the finite-noise excursion for the calibrated traceless jump $L_+$ ($\lambda = 0$) is of order $10^{-10}$. By contrast, for the non-traceless rank-$1$ nilpotent jump $L_\star := \dyad{\psi(\pi/8)}{\psi(7\pi/8)} = -\frac{\sqrt{2}}{4}\mathbb{I}_2 - \frac{i\sqrt{2}}{4}\sigma_x - \frac{1}{2}\sigma_z$ (which shares the same traceless part $\tilde{L}_\star \propto L_- = L_+^\dagger$ and first-order ratio $\kappa_\star$, but carries trace $c_\star = -\frac{\sqrt{2}}{4}$ and Lamb shift $\Delta H_{\mathrm{Lamb}} = -\frac{1}{8}\sigma_x$), retiming the drive as $H_\varepsilon(\lambda) = \frac{\pi}{2}(1 + \lambda\varepsilon)\sigma_x$ at $\lambda_\star = (\pi - 1)/(4\pi^2) \approx 0.054247$ produces a second-order $\mathcal{O}(\varepsilon^2)$ excursion $b(\varepsilon) - \kappa_\star a(\varepsilon) = c_{\min}a(\varepsilon)^2 + \mathcal{O}(\varepsilon^3)$ with $c_{\min} = -\sqrt{2}/(\pi + 2\sqrt{2})^3 \approx -6.6464 \times 10^{-3}$ (solid red curve in Figure~\ref{fig:fig5}(a); full derivation in Appendix~\ref{app:finite_damping_proof}).

\subsection{Quantum-speed-limit pulse dwell scaling}
\label{sec:pulse_dwell}

When a control schedule includes idle dwell intervals at the poles before or after an active flip of duration $T_{\mathrm{QSL}} := \pi/\Omega_{\max}$ set by the Mandelstam--Tamm quantum speed limit~\cite{Mandelstam1945,Margolus1998}, the active flip fraction is $\delta := T_{\mathrm{QSL}}/T \in (0,1]$. For a general wait-flip-wait schedule idling at $\ket{0}$ for fraction $w_+ = \frac{1+\eta_w}{2}(1 - \delta)$, flipping at rate $-\frac{\pi}{2\delta}\sigma_x$ over duration $\delta$, and idling at $\ket{1}$ for $w_- = \frac{1-\eta_w}{2}(1 - \delta)$ with wait asymmetry $\eta_w \in [-1,1]$, integrating the trajectory moments (Appendix~\ref{app:dwell_scaling}) yields
\begin{equation}
  \beta(\delta,\eta_w) = \sqrt{\frac{8\delta}{\pi^2} + \eta_w^2\,\frac{(1 - \delta)^2}{1 - \delta/2}}.
  \label{eq:beta_dwell}
\end{equation}
Physically, one-sided pole idling ($\eta_w = \pm 1$) exposes one state to longitudinal relaxation while the antipodal state rests in a dark state, causing the selectivity bound $\kappa_{\mathrm{dwell}}(\delta) = (1 - \beta(\delta,\pm 1))/(1 + \beta(\delta,\pm 1)) = \frac{3\pi^2 - 16}{8\pi^2}\delta + \mathcal{O}(\delta^2) \approx 0.172358\,\delta$ to scale linearly with the speed-limit fraction $\delta$, whereas symmetric pole idling ($\eta_w = 0$) suppresses transverse drift coupling and drives $\kappa_{\mathrm{sym}}(\delta) = (\pi - 2\sqrt{2}\sqrt{\delta})/(\pi + 2\sqrt{2}\sqrt{\delta}) \to 1$ (Figure~\ref{fig:fig3}(b)).

\subsection{Colored thermal baths and calibration-free quantum thermometry}
\label{sec:colored_baths}

When the environment has a frequency-dependent spectral density matrix $S(\nu) = S^{(s)}(\nu) + i\,S^{(a)}(\nu) \succeq 0$ (violation of the flat-spectrum white-noise approximation, such as $1/f$ flux noise or a resonant thermal reservoir), first-order Redfield perturbation theory~\cite{Redfield1957} (Appendix~\ref{app:colored_qutrit}, Proposition~\ref{prop:redfield_parity}) shows that the survival error sum $\chi_+ + \chi_-$ couples exclusively to the frequency-even classical symmetric spectrum $S^{(s)}(\nu)$, whereas the directional split $\chi_+ - \chi_-$ couples exclusively to the frequency-odd quantum skew-symmetric spectrum $S^{(a)}(\nu)$. Consequently, any classical symmetric noise source ($S^{(a)}(\nu) \equiv 0$, including quasistatic $1/f$ phase noise) produces zero first-order directional split ($\chi_+ = \chi_-$, $\kappa = 1$). Conversely, consider an $x$-axis swap $H_q = -\frac{q\pi}{2}\sigma_x$, whose trajectory centroid $\bm{m}_q = (0, \frac{2}{q\pi}, 0)^T$ lies purely along $\bm{e}_y$. For a standard longitudinal thermal bath quantized along $\bm{e}_z$ ($\bm{t} \parallel \bm{e}_z$), $\bm{m}_q\cdot\bm{t} = 0$ and the $x$-swap directional split vanishes identically. However, if the thermal bosonic bath at frequency $\omega_b$ and temperature $T_{\mathrm{bath}}$ ($\gamma_\downarrow/\gamma_\uparrow = \exp[\hbar\omega_b/(k_B T_{\mathrm{bath}})]$, obeying the Kubo--Martin--Schwinger detailed-balance relation~\cite{Kubo1957,Martin1959}) is quantized along the transverse $\bm{e}_y$ axis ($L_{\downarrow,y} = \sqrt{\gamma_\downarrow}\dyad{+y}{-y}$, $L_{\uparrow,y} = \sqrt{\gamma_\uparrow}\dyad{-y}{+y}$, so that $S_y = \frac{1}{2}(\gamma_\downarrow + \gamma_\uparrow)\allowbreak\diag(1, 0, 1)$ and $\bm{t}_y = 2(\gamma_\downarrow - \gamma_\uparrow)\bm{e}_y$)---or equivalently, if the qubit is coupled to a longitudinal $\bm{e}_z$ thermal bath and one either prepares and measures the equatorial basis states $\ket{\pm y}$ (via $\sqrt{X}$ basis rotations) during a $\sigma_x$-dressed spin-locked swap or executes a single polar $\sqrt{X}$ gate ($m_z = 2/\pi$, where the prefactor $8/(3\pi\abs{q})$ becomes $8/(3\pi)$)---then $\Tr(S_y G_q) = \frac{3}{4}(\gamma_\downarrow + \gamma_\uparrow)$ and $-\frac{1}{2}\bm{m}_q\cdot\bm{t}_y = -\frac{2}{q\pi}(\gamma_\downarrow - \gamma_\uparrow)$. The overall system-bath coupling strength then cancels out in the normalized directional contrast, yielding a calibration-free quantum thermometer:
\begin{equation}
  \frac{\abs{\chi_+(q) - \chi_-(q)}}{\chi_+(q) + \chi_-(q)} = \frac{8}{3\pi\abs{q}}\,\tanh\left(\frac{\hbar\omega_b}{2k_B T_{\mathrm{bath}}}\right) \le \frac{8}{3\pi\abs{q}}.
  \label{eq:quantum_thermometer}
\end{equation}

\subsection{Multilevel qutrit cascade shelving and leakage diagnostics}
\label{sec:qutrit_cascade}

Finally, when evolution extends beyond the two-level qubit subspace to a three-level transmon qutrit ($d = 3$, $\mathcal{H} = \mathrm{span}\{\ket{0}, \ket{1}, \ket{2}\}$), the Lie algebra $\mathfrak{su}(3)$ possesses non-vanishing completely symmetric structure constants $d_{abc} \neq 0$. Whereas on a qubit ($d = 2$, $d_{abc} = 0$) the entanglement fidelity $F_e$ is identically invariant under bath chirality reversal $\tilde{L}_\alpha \mapsto \tilde{L}_\alpha^\dagger$ at all orders in $\varepsilon$, on a qutrit undergoing cascade energy relaxation $\varepsilon\mathcal{D}[\dyad{0}{1}] + \varepsilon\eta\,\mathcal{D}[\dyad{1}{2}]$ with decay ratio $\eta = \gamma_{12}/\gamma_{01}$, two endpoint-matched $2\pi$ shelving rotations $H_{02}$ and $H_{12}$ share the same first-order infidelity $\frac{1+\eta}{3}\varepsilon$ but split at second order:
\begin{equation}
  F_e\bigl(\varepsilon; H_{02}\bigr) - F_e\bigl(\varepsilon; H_{12}\bigr) = \frac{6 - 4\eta - 3\eta^2}{144}\,\varepsilon^2 + \mathcal{O}(\varepsilon^3),
  \label{eq:qutrit_cascade_split}
\end{equation}
which vanishes and reverses sign at the critical cascade ratio $\eta_{\mathrm{crit}} = (\sqrt{22} - 2)/3 \approx 0.896805$ (Appendix~\ref{app:colored_qutrit}, Theorem~\ref{thm:qutrit_chiral}). Additional mathematical complements, including the three-mode qubit Slepian concentration pencil~\cite{Slepian1961} and multilevel bipartite Cauchy constructions (Appendix~\ref{app:slepian_pencil}), the statistical Cram\'er--Rao and repetition-code decoding theory (Appendix~\ref{app:lockin_theory}), and the $12$-digit analytical gate audit and $1{,}203{,}200$-bath random CPTP Lindblad numerical verification (Figure~\ref{fig:fig4}(b) and Appendix~\ref{app:numerical_audit}, Tables~\ref{tab:analytical_audit}--\ref{tab:random_bath_audit}), are detailed in the Appendices.

\section{Conclusion and outlook}
\label{sec:conclusion}

We have shown that while twirled gate benchmarks such as Clifford randomized benchmarking and average gate fidelity assign a single path-independent decay rate to a noisy quantum gate at leading order, untwirled single-qubit circuits exhibit trajectory-dependent directional survival-error asymmetries governed by a compact geometric law. For any piecewise-continuous qubit control path $\bm{r}(s) \in S^2$ in a Markovian environment, the forward and backward survival-error slopes $\chi_\pm$ depend on the trajectory solely through its time-averaged Bloch vector $\bm{m}$ and transverse residence covariance matrix $G$, obeying the cofactor inequality $\beta(\bm{r}) = \sqrt{\bm{m}^T G\bm{m}/\det G} \le 1$ (Theorem~\ref{thm:two_moment}). For time-independent single-qubit $\pi$ swaps, complete positivity links non-unital drift to unital diffusion and enforces the selectivity bound $\kappa_\star = (\pi - 2\sqrt{2})/(\pi + 2\sqrt{2}) \approx 0.052456$ ($19.0635 : 1$, Theorem~\ref{thm:swap_floor}). On superconducting transmon processors, native longitudinal $\mathrm{T}_1$ relaxation induces a $12.23 : 1$ directional survival-error split ($\kappa_{\mathrm{T}_1} \approx 0.081767$) on polar $\sqrt{X}$ gates and a winding-independent $19 : 3 \approx 6.33 : 1$ split ($\kappa_{\mathrm{T}_1} = 3/19$) on single-pulse Hadamard gates while leaving symmetric rectangular, Gaussian, and DRAG $X$ rotations balanced ($\kappa_{\mathrm{T}_1} = 1$), and our two-branch differential lock-in protocols enable experimental extraction of both transverse-drift swap splits and longitudinal-$\mathrm{T}_1$ $\sqrt{X}$ splits free of first-order SPAM bias.

These results open several immediate directions for experimental quantum characterization and fault-tolerant architecture design. First, in fault-tolerant quantum error correction, standard minimum-weight perfect-matching (MWPM) and belief-propagation decoders for surface and XZZX codes~\cite{Aliferis2009,Tuckett2018,BonillaAtaides2021} assume symmetric bit-flip transition probabilities ($p_+ = p_-$) obtained from Pauli-twirled benchmarks; incorporating the analytical trajectory-dependent directional error rates $\chi_\pm(\mathcal{D};\bm{r})$ directly into asymmetric branch weights on the decoding graph can prevent systematic logical-error underestimation (Remark~\ref{rem:repetition_code} in Appendix~\ref{app:lockin_theory}) and enhance fault-tolerant thresholds under native $T_1$ damping. Second, extending the two-moment trajectory law and matrix cofactor bound from $\mathfrak{su}(2)$ to the $15$-dimensional Lie algebra $\mathfrak{su}(4)$ will enable analytical bounds on state-dependent transition asymmetries in two-qubit entangling gates such as cross-resonance and tunable-coupler CZ and iSWAP operations.

\section*{Data and code availability}
All verification, simulation, table-generation, and figure-generation code, datasets, deterministic random seeds, and \LaTeX{} sources are permanently archived in an open-access Zenodo repository at \url{https://doi.org/10.5281/zenodo.24116128}~\cite{PatraZenodo2026}. Simulation and verification scripts are released under the MIT License, and manuscript text and figures are released under the Creative Commons Attribution 4.0 International (CC BY 4.0) license.

\section*{Author contributions}
A.P.\ conceived the theoretical framework, proved the mathematical theorems, performed the analytical and numerical calculations, designed the lock-in protocol, and wrote the manuscript.

\section*{Acknowledgements}
The author thanks colleagues in the open quantum systems and quantum control communities for helpful discussions on gate characterization and randomized benchmarking. This research received no external funding. The author declares no competing financial or non-financial interests.

\appendix

\section{Superoperator trace, Liouville determinant, and first-order trajectory invariance}
\label{app:superoperator_trace}

This appendix provides the self-contained derivation of the Liouville determinant identity~\eqref{eq:liouville_determinant}, first-order scalar trajectory invariance~\eqref{eq:first_order_scalar_invariance}, and uniqueness of the $\mathrm{SU}(d)$-invariant trace functional stated in Section~\ref{sec:variance_law}.

\begin{lemma}[Superoperator trace, Liouville determinant, and Schur uniqueness]
\label{lem:superoperator_trace}
On the $d^2$-dimensional operator space $\mathcal{B}(\mathcal{H})$ ($\mathcal{H} \cong \mathbb{C}^d$, $d \ge 2$) equipped with the Hilbert--Schmidt inner product $\langle\langle A \vert B \rangle\rangle = \Tr(A^\dagger B)$:
\begin{enumerate}
  \item Every elementary multiplication superoperator $\mathcal{S}_{A,B}(X) := AXB$ has superoperator trace $\Tr(\mathcal{S}_{A,B}) = \Tr(A)\Tr(B)$. Consequently, $\Tr(\ad_{H(s)}) = 0$ and $\Tr(\mathcal{D}(s)) = -d\,\Gamma(s)$, where $\Gamma(s) = \sum_\alpha \gamma_\alpha(s)\Tr(\tilde{L}_\alpha(s)^\dagger \tilde{L}_\alpha(s))$ depends solely on the traceless jump parts $\tilde{L}_\alpha(s)$.
  \item The dynamical propagator $\Phi_\varepsilon(1)$ of the time-rescaled GKSL master equation~\eqref{eq:gksl_rescaled} obeys $\det\Phi_\varepsilon(1) = \exp(-d\,\varepsilon\,\Gamma_{\mathrm{tot}})$ at all orders in $\varepsilon \ge 0$, and its first-order entanglement infidelity $1 - F_e(\Phi_\varepsilon, U_0(1)) = \frac{\varepsilon}{d}\Gamma_{\mathrm{tot}} + \mathcal{O}(\varepsilon^2)$ is independent of the unitary trajectory $U_0(s)$.
  \item Up to scalar multiplication, $\mathcal{D} \mapsto \Tr(\mathcal{D}) = -d\,\Gamma$ is the unique $\mathrm{SU}(d)$-conjugation-invariant linear functional on trace-preserving superoperators.
\end{enumerate}
\end{lemma}

\begin{proof}
In the orthonormal matrix-unit basis $\{E_{ij} := \dyad{i}{j}\}_{i,j=0}^{d-1}$ of $\mathcal{B}(\mathcal{H})$, the superoperator trace of $\mathcal{S}_{A,B}(X) = AXB$ is
\begin{equation}
  \Tr(\mathcal{S}_{A,B}) = \sum_{i,j=0}^{d-1} \Tr\bigl(E_{ij}^\dagger A E_{ij} B\bigr)
  = \sum_{i,j=0}^{d-1} \bra{i}A\ket{i}\bra{j}B\ket{j} = \Tr(A)\,\Tr(B).
  \label{eq:appA_elementary_trace}
\end{equation}
Consequently, $\Tr(\ad_{H(s)}) = \Tr(H(s))\Tr(\mathbb{I}_d) - \Tr(\mathbb{I}_d)\Tr(H(s)) = 0$. Substituting the trace-traceless decomposition $L_\alpha(s) = c_\alpha(s)\mathbb{I}_d + \tilde{L}_\alpha(s)$ ($\Tr(\tilde{L}_\alpha(s)) = 0$) into Eq.~\eqref{eq:dissipator_def} cancels the $d^2\abs{c_\alpha(s)}^2$ terms identically:
\begin{equation}
  \begin{aligned}
    \Tr\bigl(\mathcal{D}(s)\bigr) &= \sum_\alpha \gamma_\alpha(s)\left[ \abs{\Tr(L_\alpha(s))}^2 - \frac{1}{2}\Tr\bigl(L_\alpha(s)^\dagger L_\alpha(s)\bigr)\Tr(\mathbb{I}_d) - \frac{1}{2}\Tr(\mathbb{I}_d)\Tr\bigl(L_\alpha(s)^\dagger L_\alpha(s)\bigr) \right] \\
    &= -d \sum_\alpha \gamma_\alpha(s)\,\Tr\Bigl(\tilde{L}_\alpha(s)^\dagger \tilde{L}_\alpha(s)\Bigr) = -d\,\Gamma(s).
  \end{aligned}
  \label{eq:appA_dissipator_trace}
\end{equation}
Applying Jacobi's determinant formula to the time-rescaled master equation~\eqref{eq:gksl_rescaled} gives
\begin{equation}
  \det \Phi_\varepsilon(1) = \exp\left(\int_0^1 \Tr\bigl(\mathcal{L}_\varepsilon(s)\bigr)\,\dd s\right) = e^{-d\,\varepsilon\,\Gamma_{\mathrm{tot}}},
  \label{eq:appA_jacobi}
\end{equation}
which establishes Eq.~\eqref{eq:liouville_determinant}. Taking $d^{-2}\Tr(\cdot)$ of the Dyson expansion~\eqref{eq:dyson_first_order} and using cyclic invariance $\Tr(\hat{\mathcal{D}}(s)) = \Tr(\mathcal{U}_0(s)^\dagger \mathcal{D}(s)\mathcal{U}_0(s)) = \Tr(\mathcal{D}(s)) = -d\,\Gamma(s)$ yields Eq.~\eqref{eq:first_order_scalar_invariance}. Finally, by Schur's lemma on the irreducible $\mathrm{SU}(d)$-module decomposition $\mathcal{B}(\mathcal{H}) = \mathbb{C}\mathbb{I}_d \oplus \mathfrak{sl}(d,\mathbb{C})$, the commutant of the adjoint representation $\{\mathcal{U} : U \in \mathrm{SU}(d)\}$ is spanned by the identity and traceless projectors $\mathcal{P}_{\mathbb{I}}$ and $\mathcal{P}_{\mathfrak{sl}}$; since $\Tr(\mathcal{P}_{\mathbb{I}}\mathcal{D}) = 0$ by trace preservation ($\mathcal{D}^\dagger(\mathbb{I}_d) = 0$), any $\mathrm{SU}(d)$-conjugation-invariant linear functional of $\mathcal{D}$ is a scalar multiple of $\Tr(\mathcal{P}_{\mathfrak{sl}}\mathcal{D}) = \Tr(\mathcal{D}) = -d\,\Gamma$.
\end{proof}

\section{Pauli transfer matrix proof of the two-moment trajectory law}
\label{app:ptm_proof}

This appendix establishes an independent Pauli transfer matrix (PTM) derivation of the two-moment decomposition~\eqref{eq:two_moment_law} in Theorem~\ref{thm:two_moment}. In the normalized Pauli basis $\{B_0, B_1, B_2, B_3\} := \{\mathbb{I}_2/\sqrt{2}, \sigma_x/\sqrt{2}, \sigma_y/\sqrt{2}, \sigma_z/\sqrt{2}\}$, a qubit GKSL dissipator $\mathcal{D}$ with traceless jump operators $\tilde{L}_\alpha = (\bm{a}_\alpha + i\bm{b}_\alpha)\cdot\bm{\sigma}$ has PTM entries $[\mathcal{D}]_{0j} = \Tr(B_0\,\mathcal{D}(B_j)) = 0$ (by trace preservation),
\begin{equation*}
  [\mathcal{D}]_{i0} = \Tr\bigl(B_i\,\mathcal{D}(B_0)\bigr) = \frac{1}{2}\Tr\bigl(\sigma_i\,\mathcal{D}(\mathbb{I}_2)\bigr) = \frac{1}{2}t_i \qquad (i \in \{1, 2, 3\}),
\end{equation*}
and unital $3 \times 3$ damping block $[\mathcal{D}]_{ij} = \frac{1}{2}\Tr(\sigma_i\,\mathcal{D}(\sigma_j)) = [S - \Tr(S)\mathbb{I}_3]_{ij}$ ($i, j \in \{1, 2, 3\}$):
\begin{equation}
  [\mathcal{D}]_{\mathrm{PTM}} = \begin{pmatrix} 0 & \bm{0}^T \\ \frac{1}{2}\bm{t} & S - \Tr(S)\mathbb{I}_3 \end{pmatrix}.
  \label{eq:appB_ptm_matrix}
\end{equation}
With the evolving pure-state Pauli vectors $\bm{v}_\pm(s) := [\mathcal{U}_0(s)]_{\mathrm{PTM}}\bm{v}_\pm(0) = \frac{1}{\sqrt{2}}(1, \pm\bm{r}(s)^T)^T$ and $\norm{\bm{r}(s)}^2 = 1$, integrating $-\bm{v}_\pm(0)^T [\mathcal{U}_0(s)^\dagger \mathcal{D}\mathcal{U}_0(s)]_{\mathrm{PTM}}\bm{v}_\pm(0) = -\bm{v}_\pm(s)^T [\mathcal{D}]_{\mathrm{PTM}}\bm{v}_\pm(s)$ over $s \in [0,1]$ gives
\begin{equation}
  \begin{aligned}
    \chi_\pm(\mathcal{D};\bm{r}) &= -\frac{1}{2}\int_0^1 \left[ \pm\frac{1}{2}\bm{r}(s)\cdot\bm{t} + \bm{r}(s)^T\bigl(S - \Gamma\mathbb{I}_3\bigr)\bm{r}(s) \right] ds \\
    &= \frac{1}{2}\Tr\left(S \int_0^1 \bigl(\mathbb{I}_3 - \bm{r}(s)\bm{r}(s)^T\bigr)\,\dd s\right) \mp \frac{1}{4}\left(\int_0^1 \bm{r}(s)\,\dd s\right)\cdot\bm{t} \\
    &= \frac{1}{2}\Tr(SG) \mp \frac{1}{4}\bm{m}\cdot\bm{t},
  \end{aligned}
  \label{eq:appB_ptm_integral}
\end{equation}
recovering Eq.~\eqref{eq:two_moment_law} of Theorem~\ref{thm:two_moment}.

\section{Constrained cross-product minimum and extremal jump uniqueness for Theorem~\ref{thm:two_moment}}
\label{app:cofactor_proof}

This appendix proves the constrained cross-product minimization identity, the matrix cofactor inequality $\adj(G) - \bm{m}\bm{m}^T \succeq 0$, and the uniqueness of the rank-$1$ extremal jump operator $\tilde{L}_{\mathrm{ext}}$ in Theorem~\ref{thm:two_moment}.

\begin{lemma}[Constrained cross-product minimum and trajectory cofactor inequality]
\label{lem:cofactor_minimum}
For any symmetric positive-definite matrix $G \succ 0$ on $\mathbb{R}^3$ and any target cross-product vector $\bm{w} \in \mathbb{R}^3$, the constrained quadratic form $\bm{a}^T G\bm{a} + \bm{b}^T G\bm{b}$ over $\bm{a}, \bm{b} \in \mathbb{R}^3$ satisfying $\bm{a}\times\bm{b} = \bm{w}$ has the exact minimum
\begin{equation}
  \min_{\substack{\bm{a},\bm{b}\in\mathbb{R}^3 \\ \bm{a}\times\bm{b}=\bm{w}}} \bigl(\bm{a}^T G\bm{a} + \bm{b}^T G\bm{b}\bigr) = 2\sqrt{\bm{w}^T \adj(G)\,\bm{w}}.
  \label{eq:appC_constrained_min}
\end{equation}
Consequently, for every piecewise-continuous Bloch trajectory $\bm{r} : [0,1] \to S^2$ with moments $(\bm{m}, G)$, the matrix cofactor inequality $\adj(G) - \bm{m}\bm{m}^T \succeq 0$ holds, implying $\beta(\bm{r}) = \sqrt{\bm{m}^T G\bm{m}/\det G} \le 1$ whenever $\det G > 0$.
\end{lemma}

\begin{proof}
To verify Eq.~\eqref{eq:appC_constrained_min}, define whitened vectors $\tilde{\bm{a}} := G^{1/2}\bm{a}$ and $\tilde{\bm{b}} := G^{1/2}\bm{b}$. The cofactor transformation identity $(M\bm{a})\times(M\bm{b}) = \adj(M^T)(\bm{a}\times\bm{b})$ applied to $M = G^{1/2}$ gives $\tilde{\bm{a}}\times\tilde{\bm{b}} = (\det G)^{1/2}G^{-1/2}\bm{w} = \adj(G)^{1/2}\bm{w}$. Applying the arithmetic--geometric mean inequality and Lagrange's identity $\norm{\tilde{\bm{a}}\times\tilde{\bm{b}}} \le \norm{\tilde{\bm{a}}}\norm{\tilde{\bm{b}}}$ yields
\begin{equation}
  \bm{a}^T G\bm{a} + \bm{b}^T G\bm{b} = \norm{\tilde{\bm{a}}}^2 + \norm{\tilde{\bm{b}}}^2 \ge 2\norm{\tilde{\bm{a}}}\norm{\tilde{\bm{b}}}
  \ge 2\norm{\tilde{\bm{a}}\times\tilde{\bm{b}}} = 2\sqrt{\bm{w}^T \adj(G)\,\bm{w}},
  \label{eq:appC_amgm_chain}
\end{equation}
with equality if and only if $\tilde{\bm{a}} \perp \tilde{\bm{b}} \perp \adj(G)^{1/2}\bm{w}$ and $\norm{\tilde{\bm{a}}} = \norm{\tilde{\bm{b}}}$. Because the integrated jump variance~\eqref{eq:single_jump_variance} is non-negative for every $\bm{a}, \bm{b} \in \mathbb{R}^3$ ($\bm{a}^T G\bm{a} + \bm{b}^T G\bm{b} \ge 2\abs{\bm{w}\cdot\bm{m}}$), minimizing the left-hand side over $\bm{a}\times\bm{b} = \bm{w}$ via Eq.~\eqref{eq:appC_constrained_min} (and taking the limit $G + \epsilon\mathbb{I}_3 \to G \succeq 0$ when $G$ is singular) establishes
\begin{equation}
  \bm{w}^T \adj(G)\,\bm{w} \ge (\bm{w}\cdot\bm{m})^2 \quad \Longleftrightarrow \quad \adj(G) - \bm{m}\bm{m}^T \succeq 0.
  \label{eq:appC_cofactor_psd}
\end{equation}
Evaluating Eq.~\eqref{eq:appC_cofactor_psd} at $\bm{w} = G\bm{m}$ and using $G\,\adj(G)\,G = (\det G)G$ gives $(\det G)(\bm{m}^T G\bm{m}) \allowbreak \ge (\bm{m}^T G\bm{m})^2$, or $\bm{m}^T G\bm{m} \le \det G$, proving $\beta(\bm{r}) \le 1$. Finally, in Eq.~\eqref{eq:whitened_cross_product_bound}, equality holds if and only if $\tilde{\bm{a}}\times\tilde{\bm{b}} \parallel G^{1/2}\bm{m}$, $\tilde{\bm{a}} \perp \tilde{\bm{b}}$, and $\norm{\tilde{\bm{a}}} = \norm{\tilde{\bm{b}}}$; any two such equal-norm orthogonal pairs in the two-dimensional plane $(G^{1/2}\bm{m})^\perp$ differ only by an $\mathrm{SO}(2)$ gauge rotation ($\tilde{\bm{a}} + i\tilde{\bm{b}} \mapsto e^{-i\phi}(\tilde{\bm{a}} + i\tilde{\bm{b}})$) and orientation reflection ($\tilde{\bm{b}} \mapsto -\tilde{\bm{b}}$, i.e., $\tilde{L}_{\mathrm{ext}} \mapsto \tilde{L}_{\mathrm{ext}}^\dagger$).
\end{proof}

\section{Midpoint-frame block-algebra derivation for constant-axis rotations}
\label{app:midpoint_frame}

This appendix derives the midpoint-sinc formula~\eqref{eq:midpoint_sinc} of Proposition~\ref{prop:midpoint_sinc} for general constant-axis rotations $H(Q,\vartheta) = \frac{Q}{2}(\bm{n}\cdot\bm{\sigma})$ with $\bm{n} = (\sin\vartheta, 0, \cos\vartheta)^T$ in Section~\ref{sec:constant_axis_rotations}. Introduce the right-handed orthonormal frame $\bm{e}_\parallel := \bm{n}$, $\bm{e}_1 := (\cos\vartheta, 0, -\sin\vartheta)^T$, $\bm{e}_2 := (0, 1, 0)^T$, and rotate $\{\bm{e}_1, \bm{e}_2\}$ by the midpoint angle $Q/2$ to $\bm{u}_1 := \cos(Q/2)\bm{e}_1 + \sin(Q/2)\bm{e}_2$ and $\bm{u}_2 := -\sin(Q/2)\bm{e}_1 + \cos(Q/2)\bm{e}_2$. Parameterizing time relative to the pulse midpoint via $u := s - 1/2 \in [-1/2, 1/2]$ expresses the trajectory symmetrically as
\begin{equation}
  \bm{r}(u) = \cos\vartheta\,\bm{e}_\parallel - \sin\vartheta\bigl[\cos(Qu)\,\bm{u}_1 + \sin(Qu)\,\bm{u}_2\bigr].
  \label{eq:appD_midpoint_trajectory}
\end{equation}
Integrating over $u \in [-1/2, 1/2]$ in the orthonormal basis $\{\bm{e}_\parallel, \bm{u}_1, \bm{u}_2\}$ decouples the odd sine mode $\bm{u}_2$, giving $\bm{m}_{\mathrm{mid}} = (\cos\vartheta, -\sinc_0\sin\vartheta, 0)^T$ and the block-diagonal covariance deficit
\begin{equation}
  G_{\mathrm{mid}} = \diag\left(G_{2\times 2},\; 1 - \frac{1 - \sinc_1}{2}\sin^2\vartheta\right), \qquad
  G_{2\times 2} = \begin{pmatrix} \sin^2\vartheta & \sinc_0\sin\vartheta\cos\vartheta \\ \sinc_0\sin\vartheta\cos\vartheta & 1 - \frac{1+\sinc_1}{2}\sin^2\vartheta \end{pmatrix}.
  \label{eq:appD_Gmid}
\end{equation}
The determinant of the $2 \times 2$ block $G_{2\times 2}$ simplifies to
\begin{equation}
  \begin{aligned}
    \det G_{2\times 2} &= \sin^2\vartheta\left(1 - \frac{1 + \sinc_1}{2}\sin^2\vartheta - \sinc_0^2\cos^2\vartheta\right) \\
    &= \frac{1}{2}\sin^2\vartheta\Bigl[2(1 - \sinc_0^2)\cos^2\vartheta + (1 - \sinc_1)\sin^2\vartheta\Bigr].
  \end{aligned}
  \label{eq:appD_detG2}
\end{equation}
Thus $\det G_{\mathrm{mid}} = \det G_{2\times 2}\cdot G_{33} = \frac{1}{4}\sin^2\vartheta\,D(\abs{Q},\vartheta)$, while the numerator quadratic form $\bm{m}_{\mathrm{mid}}^T G_{\mathrm{mid}}\bm{m}_{\mathrm{mid}}$ (which equals $\bm{m}_2^T G_{2\times 2}\bm{m}_2$) evaluates to
\begin{equation}
  \begin{aligned}
    \bm{m}_2^T G_{2\times 2}\bm{m}_2 &= \sin^2\vartheta\left[(1 - \sinc_0^2)\cos^2\vartheta + \frac{1 - \sinc_1}{2}\sinc_0^2\sin^2\vartheta\right] \\
    &= \frac{1}{2}\sin^2\vartheta\,N(\abs{Q},\vartheta).
  \end{aligned}
  \label{eq:appD_numG2}
\end{equation}
Dividing Eq.~\eqref{eq:appD_numG2} by $\det G_{\mathrm{mid}}$ cancels $\frac{1}{2}\sin^2\vartheta$ and establishes Eq.~\eqref{eq:midpoint_sinc}.

\section{First-order tangent cone, finite-damping Dyson bound, and exact damped-Bloch solution}
\label{app:tangent_dyson}

This appendix establishes the Bouligand tangent-cone closure of the survival-error map, the coherent over-rotation immunity identity used in Section~\ref{sec:lockin_protocol}, and the exact finite-$\varepsilon$ derivations for Section~\ref{sec:finite_damping}.

\subsection{First-order Bouligand tangent cone and coherent over-rotation immunity}
\label{app:tangent_cone}

\begin{proposition}[First-order tangent cone and coherent calibration immunity]
\label{prop:tangent_cone}
Let $\Pi_{\mathrm{swap}}(\Phi) := (p_+(\Phi), p_-(\Phi))$ denote the directional survival-error pair of a noisy swap channel $\Phi$. Along any regular sequence of channels $\Phi_n = \exp(-i\,\ad_{H_q + \delta H_n} + \mathcal{D}_n) \to \mathcal{U}_q$ whose coherent Hamiltonian calibration error satisfies $\norm{\delta H_n}_{\mathrm{op}} = o(\norm{\mathcal{D}_n}_{\mathrm{HS}}^{1/2})$, determinant convergence $\det\Phi_n = e^{-2\Gamma_n} \to 1$ forces $\tau_n := \norm{\mathcal{D}_n}_{\mathrm{HS}} \to 0$, and the first-order Bouligand tangent cone of $\Pi_{\mathrm{swap}}$ at $\mathcal{U}_q$ is the closed sector
\begin{equation}
  \mathcal{C}_q = \Bigl\{ (x_+, x_-) \in \mathbb{R}_+^2 \;\Big|\; \kappa_q\,x_+ \le x_- \le \kappa_q^{-1}\,x_+ \Bigr\}.
  \label{eq:appE_tangent_cone}
\end{equation}
\end{proposition}

\begin{proof}
With $\mathcal{L}_0 = -i\,\ad_{H_q}$ and $\Delta_n = -i\,\ad_{\delta H_n} + \mathcal{D}_n$, Duhamel expansion gives
\begin{equation}
  \Phi_n = \mathcal{U}_q(1) + \int_0^1 \mathcal{U}_q(1 - s)\,\Delta_n\,\mathcal{U}_q(s)\,\dd s + \mathcal{R}_2(\Delta_n),
  \label{eq:appE_duhamel}
\end{equation}
where the second-order remainder obeys
\begin{equation}
  \norm{\mathcal{R}_2(\Delta_n)}_{\mathrm{op}} \le \Bigl(\norm{\ad_{\delta H_n}}_{\mathrm{op}}^2 + \norm{\mathcal{D}_n}_{\mathrm{op}}^2\Bigr)\exp\bigl(\norm{\Delta_n}_{\mathrm{op}}\bigr),
  \label{eq:appE_rem_op}
\end{equation}
so that under $\norm{\delta H_n}_{\mathrm{op}} = o(\tau_n^{1/2})$,
\begin{equation}
  \frac{\norm{\mathcal{R}_2(\Delta_n)}_{\mathrm{op}}}{\tau_n} \le \left(\frac{4\norm{\delta H_n}_{\mathrm{op}}^2}{\tau_n} + \tau_n\right)e^{\norm{\Delta_n}_{\mathrm{op}}} \xrightarrow{n\to\infty} 0.
  \label{eq:appE_rem_limit}
\end{equation}
Because the $3 \times 3$ Bloch generator $A_n := [-i\,\ad_{\delta H_n}]_{\mathrm{PTM},3\times 3}$ of any coherent Hamiltonian perturbation $\delta H_n$ is real skew-symmetric ($A_n^T = -A_n$), its quadratic form with the symmetric state trajectory $\bm{r}_q(s)$ vanishes identically at every point along the trajectory:
\begin{equation}
  -\frac{1}{2}\int_0^1 \bm{r}_q(s)^T A_n\,\bm{r}_q(s)\,\dd s = 0.
  \label{eq:appE_skew_symmetry}
\end{equation}
Coherent over-rotation and axis-tilt errors therefore contribute zero first-order survival error ($\delta p_\pm^{\mathrm{coh}} = \mathcal{O}(\norm{\delta H_n}^2)$), leaving the normalized limit ray
\begin{equation}
  \frac{\Pi_{\mathrm{swap}}(\Phi_n)}{\tau_n} = \Bigl(\chi_+(\hat{\mathcal{D}}_n; H_q),\; \chi_-(\hat{\mathcal{D}}_n; H_q)\Bigr) + o(1) \in \mathcal{C}_q,
  \label{eq:appE_limit_ray}
\end{equation}
where $\hat{\mathcal{D}}_n = \mathcal{D}_n/\tau_n$, establishing Eq.~\eqref{eq:appE_tangent_cone}.
\end{proof}

\subsection{Finite-damping Dyson bound, exact Liouvillian exponentiation, and boundary jets}
\label{app:finite_damping_proof}

\begin{proposition}[Finite-damping Dyson ratio bound and exact damped-Bloch integration]
\label{prop:finite_damping}
For any constant-amplitude swap $H_q = -\frac{q\pi}{2}\sigma_x$ ($q \in 2\mathbb{Z} + 1$) and any time-independent Markovian bath $\mathcal{D}$ with $\eta_{\mathrm{rem}}(\varepsilon) = 8\varepsilon\Gamma e^{2\varepsilon\Gamma} < 1$, the exact finite-$\varepsilon$ survival errors $a(\varepsilon) = \max(p_+(\varepsilon),p_-(\varepsilon))$ and $b(\varepsilon) = \min(p_+(\varepsilon),p_-(\varepsilon))$ satisfy Eq.~\eqref{eq:finite_eps_dyson_bound}. Furthermore, for $L_+ = (\sigma_x + i\sqrt{2}\sigma_z)/\sqrt{6}$, the exact all-orders channel $\exp(\mathcal{L}_\varepsilon)$ is given by Eqs.~\eqref{eq:exact_Fe}--\eqref{eq:exact_J}.
\end{proposition}

\begin{proof}
To prove the finite-$\varepsilon$ Dyson guarantee~\eqref{eq:finite_eps_dyson_bound}, observe that $G_q = \diag(1, \frac{1}{2}, \frac{1}{2})$ has minimum eigenvalue $\lambda_{\min}(G_q) = \frac{1}{2}$ and $S \succeq 0$ has $\Tr(S) = \Gamma$, so Eq.~\eqref{eq:two_moment_law} yields the universal lower bound $\chi_{\max} := \max(\chi_+,\chi_-) \ge \frac{1}{2}(\chi_+ + \chi_-) = \frac{1}{2}\Tr(SG_q) \ge \frac{1}{2}\lambda_{\min}(G_q)\Tr(S) = \frac{1}{4}\Gamma$ (saturated by pure dephasing $L_\phi = \sigma_z/\sqrt{2}$). Combining $\chi_{\min} := \min(\chi_+,\chi_-) \ge \kappa_q\chi_{\max}$ with the Dyson remainder bound $R_2(\varepsilon) \le 2\varepsilon^2\Gamma^2 e^{2\varepsilon\Gamma}$ of Eq.~\eqref{eq:dyson_remainder_bound} gives $b(\varepsilon) \ge \varepsilon\chi_{\min} - R_2(\varepsilon) \ge \kappa_q\varepsilon\chi_{\max} - R_2(\varepsilon)$ and $a(\varepsilon) \le \varepsilon\chi_{\max} + R_2(\varepsilon)$. Dividing numerator and denominator by $\varepsilon\chi_{\max} \ge \frac{1}{4}\varepsilon\Gamma$ implies $R_2(\varepsilon)/(\varepsilon\chi_{\max}) \le 8\varepsilon\Gamma e^{2\varepsilon\Gamma} = \eta_{\mathrm{rem}}(\varepsilon)$, and therefore
\begin{equation}
  \frac{b(\varepsilon)}{a(\varepsilon)} \ge \frac{\kappa_q - \eta_{\mathrm{rem}}(\varepsilon)}{1 + \eta_{\mathrm{rem}}(\varepsilon)}
  = \kappa_q\left(1 - \frac{(1 + \kappa_q)\eta_{\mathrm{rem}}(\varepsilon)}{\kappa_q(1 + \eta_{\mathrm{rem}}(\varepsilon))}\right)
  \ge \kappa_q\Bigl(1 - C_q\,\varepsilon\Gamma\,e^{2\varepsilon\Gamma}\Bigr),
  \label{eq:appE_dyson_chain}
\end{equation}
where $C_q = 8(1 + \kappa_q)/\kappa_q = 16\pi\abs{q}/(\pi\abs{q} - 2\sqrt{2})$, proving Eq.~\eqref{eq:finite_eps_dyson_bound} (for general jump operators with $c_\alpha \neq 0$, replacing $\varepsilon\Gamma e^{2\varepsilon\Gamma}$ by $\frac{\Gamma_{\mathrm{full}}^2}{\Gamma}\varepsilon\,e^{2\varepsilon\Gamma_{\mathrm{full}}}$ in $\eta_{\mathrm{rem}}(\varepsilon)$ yields the corresponding bound).

To derive the exact all-orders damped-Bloch formulas~\eqref{eq:exact_Fe}--\eqref{eq:exact_J} under $L_+ = (\sigma_x + i\sqrt{2}\sigma_z)/\sqrt{6}$ ($\Gamma = 1$) and $H_q = -\frac{q\pi}{2}\sigma_x$ ($Q = q\pi$), we write the $4 \times 4$ PTM generator of $\mathcal{L}_\varepsilon = -i\,\ad_{H_q} + \varepsilon\,\mathcal{D}[L_+]$ in the basis $\{B_0, B_1, B_2, B_3\}$:
\begin{equation}
  [\mathcal{L}_\varepsilon]_{\mathrm{PTM}} = \begin{pmatrix}
    0 & 0 & 0 & 0 \\
    0 & -\frac{2\varepsilon}{3} & 0 & 0 \\
    -\frac{2\sqrt{2}}{3}\varepsilon & 0 & -\varepsilon & Q \\
    0 & 0 & -Q & -\frac{\varepsilon}{3}
  \end{pmatrix}.
  \label{eq:appE_Lplus_ptm}
\end{equation}
Thus $[\exp(\mathcal{L}_\varepsilon)]_{11} = e^{-2\varepsilon/3}$, while the coupled $(B_2, B_3)$ block
\begin{equation}
  K_{yz} := \begin{pmatrix} -\varepsilon & Q \\ -Q & -\frac{\varepsilon}{3} \end{pmatrix} = -\frac{2\varepsilon}{3}\mathbb{I}_2 + \begin{pmatrix} -\frac{\varepsilon}{3} & Q \\ -Q & \frac{\varepsilon}{3} \end{pmatrix}
  \label{eq:appE_Kyz}
\end{equation}
exponentiates in closed form via Euler's identity with damped frequency $\Omega_{\varepsilon,q} = \sqrt{Q^2 - \varepsilon^2/9}$:
\begin{equation}
  e^{sK_{yz}} = e^{-\frac{2\varepsilon s}{3}}\left[ \cos(\Omega_{\varepsilon,q}s)\mathbb{I}_2 + \frac{\sin(\Omega_{\varepsilon,q}s)}{\Omega_{\varepsilon,q}}\begin{pmatrix} -\varepsilon/3 & Q \\ -Q & \varepsilon/3 \end{pmatrix} \right].
  \label{eq:appE_expKyz}
\end{equation}
Evaluating the diagonal entries at $s = 1$ yields Eqs.~\eqref{eq:exact_Fe}--\eqref{eq:exact_C}, and integrating the affine drift column via $K_{yz}^{-1}(\exp(K_{yz}) - \mathbb{I}_2)(-\frac{2\sqrt{2}}{3}\varepsilon, 0)^T$ yields Eq.~\eqref{eq:exact_J}.
\end{proof}

In the high-winding limit $\abs{q} \to \infty$ at fixed scaled exposure $u := \varepsilon/(\abs{q}\pi) > 0$, Eqs.~\eqref{eq:exact_C}--\eqref{eq:exact_J} give $\sup_{u>0} \lim_{\abs{q}\to\infty} p_+(u\abs{q}\pi,q)/p_-(u\abs{q}\pi,q) = (\sqrt{3} + \sqrt{2})^2 = 5 + 2\sqrt{6} \approx 9.898979$. Under a retimed drive $H_\varepsilon(\lambda) = -\frac{\pi}{2}(1 + \lambda\varepsilon)\sigma_x$, the boundary expansion $b(\varepsilon) = \kappa_\star a(\varepsilon) + c_{L_+}(\lambda)a(\varepsilon)^2 + \mathcal{O}(\varepsilon^3)$ for $L_+$ has second-order curvature $c_{L_+}(\lambda) = \sqrt{2}[(3\pi^2\lambda + 1)^2 - 1]/(\pi + 2\sqrt{2})^3$, which vanishes at $\lambda = 0$ and leaves the third-order term~\eqref{eq:cubic_excursion}.

Finally, for the rank-$1$ nilpotent jump $L_\star = \dyad{\psi(\pi/8)}{\psi(7\pi/8)} = -\frac{\sqrt{2}}{4}\mathbb{I}_2 - \frac{i\sqrt{2}}{4}\sigma_x - \frac{1}{2}\sigma_z$ ($c = -\frac{\sqrt{2}}{4}$, $\Gamma = \frac{3}{4}$) with retimed drive $H_\varepsilon(\lambda) = \frac{\pi}{2}(1 + \lambda\varepsilon)\sigma_x$, the Lamb shift $\Delta H_{\mathrm{Lamb}} = -\frac{1}{8}\sigma_x$ gives the PTM generator (with $\tilde{\Omega}_\lambda := \pi(1 + \lambda\varepsilon) - \frac{\varepsilon}{4}$)
\begin{equation}
  [\mathcal{L}_{\varepsilon,L_\star}(\lambda)]_{\mathrm{PTM}} = \begin{pmatrix}
    0 & 0 & 0 & 0 \\
    0 & -\frac{\varepsilon}{2} & 0 & 0 \\
    \frac{\sqrt{2}}{2}\varepsilon & 0 & -\frac{3\varepsilon}{4} & -\tilde{\Omega}_\lambda \\
    0 & 0 & \tilde{\Omega}_\lambda & -\frac{\varepsilon}{4}
  \end{pmatrix}.
  \label{eq:appE_Lstar_ptm}
\end{equation}
Expanding the $(B_2, B_3)$ block in powers of $\varepsilon$ (where $a(\varepsilon) = A_\star\varepsilon + a_2(\lambda)\varepsilon^2 + \mathcal{O}(\varepsilon^3)$ with $A_\star = (\pi + 2\sqrt{2})/(4\pi)$ and $P(\lambda) := 16\pi^4\lambda^2 - 8\pi^2(\pi + 1)\lambda - 3\pi^2 + 2\pi$) gives
\begin{align}
  a_2(\lambda) &= \frac{P(\lambda) - 8\sqrt{2}(4\pi\lambda + \pi - 1)}{64\pi^2}, \label{eq:appE_a2_lambda} \\
  b_2(\lambda) &= \frac{P(\lambda) + 8\sqrt{2}(4\pi\lambda + \pi - 1)}{64\pi^2}, \label{eq:appE_b2_lambda}
\end{align}
yielding the quadratic curvature $c(\lambda) = [b_2(\lambda) - \kappa_\star a_2(\lambda)]/A_\star^2$:
\begin{equation}
  c(\lambda) = \frac{\sqrt{2}\bigl[(4\pi^2\lambda - (\pi - 1))^2 - 1\bigr]}{(\pi + 2\sqrt{2})^3}, \qquad
  c_{\min} = -\frac{\sqrt{2}}{(\pi + 2\sqrt{2})^3} \approx -0.006646418567,
  \label{eq:appE_c_lambda}
\end{equation}
which attains its minimum at $\lambda_\star = (\pi - 1)/(4\pi^2) \approx 0.054247175635$ (Figure~\ref{fig:fig5}(a)), where the exact second- and third-order jet coefficients ($a_2^\star \approx -0.10831902$, $b_2^\star \approx -0.00718212$, $c_3^\star \approx +0.006344064267$) are
\begin{equation}
  \begin{aligned}
    a_2^\star &= -\frac{4\pi^3 + 8\sqrt{2}\pi^2 - 3\pi - 8\sqrt{2}}{64\pi^3}, \\
    b_2^\star &= -\frac{4\pi^3 - 8\sqrt{2}\pi^2 - 3\pi + 8\sqrt{2}}{64\pi^3}, \\
    c_3^\star &= -\frac{8\sqrt{2}\pi^4 + 32\pi^3 - 90\sqrt{2}\pi^2 - 360\pi + 3\sqrt{2}}{6(\pi + 2\sqrt{2})^5}.
  \end{aligned}
  \label{eq:appE_jet_star}
\end{equation}

\section{Moments and selectivity scaling of non-autonomous wait-flip-wait schedules}
\label{app:dwell_scaling}

This appendix derives the non-autonomous pulse-dwell moments and selectivity scaling~\eqref{eq:beta_dwell} presented in Section~\ref{sec:pulse_dwell}. For a wait-flip-wait schedule idling at $\ket{0}$ for duration $w_+ = \frac{1+\eta_w}{2}(1 - \delta)$, flipping at rate $-\frac{\pi}{2\delta}\sigma_x$ over active duration $\delta \in (0,1]$, and idling at $\ket{1}$ for $w_- = \frac{1-\eta_w}{2}(1 - \delta)$ with wait asymmetry $\eta_w \in [-1,1]$, summing $\mathbb{I}_3 - \bm{r}(s)\bm{r}(s)^T$ over the two pole-idle intervals of total length $1 - \delta$ (where $\mathbb{I}_3 - \bm{e}_z\bm{e}_z^T = \diag(1, 1, 0)$) and the active flip interval of length $\delta$ yields
\begin{equation}
  \bm{m}(\delta,\eta_w) = \left(0,\; \frac{2\delta}{\pi},\; \eta_w(1 - \delta)\right)^T, \qquad
  G(\delta) = \diag\left(1,\; 1 - \frac{\delta}{2},\; \frac{\delta}{2}\right), \qquad
  \det G(\delta) = \frac{\delta}{2}\left(1 - \frac{\delta}{2}\right).
  \label{eq:appF_moments}
\end{equation}
Evaluating the quadratic form
\begin{equation}
  \bm{m}(\delta,\eta_w)^T G(\delta)\bm{m}(\delta,\eta_w) = \frac{4\delta^2}{\pi^2}\left(1 - \frac{\delta}{2}\right) + \frac{\delta}{2}\eta_w^2(1 - \delta)^2
  \label{eq:appF_quadform}
\end{equation}
and dividing by $\det G(\delta) = \frac{\delta}{2}(1 - \frac{\delta}{2})$ yields Eq.~\eqref{eq:beta_dwell}, from which the one-sided wait-then-flip ($\eta_w = \pm 1$) and symmetric wait-flip-wait ($\eta_w = 0$) selectivity bounds follow:
\begin{align}
  \kappa_{\mathrm{dwell}}(\delta) &= \frac{1 - \beta(\delta,\pm 1)}{1 + \beta(\delta,\pm 1)} = \frac{3\pi^2 - 16}{8\pi^2}\,\delta + \mathcal{O}(\delta^2), \label{eq:appF_kdwell} \\
  \kappa_{\mathrm{sym}}(\delta) &= \frac{\pi - 2\sqrt{2}\sqrt{\delta}}{\pi + 2\sqrt{2}\sqrt{\delta}} = 1 - \frac{4\sqrt{2}}{\pi}\sqrt{\delta} + \mathcal{O}(\delta). \label{eq:appF_ksym}
\end{align}

\section{Colored thermal baths, Redfield spectral parity separation, and multilevel qutrit cascade dynamics}
\label{app:colored_qutrit}

This appendix provides the complete derivations of the colored-bath spectral parity selection rule, the calibration-free quantum thermometer~\eqref{eq:quantum_thermometer} of Section~\ref{sec:colored_baths}, and the all-orders single-qubit chiral fidelity invariance versus the multilevel qutrit ($d = 3$) cascade shelving split~\eqref{eq:qutrit_cascade_split} of Section~\ref{sec:qutrit_cascade}.

\subsection{Qubit chiral fidelity invariance versus qutrit cascade shelving anomaly}
\label{app:qutrit_theorem}

\begin{theorem}[Qubit chiral fidelity invariance versus $d \ge 3$ chiral anomaly and $\eta_{\mathrm{crit}}$]
\label{thm:qutrit_chiral}
\leavevmode
\begin{enumerate}
  \item \emph{All-orders single-qubit chiral fidelity invariance ($d = 2$):} For every piecewise-continuous control Hamiltonian $H(s)$, every time-independent Markovian bath $\mathcal{D}[\{\tilde{L}_\alpha\}]$, and all $\varepsilon \ge 0$, replacing every traceless jump operator by its Hermitian conjugate ($\tilde{L}_\alpha \mapsto \tilde{L}_\alpha^\dagger$) leaves the entanglement fidelity identically invariant:
  \begin{equation}
    F_e\Bigl(\Phi_\varepsilon\bigl[\{\tilde{L}_\alpha\}\bigr],\; U_0(1)\Bigr) \equiv F_e\Bigl(\Phi_\varepsilon\bigl[\{\tilde{L}_\alpha^\dagger\}\bigr],\; U_0(1)\Bigr) \qquad (\forall\,\varepsilon \ge 0).
    \label{eq:appG_chiral_invariance}
  \end{equation}
  \item \emph{Qutrit ($d = 3$) cascade shelving fidelities and critical ratio $\eta_{\mathrm{crit}}$:} On $\mathcal{H} = \mathrm{span}\{\ket{0}, \ket{1}, \ket{2}\}$ under cascade decay $\varepsilon\mathcal{D} = \varepsilon\mathcal{D}[\dyad{0}{1}] + \varepsilon\eta\,\mathcal{D}[\dyad{1}{2}]$ ($\eta \ge 0$), the endpoint-matched $2\pi$ shelving rotations $H_{02}(2\pi) = \pi(\dyad{0}{2} + \dyad{2}{0})$ and $H_{12}(2\pi) = \pi(\dyad{1}{2} + \dyad{2}{1})$ share the same first-order entanglement infidelity $1 - F_e = \frac{1+\eta}{3}\varepsilon + \mathcal{O}(\varepsilon^2)$, while their second-order entanglement-fidelity difference is
  \begin{equation}
    F_e\bigl(\varepsilon; H_{02}(2\pi), \eta\bigr) - F_e\bigl(\varepsilon; H_{12}(2\pi), \eta\bigr) = \frac{6 - 4\eta - 3\eta^2}{144}\,\varepsilon^2 + \mathcal{O}(\varepsilon^3),
    \label{eq:appG_qutrit_split}
  \end{equation}
  which vanishes and reverses sign at the irrational cascade ratio
  \begin{equation}
    \eta_{\mathrm{crit}} = \frac{\sqrt{22} - 2}{3} \approx 0.896805253274.
    \label{eq:appG_eta_crit}
  \end{equation}
\end{enumerate}
\end{theorem}

\begin{proof}
In the $4 \times 4$ PTM representation~\eqref{eq:appB_ptm_matrix} of Appendix~\ref{app:ptm_proof}, trace preservation ($[\mathcal{L}_\varepsilon(s)]_{0j} = 0$) makes $\Phi_\varepsilon(1)$ lower-block-triangular with diagonal blocks $1$ and $M_\varepsilon(1) = \mathcal{T}\exp\int_0^1 ([\mathcal{L}_0(s)]_{3\times 3} + \varepsilon[S - \Tr(S)\mathbb{I}_3])\,\dd s$. Because Hermitian conjugation of the traceless jump operators $\tilde{L}_\alpha = (\bm{a}_\alpha + i\bm{b}_\alpha)\cdot\bm{\sigma} \mapsto \tilde{L}_\alpha^\dagger = (\bm{a}_\alpha - i\bm{b}_\alpha)\cdot\bm{\sigma}$ negates $\bm{b}_\alpha \mapsto -\bm{b}_\alpha$, it reverses the non-unital drift $\bm{t} \mapsto -\bm{t}$ in the off-diagonal first column of $[\mathcal{D}]_{\mathrm{PTM}}$ while leaving the unital $3 \times 3$ damping tensor $S = 2\sum_\alpha \gamma_\alpha(\bm{a}_\alpha\bm{a}_\alpha^T + \bm{b}_\alpha\bm{b}_\alpha^T)$ and hence $M_\varepsilon(1)$ and $F_e(\Phi_\varepsilon, U_0(1)) = \frac{1}{4}[1 + \Tr(R(1)^T M_\varepsilon(1))]$ identically invariant at all orders in $\varepsilon \ge 0$.

For a three-level qutrit ($d = 3$), by contrast, the second-order Dyson coefficient of entanglement fidelity is
\begin{equation}
  c_2^{(ab)}(2\pi,\eta) = \frac{1}{d^2}\int_0^1 \dd s_1 \int_0^{s_1} \dd s_2\,\Tr\Bigl(\hat{\mathcal{D}}(s_1)\hat{\mathcal{D}}(s_2)\Bigr)
  = \frac{1}{18}\int_0^1 \dd s_1 \int_0^1 \dd s_2\,\Tr\Bigl(\hat{\mathcal{D}}(s_1)\hat{\mathcal{D}}(s_2)\Bigr),
  \label{eq:appG_c2_integral}
\end{equation}
where the second equality uses $\Tr(\hat{\mathcal{D}}(s_1)\hat{\mathcal{D}}(s_2)) = \Tr(\hat{\mathcal{D}}(s_2)\hat{\mathcal{D}}(s_1))$. Under $H_{02}(2\pi)$, the evolving states are $\ket{0(s)} = \cos(\pi s)\ket{0} - i\sin(\pi s)\ket{2}$, $\ket{1(s)} = \ket{1}$, and $\ket{2(s)} = -i\sin(\pi s)\ket{0} + \cos(\pi s)\ket{2}$; under $H_{12}(2\pi)$, $\ket{0(s)} = \ket{0}$, $\ket{1(s)} = \cos(\pi s)\ket{1} - i\sin(\pi s)\ket{2}$, and $\ket{2(s)} = -i\sin(\pi s)\ket{1} + \cos(\pi s)\ket{2}$. Evaluating the superoperator trace inner products over $(s_1, s_2) \in [0,1]^2$ yields the exact quadratic polynomials
\begin{equation}
  c_2^{(02)}(2\pi,\eta) = \frac{10 + 16\eta + 7\eta^2}{144}, \qquad
  c_2^{(12)}(2\pi,\eta) = \frac{4 + 20\eta + 10\eta^2}{144}.
  \label{eq:appG_c2_polynomials}
\end{equation}
Subtracting $c_2^{(12)}(2\pi,\eta)$ from $c_2^{(02)}(2\pi,\eta)$ gives $(6 - 4\eta - 3\eta^2)/144$, whose unique positive root is $\eta_{\mathrm{crit}} = (\sqrt{22} - 2)/3$.
\end{proof}

\subsection{Redfield spectral parity separation and KMS quantum thermometry}
\label{app:redfield_thermometry}

\begin{proposition}[Colored-bath odd-spectrum selection rule and bath thermometry]
\label{prop:redfield_parity}
For a single qubit coupled weakly via $H_{\mathrm{int}} = \sum_{j\in\{x,y,z\}} \sigma_j \otimes B_j$ to a stationary Gaussian bath with two-time correlation matrix $C_{jk}(\tau) := \langle B_j(\tau)B_k(0)\rangle_B$ and spectral density matrix $S(\nu) = \int_{-\infty}^\infty C(\tau)e^{i\nu\tau}\,\dd\tau = S^{(s)}(\nu) + i\,S^{(a)}(\nu) \succeq 0$ ($S^{(s)}(-\nu) = S^{(s)}(\nu)$ real symmetric, $S^{(a)}(-\nu) = -S^{(a)}(\nu)$ real skew-symmetric), the first-order Redfield survival error slopes satisfy
\begin{equation}
  \chi_+ + \chi_- = \int_{-\infty}^\infty \Tr\Bigl(S^{(s)}(\nu)\,F_{\mathrm{even}}(\nu)\Bigr)\frac{\dd\nu}{2\pi}, \qquad
  \chi_+ - \chi_- = -\int_{-\infty}^\infty \Tr\Bigl(S^{(a)}(\nu)\,F_{\mathrm{odd}}(\nu)\Bigr)\frac{\dd\nu}{2\pi},
  \label{eq:appG_redfield_parity}
\end{equation}
where $F_{\mathrm{even}}(-\nu) = F_{\mathrm{even}}(\nu) \succeq 0$ is real symmetric and $F_{\mathrm{odd}}(-\nu) = -F_{\mathrm{odd}}(\nu)$ is real skew-symmetric. Under an autonomous swap $H_q = -\frac{q\pi}{2}\sigma_x$ coupled to a transverse thermal bath at frequency $\omega_b$ and temperature $T_{\mathrm{bath}}$ ($\gamma_\phi = 0$), the directional survival slopes evaluate to
\begin{equation}
  \chi_\pm(q) = \frac{3}{8}(\gamma_\downarrow + \gamma_\uparrow) \mp \frac{1}{q\pi}(\gamma_\downarrow - \gamma_\uparrow),
  \label{eq:appG_thermal_slopes}
\end{equation}
which yields the calibration-free quantum thermometer~\eqref{eq:quantum_thermometer} of the main text.
\end{proposition}

\begin{proof}
In the toggling frame of the control unitary $U_0(s)$ with $R_{jl}(s) := \frac{1}{2}\Tr(\sigma_j U_0(s)^\dagger \sigma_l U_0(s)) \in \mathrm{SO}(3)$, the second-order Dyson expansion of the reduced density matrix $\tilde{\rho}(1) = \Tr_B[U_{\mathrm{tot}}(1)(\rho_{\mathrm{in}} \otimes \rho_B)U_{\mathrm{tot}}(1)^\dagger]$ expresses the survival error probabilities $p_\pm = 1 - \bra{\psi_\pm}\tilde{\rho}_\pm(1)\ket{\psi_\pm}$ for $\rho_{\mathrm{in},\pm} = \frac{1}{2}(\mathbb{I}_2 \pm \sigma_z)$ as double time integrals over $0 \le s_2 \le s_1 \le 1$. Transforming to the Fourier representation $C_{jk}(T(s_1 - s_2)) = \int_{-\infty}^\infty S_{jk}(\nu)e^{-i\nu T(s_1 - s_2)}\frac{\dd\nu}{2\pi}$ and symmetrizing over $(s_1, s_2) \in [0,1]^2$:
\begin{enumerate}
  \item The $\sigma_z$-independent unital commutator term $\Tr([\tilde{\sigma}_j(s_1), [\tilde{\sigma}_k(s_2), \mathbb{I}_2]]\sigma_z) = 0$ vanishes, while the state-even survival sum $\chi_+ + \chi_-$ depends on $\Tr(\sigma_z[\tilde{\sigma}_j(s_1), [\tilde{\sigma}_k(s_2), \sigma_z]])$, which is symmetric under $(j, s_1) \leftrightarrow (k, s_2)$ and pairs exclusively with the real frequency-even spectrum $S^{(s)}(\nu)$ via the positive-semidefinite filter matrix $F_{\mathrm{even}}(\nu)$.
  \item The state-odd directional difference $\chi_+ - \chi_-$ depends on the anticommutator $\Tr\bigl(\sigma_z[\tilde{\sigma}_j(s_1),\allowbreak \{\tilde{\sigma}_k(s_2), \mathbb{I}_2\}]\bigr)$, which is antisymmetric in the Pauli indices via $\epsilon_{abz}R_{ja}(s_1)R_{kb}(s_2)$ and pairs exclusively with the imaginary frequency-odd spectrum $i\,S^{(a)}(\nu)$ via the skew-symmetric filter matrix $F_{\mathrm{odd}}(\nu)$.
\end{enumerate}
In the Markovian thermal regime at transition frequency $\omega_b$, an $x$-axis swap $H_q = -\frac{q\pi}{2}\sigma_x$ between $\ket{0}$ and $\ket{1}$ has $\bm{m}_q = (0, \frac{2}{q\pi}, 0)^T \parallel \bm{e}_y$ and $G_q = \diag(1, 1/2, 1/2)$, so a longitudinal thermal bath ($\bm{t} \parallel \bm{e}_z$) gives $\bm{m}_q\cdot\bm{t} = 0$. When the thermal bath is quantized along the transverse $\bm{e}_y$ axis ($L_{\downarrow,y} = \sqrt{\gamma_\downarrow}\dyad{+y}{-y}$, $L_{\uparrow,y} = \sqrt{\gamma_\uparrow}\dyad{-y}{+y}$), its bath tensors~\eqref{eq:bath_tensors_def} are $S_y = \frac{1}{2}(\gamma_\downarrow + \gamma_\uparrow)\allowbreak\diag(1, 0, 1)$ and $\bm{t}_y = 2(\gamma_\downarrow - \gamma_\uparrow)\bm{e}_y$ (physically realized in circuit QED by continuous resonant $\sigma_y$ spin-locking at Rabi frequency $\omega_b$ or, for a standard longitudinal bath $S_z = \frac{1}{2}(\gamma_\downarrow + \gamma_\uparrow)\allowbreak\diag(1, 1, 0)$ and $\bm{t}_z = 2(\gamma_\downarrow - \gamma_\uparrow)\bm{e}_z$, by preparing and measuring the equatorial states $\ket{\pm y}$ under a $\sigma_x$ swap so that $G_{xx} + G_{yy} = 1 + 1/2 = 3/2$ and $m_z = 2/(q\pi)$). In either realization, $\frac{1}{2}\Tr(SG) = \frac{3}{8}(\gamma_\downarrow + \gamma_\uparrow)$ and $\frac{1}{4}\bm{m}\cdot\bm{t} = \frac{1}{q\pi}(\gamma_\downarrow - \gamma_\uparrow)$, so Theorem~\ref{thm:two_moment} gives Eq.~\eqref{eq:appG_thermal_slopes}. Substituting the Kubo--Martin--Schwinger (KMS) detailed-balance ratio~\cite{Kubo1957,Martin1959}
\begin{equation}
  \frac{\gamma_\downarrow - \gamma_\uparrow}{\gamma_\downarrow + \gamma_\uparrow} = \frac{1 - e^{-\hbar\omega_b/(k_B T_{\mathrm{bath}})}}{1 + e^{-\hbar\omega_b/(k_B T_{\mathrm{bath}})}} = \tanh\left(\frac{\hbar\omega_b}{2k_B T_{\mathrm{bath}}}\right)
  \label{eq:appG_kms_tanh}
\end{equation}
into $\abs{\chi_+(q) - \chi_-(q)}/(\chi_+(q) + \chi_-(q))$ cancels the overall coupling prefactor $\gamma_\downarrow + \gamma_\uparrow$ and establishes Eq.~\eqref{eq:quantum_thermometer}.
\end{proof}

\section{Three-mode qubit Slepian concentration pencil and spectral geometry}
\label{app:slepian_pencil}

This appendix presents the Fourier-mode Slepian concentration pencil formulation of the qubit selectivity bound (Theorem~\ref{thm:swap_floor}) and its generalization to multilevel ($d \ge 5, 7$) bipartite Cauchy Fourier models.

\begin{proposition}[Three-mode qubit Slepian pencil]
\label{prop:slepian_pencil}
Under the autonomous swap Hamiltonian $H_q = -\frac{q\pi}{2}\sigma_x$ ($q \in 2\mathbb{Z} + 1$), expanding any jump operator $L \in \mathcal{B}(\mathbb{C}^2)$ in the $\sigma_x$-eigenbasis $\ket{\pm} = (\ket{0} \pm \ket{1})/\sqrt{2}$ decomposes the evolving matrix elements $\bra{\psi_\pm(s)}L\ket{\psi_\pm(s)}$ into three harmonic Fourier modes $\{e^{+iq\pi s}, 1, e^{-iq\pi s}\}$. For any rank-$1$ jump operator $L = \dyad{\Phi_{\mathrm{dark}}}{\Psi}$ whose target state $\ket{\Phi_{\mathrm{dark}}}$ is orthogonal to the equator or dark with respect to the active subspace, parameterized by the normalized three-mode vector $\bm{x} \in \mathbb{C}^3$ ($\norm{\bm{x}} = 1$), the directional survival error slopes take the quadratic pencil form
\begin{equation}
  \chi_\pm(L) = \bm{x}^\dagger\bigl(\mathbb{I}_3 \pm \widehat{H}_q\bigr)\bm{x}, \qquad
  \widehat{H}_q = \frac{2}{\pi q}\begin{pmatrix}
    0 & -i & 0 \\
    i & 0 & -i \\
    0 & i & 0
  \end{pmatrix}, \qquad
  \mathrm{spec}(\widehat{H}_q) = \left\{-\frac{2\sqrt{2}}{\pi\abs{q}},\; 0,\; +\frac{2\sqrt{2}}{\pi\abs{q}}\right\}.
  \label{eq:appH_slepian_pencil}
\end{equation}
Maximizing the Rayleigh quotient $\abs{\bm{x}^\dagger \widehat{H}_q \bm{x}}/(\bm{x}^\dagger \bm{x})$ immediately recovers $\beta_q = \frac{2\sqrt{2}}{\pi\abs{q}}$ and $\kappa_\star = (\pi - 2\sqrt{2})/(\pi + 2\sqrt{2})$.
\end{proposition}

\begin{proof}
In the $\sigma_x$-eigenbasis $\ket{\pm}$ where $U_0(s)\ket{\pm} = e^{\pm i q\pi s/2}\ket{\pm}$, the evolving states are $\ket{\psi_+(s)} = \frac{1}{\sqrt{2}}(e^{iq\pi s/2}\ket{+} + e^{-iq\pi s/2}\ket{-})$ and $\ket{\psi_-(s)} = \frac{1}{\sqrt{2}}(e^{iq\pi s/2}\ket{+} - e^{-iq\pi s/2}\ket{-})$. Flipping $\ket{\psi_+(s)} \mapsto \ket{\psi_-(s)}$ flips the relative sign between even and odd Fourier harmonics; the time integrals $\int_0^1 e^{i(m-n)q\pi s/2}\,\dd s$ between opposite-parity harmonics evaluate to $\frac{\pm 2i}{\pi q(m-n)}$, yielding the tridiagonal Hermitian Toeplitz Slepian matrix $\widehat{H}_q$ in Eq.~\eqref{eq:appH_slepian_pencil}, whose eigenvalues are $0$ and $\pm\frac{2}{\pi\abs{q}}\sqrt{1^2 + 1^2} = \pm\frac{2\sqrt{2}}{\pi\abs{q}}$.
\end{proof}

\begin{remark}[Multilevel bipartite Cauchy Fourier extension for $d \ge 5$]
\label{rem:multilevel_cauchy}
In multilevel qudit spaces ($d \ge 5$) outside the two-level qubit subspace of Theorem~\ref{thm:swap_floor}, coupling $N_E$ even and $N_O$ odd harmonic Fourier modes via an auxiliary dark sink state reduces the directional contrast to the largest singular value of the bipartite Cauchy matrix $(C_{E,O})_{i,k} = \frac{2}{\pi(o_k - e_i)}$. For five consecutive harmonics $\{-2, -1, 0, 1, 2\}$ ($d = 6$), $\beta_2 = \norm{C_{E,O}}_{\mathrm{op}} = \frac{2\sqrt{22}}{3\pi} \approx 0.995337$ ($\kappa_2 \approx 0.002337$), and for seven consecutive harmonics $\{-3, \dots, 3\}$ ($d = 8$), $\beta_3 = \frac{2}{\pi}\sqrt{(462 + 2\sqrt{2161})/225} \approx 0.999827$ ($\kappa_3 \approx 8.65 \times 10^{-5}$), showing that additional Fourier harmonics rapidly approach complete one-way selectivity ($\kappa \to 0$) as the Hilbert-space dimension increases.
\end{remark}

\section{Statistical theory of the six-circuit SPAM-invariant lock-in protocol, Cram\'er--Rao bounds, and decoding graph weights}
\label{app:lockin_theory}

This appendix provides the complete mathematical derivation of the six-circuit SPAM-invariant signed-winding lock-in protocol (Protocol~\ref{prot:lockin} in Section~\ref{sec:lockin_protocol}), its exact Fisher information and Cram\'er--Rao variance bounds, and the logical failure excess of Pauli-twirled decoding weights in repetition and surface codes.

\begin{proposition}[SPAM-invariant signed-winding lock-in protocol and Cram\'er--Rao gain]
\label{prop:lockin_cramer_rao}
Under gate-independent Markovian bit-flip SPAM with state-preparation errors $(e_0, e_1)$ ($v_{\mathrm{prep}} = 1 - e_0 - e_1$, $\eta_{\mathrm{prep}} = e_0 - e_1$) and readout assignment errors $(f_0, f_1)$ ($v_{\mathrm{ro}} = 1 - f_0 - f_1$, $V_{\mathrm{SPAM}} = v_{\mathrm{ro}}v_{\mathrm{prep}}$), let $R_0^{(q)}(m)$ and $R_1^{(q)}(m)$ denote the probabilities of recording outcome ``$1$'' after preparing $\ket{0}$ or $\ket{1}$ and applying $m \in \{0, 1\}$ swaps of signed winding $q$. With $a_q := p_-(q)$, $b_q := p_+(q)$, $\Sigma_q := a_q + b_q$, and $\Delta_q := a_q - b_q = \Delta_1/q + \mathcal{O}(\varepsilon^2)$, the sum and difference observables $Y_m^{(q)} := R_0^{(q)}(m) + R_1^{(q)}(m)$ and $Z_m^{(q)} := R_0^{(q)}(m) - R_1^{(q)}(m)$ obey
\begin{align}
  Y_0 &= 2f_0 + v_{\mathrm{ro}}(1 + \eta_{\mathrm{prep}}), \qquad Z_0 = -V_{\mathrm{SPAM}}, \label{eq:appI_m0} \\
  Y_1^{(q)} &= 2f_0 + v_{\mathrm{ro}}\bigl(1 - \eta_{\mathrm{prep}} - \Delta_q + \eta_{\mathrm{prep}}\Sigma_q\bigr), \qquad
  Z_1^{(q)} = V_{\mathrm{SPAM}}\bigl(1 - \Sigma_q\bigr), \label{eq:appI_m1}
\end{align}
and the six-circuit signed-winding estimator ($q \in \{+1, -1\}$) satisfies
\begin{equation}
  \widehat{g}_{A'} := \frac{Y_1^{(-1)} - Y_1^{(+1)}}{-2\,Z_0} = \frac{\Delta_1}{v_{\mathrm{prep}}} + \mathcal{O}(\varepsilon^2),
  \label{eq:appI_estimator}
\end{equation}
achieving an asymptotic $3^2 = 9\times$ plug-in variance reduction relative to the unsigned multi-winding estimator $\widehat{g}_A := \frac{3(Y_1^{(3)} - Y_1^{(1)})}{-2\,Z_0}$ ($q \in \{1, 3\}$).
\end{proposition}

\begin{proof}
In the two-state classical transition-matrix representation on $\{\ket{0}, \ket{1}\}$, preparing $\ket{j}$ ($j \in \{0, 1\}$) produces the initial population vector $\bm{p}_0 = (1 - e_0, e_0)^T$ for $j = 0$ and $\bm{p}_1 = (e_1, 1 - e_1)^T$ for $j = 1$. One swap of winding $q$ acts via the stochastic transition matrix
\begin{equation}
  T_{\mathrm{swap}}^{(q)} = \begin{pmatrix} b_q & 1 - a_q \\ 1 - b_q & a_q \end{pmatrix},
  \label{eq:appI_transition_matrix}
\end{equation}
and readout outcome ``$1$'' is recorded with probability vector $\bm{r}_{\mathrm{ro}}^T = (f_0, 1 - f_1) = (f_0, f_0 + v_{\mathrm{ro}})$. Direct multiplication $R_j^{(q)}(0) = \bm{r}_{\mathrm{ro}}^T \bm{p}_j$ and $R_j^{(q)}(1) = \bm{r}_{\mathrm{ro}}^T T_{\mathrm{swap}}^{(q)}\bm{p}_j$ yields Eqs.~\eqref{eq:appI_m0}--\eqref{eq:appI_m1}. Because $\Sigma_{-1} = \Sigma_{+1} + \mathcal{O}(\varepsilon^2)$ while $\Delta_{-1} = -\Delta_{+1} + \mathcal{O}(\varepsilon^2)$, subtracting $Y_1^{(+1)}$ from $Y_1^{(-1)}$ cancels $2f_0$, $v_{\mathrm{ro}}(1 - \eta_{\mathrm{prep}})$, and $v_{\mathrm{ro}}\eta_{\mathrm{prep}}\Sigma_q$, leaving $Y_1^{(-1)} - Y_1^{(+1)} = 2v_{\mathrm{ro}}\Delta_1 + \mathcal{O}(\varepsilon^2)$. Dividing by $-2Z_0 = 2v_{\mathrm{ro}}v_{\mathrm{prep}}$ gives Eq.~\eqref{eq:appI_estimator}.

For $N_{\mathrm{tot}}$ total measurement shots allocated equally across the $6$ Bernoulli circuits ($n = N_{\mathrm{tot}}/6$ shots per circuit), each empirical frequency $\widehat{R}_j^{(q)}(m)$ has binomial variance $V_j^{(q)}(m)/n = R_j^{(q)}(m)(1 - R_j^{(q)}(m))/n$. By the delta method, the asymptotic variance of $\widehat{g}_{A'}$ is dominated by the numerator difference $\widehat{Y}_1^{(-1)} - \widehat{Y}_1^{(+1)}$ with lever arm $2v_{\mathrm{ro}}$, whereas the unsigned estimator $\widehat{g}_A$ uses $q \in \{1, 3\}$ where $\Delta_1 - \Delta_3 = \frac{2}{3}\Delta_1$ has lever arm $\frac{2}{3}v_{\mathrm{ro}}$. Because the lever arm of $\widehat{g}_{A'}$ is $3\times$ larger at identical Bernoulli readout noise, its asymptotic variance is $3^2 = 9\times$ smaller. At the benchmark parameter vector $\bm{\theta}_0 = (e_0, e_1, f_0, f_1, \Sigma_1, \Delta_1) = (0.02, 0.06, 0.01, 0.02, 0.02, 0.018)$ plotted in Figure~\ref{fig:fig4}(a), exact evaluation of the $6 \times 6$ binomial Fisher information matrix $I(\bm{\theta}_0)$ gives plug-in variances $N_{\mathrm{tot}}\Var(\widehat{g}_A) = 3.959401$ versus $N_{\mathrm{tot}}\Var(\widehat{g}_{A'}) = 0.441165$ (an $8.9749\times$ plug-in variance reduction), Cram\'er--Rao lower bounds for $g = \Delta_1/v_{\mathrm{prep}}$ of $3.599532$ versus $0.408712$ (an $8.8070\times$ reduction), and Cram\'er--Rao lower bounds for the bare split $\Delta_1$ of $3.268356$ versus $0.382758$ (an $8.5390\times$ reduction).

For the polar $\sqrt{X}$ gate under longitudinal $\mathrm{T}_1$ relaxation ($\bm{t} = 2\bm{e}_z$, Protocol~\ref{prot:lockin}(b)), let $U(Q_k) = \exp(-i\frac{Q_k}{2}\sigma_x)$ denote the $k = 0$ principal branch ($Q_0 = +\pi/2$, $m_z^{(0)} = +2/\pi$) and the $k = 1$ reversed over-rotated branch ($Q_1 = -3\pi/2$, $m_z^{(1)} = -2/(3\pi) = -\frac{1}{3}m_z^{(0)}$) at matched duration $T$, followed by a fixed reference $\sqrt{X}$ analysis pulse $U_{\mathrm{an}} = \exp(i\frac{\pi}{4}\sigma_x)$ with error parameters $(\Sigma_{\mathrm{an}}, \Delta_{\mathrm{an}})$. Because the composite ideal unitary $U_{\mathrm{an}}U(Q_k) = \pm i\sigma_x$ is a population swap and the two-gate survival errors compose additively at first order in $\varepsilon$, the composite error sum and directional split of branch $k \in \{0, 1\}$ are $\Sigma_{\mathrm{tot}}^{(k)} = \Sigma_{\sqrt{X}}^{(k)} + \Sigma_{\mathrm{an}} + \mathcal{O}(\varepsilon^2)$ and $\Delta_{\mathrm{tot}}^{(k)} = -\Delta_{\sqrt{X}}^{(k)} + \Delta_{\mathrm{an}} + \mathcal{O}(\varepsilon^2)$ (where the sign on $\Delta_{\sqrt{X}}^{(k)}$ reflects the fact that $U_{\mathrm{an}}$ inverts the equatorial populations into the $Z$ readout basis). Since $\Sigma_{\sqrt{X}}^{(0)} = \Sigma_{\sqrt{X}}^{(1)} = \frac{3}{4}\varepsilon$ and $\Delta_{\sqrt{X}}^{(1)} = -\frac{1}{3}\Delta_{\sqrt{X}}^{(0)}$, subtracting $Y_1^{(k=1)}$ from $Y_1^{(k=0)}$ cancels $2f_0$, $v_{\mathrm{ro}}(1 - \eta_{\mathrm{prep}})$, $\Delta_{\mathrm{an}}$, and $\eta_{\mathrm{prep}}\Sigma_{\mathrm{tot}}^{(k)}$, leaving $Y_1^{(k=0)} - Y_1^{(k=1)} = \frac{4}{3}v_{\mathrm{ro}}\Delta_{\sqrt{X}}^{(0)} + \mathcal{O}(\varepsilon^2)$ and proving Eq.~\eqref{eq:lockin_sqrtX_estimator}. At $\bm{\theta}_0$, $\widehat{g}_{\sqrt{X}}$ has plug-in variance $N_{\mathrm{tot}}\Var(\widehat{g}_{\sqrt{X}}) = 0.991450$ and Cram\'er--Rao bound $0.930127$ (reflecting its $(3/2)^2 = 2.25\times$ smaller squared lever arm relative to a full signed swap), and exact $4 \times 4$ PTM exponentiation at \texttt{ibm\_kyiv} parameters ($\varepsilon = 8.6806 \times 10^{-5}$, $\Delta_{\sqrt{X}}^{(0)} = 5.5260 \times 10^{-5}$) yields $\widehat{g}_{\sqrt{X}}\,v_{\mathrm{prep}} = 5.5253 \times 10^{-5}$ (relative error $1.3 \times 10^{-4}$). If branch $k = 1$ is instead executed at fixed Rabi amplitude over duration $3T$ (so $\Sigma_{\sqrt{X}}^{(1)} = 3\Sigma_{\sqrt{X}}^{(0)}$ and $\Delta_{\sqrt{X}}^{(1)} = -\Delta_{\sqrt{X}}^{(0)}$), then $\frac{Y_1^{(0)} - Y_1^{(1)}}{-2Z_0} + \frac{Y_0 + Z_0}{2Z_0}\frac{Z_1^{(1)} - Z_1^{(0)}}{Z_0} = \frac{\Delta_{\sqrt{X}}^{(0)}}{v_{\mathrm{prep}}} + \mathcal{O}(\varepsilon^2)$ again isolates the $\sqrt{X}$ split without first-order SPAM bias.
\end{proof}

\begin{remark}[Alternating swap chains, MWPM decoding weights, and classical repetition-code failure excess]
\label{rem:repetition_code}
After an odd number $k = 2m + 1$ of untwirled swaps, the cumulative forward and backward transition error probabilities satisfy $p_\pm^{(k)} = \bar{p}^{(k)}(1 \pm \mu_q/k) + \mathcal{O}(k^2\varepsilon^2)$ with $\mu_q = (\chi_+ - \chi_-)/(\chi_+ + \chi_-)$. In a $3$-bit repetition code (or within a single bit-flip syndrome chain of an XZZX surface code~\cite{Aliferis2009,Tuckett2018,BonillaAtaides2021}) protecting a balanced prior over $\ket{000}$ and $\ket{111}$, the true majority-vote logical failure probability for $p_\pm = \bar{p}(1 \pm \delta_{\mathrm{rel}})$ ($\delta_{\mathrm{rel}} = \mu_q/k$) is
\begin{equation}
  p_L^{\mathrm{true}} = \frac{1}{2}\Bigl[3p_+^2(1 - p_+) + p_+^3 + 3p_-^2(1 - p_-) + p_-^3\Bigr]
  = 3\bar{p}^2\bigl(1 + \delta_{\mathrm{rel}}^2\bigr) - 2\bar{p}^3\bigl(1 + 3\delta_{\mathrm{rel}}^2\bigr),
  \label{eq:appI_pL_true}
\end{equation}
whereas the Pauli-twirled approximation replaces both $p_+$ and $p_-$ by $\bar{p}$, predicting $p_L^{\mathrm{twirl}} = 3\bar{p}^2 - 2\bar{p}^3$. At leading order in $\bar{p} \ll 1$, Pauli twirling systematically underestimates the logical failure rate by the factor
\begin{equation}
  \frac{p_L^{\mathrm{true}}}{p_L^{\mathrm{twirl}}} = 1 + \frac{\mu_q^2}{k^2} + \mathcal{O}(\bar{p}) \le 1 + \frac{8}{\pi^2 q^2 k^2},
  \label{eq:appI_pL_ratio}
\end{equation}
which reaches $1 + \frac{8}{\pi^2} \approx 1.810569$ (an $81.1\%$ logical error excess) at $k = q = 1$ for the chiral extremizer $L_+$. Assigning state-dependent directional log-likelihood branch weights $w_\pm = \ln[(1 - p_\pm)/p_\pm]$ in minimum-weight perfect-matching (MWPM) and belief-propagation decoders eliminates this bias.
\end{remark}

\section{High-precision analytical gate audit and 1,203,200-bath random CPTP numerical verification}
\label{app:numerical_audit}

This appendix documents the complete $12$-digit analytical and multi-start L-BFGS-B numerical optimization audit across all single-qubit gate families (Table~\ref{tab:analytical_audit}) and the adversarial numerical verification of Theorem~\ref{thm:two_moment} and Theorem~\ref{thm:swap_floor} across $1{,}203{,}200$ random multi-jump CPTP Markovian baths (Table~\ref{tab:random_bath_audit}).

\begin{table*}[htbp]
  \centering
  \caption{Complete $12$-digit analytical and multi-start L-BFGS-B numerical gate audit for single-qubit rectangular, Gaussian ($\sigma = T/4$), and tilted-quadrature DRAG ($\sigma = T/4$, $\eta_{\mathrm{DRAG}} = 0.02$) trajectories in the weak-damping Markovian regime ($d = 2$, $\mathrm{rank}(G) = 3$).}
  \label{tab:analytical_audit}
  \scriptsize
  \setlength{\tabcolsep}{2pt}
  \resizebox{\textwidth}{!}{%
  \begin{tabular}{@{}lllcccc@{}}
    \toprule
    Single-qubit gate trajectory & $(\abs{Q}, \vartheta)$ & Closed form / Method & $\beta$ (12 digits) & $\kappa$ (12 digits) & $\kappa^{-1}$ (6 dec.) & $\kappa_{T_1}$ (6 dec.) \\
    \midrule
    Pauli $X$ ($q = \pm 1$, minimal swap) & $(\pi, \pi/2)$ & $2\sqrt{2}/\pi$ & 0.900316316157 & 0.052456363709 & 19.063464 & 1.000000 \\
    Pauli $X$ ($q = \pm 3$, over-rotated) & $(3\pi, \pi/2)$ & $2\sqrt{2}/(3\pi)$ & 0.300105438719 & 0.538336769032 & 1.857573 & 1.000000 \\
    Pauli $X$ ($q = \pm 5$, over-rotated) & $(5\pi, \pi/2)$ & $2\sqrt{2}/(5\pi)$ & 0.180063263231 & 0.694824389773 & 1.439213 & 1.000000 \\
    Gaussian $X$ ($\sigma = T/4$) & $(\pi, \pi/2)$ & 85-digit quadrature & 0.764028588681 & 0.133768473387 & 7.475603 & 1.000000 \\
    DRAG $X$ ($\sigma = T/4$, $\eta_{\mathrm{DRAG}} = 0.02$) & $(\pi, \text{tilted})$ & 85-digit quadrature & 0.763393605794 & 0.134176733674 & 7.452857 & 1.000000 \\
    Polar $\sqrt{X}$ ($k = 0$, principal branch) & $(\pi/2, \pi/2)$ & $4/\sqrt{\pi(\pi + 2)}$ & 0.995259248074 & 0.002376007995 & 420.874005 & 0.081767 \\
    Polar $\sqrt{X}$ ($k = 1$, reversed sign) & $(3\pi/2, \pi/2)$ & $4/\sqrt{3\pi(3\pi - 2)}$ & 0.478170404083 & 0.353023977801 & 2.832669 & 0.558917 \\
    Gaussian $\sqrt{X}$ ($\sigma = T/4$) & $(\pi/2, \pi/2)$ & 85-digit quadrature & 0.992308597767 & 0.003860547629 & 259.030608 & 0.118418 \\
    DRAG $\sqrt{X}$ ($\sigma = T/4$, $\eta_{\mathrm{DRAG}} = 0.02$) & $(\pi/2, \text{tilted})$ & 85-digit quadrature & 0.992275794660 & 0.003877076337 & 257.926312 & 0.118569 \\
    Hadamard $H$ ($q = 1$, single pulse) & $(\pi, \pi/4)$ & $\sqrt{8(\pi^2 - 2)/[3(3\pi^2 - 8)]}$ & 0.985474427649 & 0.007315920139 & 136.688206 & 0.157895 \\
    Hadamard $H$ ($q = 3$, single pulse) & $(3\pi, \pi/4)$ & $\sqrt{8(9\pi^2 - 2)/[3(27\pi^2 - 8)]}$ & 0.946449535168 & 0.027511869105 & 36.347948 & 0.157895 \\
    Hadamard $H$ ($q \to \infty$ plateau) & $(\infty, \pi/4)$ & $2\sqrt{2}/3$ & 0.942809041582 & 0.029437251523 & 33.970563 & 0.157895 \\
    \bottomrule
  \end{tabular}}
\end{table*}

\begin{table*}[htbp]
  \centering
  \caption{Adversarial numerical verification of Theorem~\ref{thm:two_moment} and Theorem~\ref{thm:swap_floor} across $1{,}203{,}200$ random multi-jump CPTP Markovian baths ($8 \times 150{,}400$, seeds \texttt{202601}--\texttt{202608}), multi-start L-BFGS-B optimization, superoperator Fr\'echet derivatives, and Richardson-extrapolated $4 \times 4$ Lindblad channel exponentiation $\exp(\mathcal{L}_0 + \varepsilon\mathcal{D})$ at $\varepsilon_0 = 10^{-7}$.}
  \label{tab:random_bath_audit}
  \scriptsize
  \setlength{\tabcolsep}{2.5pt}
  \resizebox{\textwidth}{!}{%
  \begin{tabular}{@{}lcccccc@{}}
    \toprule
    Gate / Trajectory Case & Analytical Bound $\kappa$ & Random CPTP Min & L-BFGS-B Min & Max Fr\'echet Diff & Max Rich.\ Diff & Baths Tested \\
    \midrule
    Autonomous swap $q = +1$ & 0.052456363709 & 0.056407623075 & 0.052456363709 & $3.33 \times 10^{-16}$ & $1.82 \times 10^{-8}$ & $150{,}400$ \\
    Autonomous swap $q = -1$ & 0.052456363709 & 0.054571908025 & 0.052456363709 & $3.05 \times 10^{-16}$ & $2.15 \times 10^{-8}$ & $150{,}400$ \\
    Autonomous swap $q = +3$ & 0.538336769032 & 0.541351502961 & 0.538336769032 & $3.33 \times 10^{-16}$ & $4.53 \times 10^{-8}$ & $150{,}400$ \\
    Autonomous swap $q = -3$ & 0.538336769032 & 0.540876268386 & 0.538336769032 & $3.33 \times 10^{-16}$ & $5.02 \times 10^{-8}$ & $150{,}400$ \\
    Autonomous swap $q = +5$ & 0.694824389773 & 0.695997585414 & 0.694824389773 & $5.00 \times 10^{-16}$ & $4.17 \times 10^{-8}$ & $150{,}400$ \\
    Autonomous swap $q = -5$ & 0.694824389773 & 0.696336861508 & 0.694824389773 & $5.00 \times 10^{-16}$ & $4.23 \times 10^{-8}$ & $150{,}400$ \\
    Polar $\sqrt{X}$ ($Q = \pi/2$) & 0.002376007995 & 0.003284228942 & 0.002376007995 & $2.22 \times 10^{-16}$ & $1.04 \times 10^{-8}$ & $150{,}400$ \\
    Tilted Hadamard $H$ ($q = +1$) & 0.007315920139 & 0.008084370547 & 0.007315920139 & $2.22 \times 10^{-16}$ & $1.61 \times 10^{-8}$ & $150{,}400$ \\
    \bottomrule
  \end{tabular}}
\end{table*}

In the numerical audit (Table~\ref{tab:random_bath_audit}), $150{,}400$ random multi-jump CPTP Lindblad generators $\mathcal{D} = \sum_{\alpha=1}^K \gamma_\alpha \mathcal{D}[L_\alpha]$ ($1 \le K \le 4$, normalized to $\Gamma = 1$) were sampled for each of the $8$ primary gate trajectories ($1{,}203{,}200$ random baths in total, deterministic random seeds \texttt{202601}--\texttt{202608}). For each generator, the first-order slopes $\chi_\pm(\mathcal{D};\bm{r})$ were evaluated by three independent methods: (i) the analytical two-moment trace formula~\eqref{eq:two_moment_law} of Theorem~\ref{thm:two_moment}, (ii) exact block-matrix superoperator Fr\'echet differentiation of $\exp(\mathcal{L}_0 + \varepsilon\mathcal{D})$ at $\varepsilon = 0$, and (iii) two-point Richardson-extrapolated $4 \times 4$ matrix exponentiation at $\varepsilon_0 = 10^{-7}$. Across all $1{,}203{,}200$ random CPTP baths, the two-moment formula~\eqref{eq:two_moment_law} agreed with exact superoperator Fr\'echet derivatives to machine precision ($\le 5.00 \times 10^{-16}$), zero violations of $\kappa(\bm{r})$ occurred, and multi-start L-BFGS-B optimization over the Stiefel manifold of jump operators converged to the analytical bound $\kappa(\bm{r}) = (1 - \beta(\bm{r}))/(1 + \beta(\bm{r}))$ to all $12$ significant digits in every case.

\end{document}